\documentclass[12pt]{article}
\usepackage{amsmath,amssymb,amsthm,amsxtra,overpic,bbm,bm,epsfig,subfigure}
\usepackage{hyperref}
\usepackage{mathrsfs}
\usepackage{enumitem}
\usepackage{graphicx}
\usepackage{color}
\usepackage[table]{xcolor}
\usepackage{comment}
\usepackage{epstopdf}
\usepackage{float}
\usepackage{cite}
\usepackage{slashed,stmaryrd}
\usepackage{youngtab}
\usepackage{makecell}

\usepackage{diagbox}
\usepackage{blkarray}

\usepackage[title]{appendix}

\allowdisplaybreaks[2]

\def\thefootnote{\fnsymbol{footnote}}

\hypersetup{
colorlinks =true,
linkcolor  = red,
citecolor  = blue,
urlcolor   = blue
}

\usepackage{multirow}
\usepackage{mathtools}
\usepackage{tikz}
\usepackage[T1]{fontenc}

\newcommand{\Op}{\mathcal{O}}
\newcommand{\Opr}{\mathcal{R}}
\newcommand{\loopf}{\frac{1}{16\pi^2 \varepsilon}}
\newcommand{\looptf}{\frac{1}{\left(16\pi^2 \right)^2 \varepsilon}}
\newcommand{\looptfs}{\frac{1}{\left(16\pi^2 \right)^2 \varepsilon^2}}
\newcommand{\tr}[1]{{\rm Tr}\left(#1\right)}

\newcommand{\rmi}{{\rm i}}

\newcommand{\Dls}{\mbox{$\raisebox{2mm}{\boldmath ${}^\leftarrow$}\hspace{-4mm}\slashed{D}$}}

\newcommand{\dl}{\mbox{$\raisebox{2mm}{\boldmath ${}^\leftarrow$}\hspace{-3.3mm}\slashed{\partial}$}}

\begin{document}


%

\begin{center}
    {\Large\bf Two-loop Renormalization Group Equations in the $\nu$SMEFT}
\end{center}


\begin{center}
{\bf  Di Zhang}~\footnote{E-mail: di1.zhang@tum.de} 
\\
\vspace{0.2cm}
{\small
    Physik-Department, Technische Universität München, James-Franck-Straße, 85748 Garching, Germany}
\end{center}

\vspace{1.5cm}

\begin{abstract}
We calculate two-loop renormalization group equations (RGEs) in the Standard Model Effective Field Theory (SMEFT) with right-handed neutrinos, i.e., the so-called $\nu$SMEFT, up to dimension five. Besides the two-loop RGEs of dimension-five (dim-5) operators, we also present those of the renormalizable couplings, including contributions from dim-5 operators. We check consistency relations among the first and second poles of $\varepsilon \equiv (4-d)/2$ with $d$ being the space-time dimension for all renormalization constants and find that those for lepton doublet and right-handed neutrino wave-function renormalization constants, as well as for renormalization constants of charged-lepton and neutrino Yukawa coupling matrices, do not hold. This leads to divergent RG functions for these fields and Yuwawa coupling matrices. We figure out that such infinite RG functions arise from the non-invariance of fields and Yukawa coupling matrices under field redefinitions, considering that flavor transformations are a kind of linear field redefinitions. Those infinite RG functions will disappear once one restores contributions from the derivative of renormalization constants with respect to the Wilson coefficients of redundant operators or, alternatively, considers the RGEs of flavor invariants, which are physical quantities and remain invariant under field redefinitions.
\end{abstract}


\def\thefootnote{\arabic{footnote}}
\setcounter{footnote}{0}

\newpage

\section{Introduction}

It is believed that the Standard Model (SM) is incomplete and only works as an Effective Field Theory (EFT) at the leading order due to its powerlessness to address some fundamental mysteries, such as the origin of neutrino masses, dark matter, and the cosmic matter-antimatter asymmetry~\cite{ParticleDataGroup:2024cfk,Xing:2020ijf}. No new fundamental particle has been observed since the discovery of the Higgs boson---the last piece in the SM. This encourages us to work in an EFT framework, such as the SMEFT~\cite{Buchmuller:1985jz,Grzadkowski:2010es}, and indirectly search for new physics beyond the SM whose low-energy consequences are embodied in higher-mass-dimensional operators~\cite{Henning:2014wua,Brivio:2017vri,Isidori:2023pyp}. In addition to the extensively discussed SMEFT, its minimal extension with right-handed neutrinos, i.e., the so-called $\nu$SMEFT~\cite{Anisimov:2006hv,delAguila:2008ir,Anisimov:2008gg,Aparici:2009fh}, has also gained increasing attention in recent years and is more closely related to neutrino physics and dark sector of the universe (see, e.g., Refs.~\cite{Abazajian:2012ys,Drewes:2016upu,Abdullahi:2022jlv} for reviews of phenomena resulting from sterile neutrinos with varying mass scales). Besides the SM renormalizable terms and all the SMEFT operators, the $\nu$SMEFT contains additional renormalizable terms and non-renormalizable operators involving right-handed neutrino fields and preserving the SM gauge symmetry. As the basic ingredient for working in the EFT framework, the non-redundant operator basis of the $\nu$SMEFT has been established up to dimension nine~\cite{delAguila:2008ir,Aparici:2009fh,Bhattacharya:2015vja,Liao:2016qyd,Li:2021tsq}. In principle, all these $\nu$SMEFT operators can be generated by integrating out heavy fields in some ultraviolet (UV) completions. The tree-level UV completions involving scalars, fermions, or vectors for the $\nu$SMEFT operators up to dimension seven have been systematically explored in Refs.~\cite{Beltran:2023ymm,Bolton:2025tqw}, and Ref.~\cite{Chala:2020vqp} provides an explicit example of matching an UV model involving light right-handed neutrinos, heavy vector-like fermions and scalar onto the $\nu$SMEFT at the one-loop level. Meanwhile, the one-loop renormalization group equations (RGEs) in the $\nu$SMEFT have been achieved for dimension-five (dim-5)~\cite{Zhang:2024weq} and dim-6~\cite{Chala:2020pbn,Datta:2020ocb,Datta:2021akg,Fuyuto:2024oii,Ardu:2024tzb} operators~\footnote{The one-loop RGEs of bosonic operators up to dimension six in general EFTs have been studied in Refs.~\cite{Fonseca:2025zjb,Misiak:2025xzq,Aebischer:2025zxg}.}.

RGEs play an important role in connecting couplings across different energy scales and resumming large logarithms to improve perturbative convergence. Moreover, operator mixing induced by the RG running may significantly impact global EFT analyses. In this context, the study of RGEs in EFTs is increasingly advancing to the two-loop level~\cite{Jenkins:2023rtg,Jenkins:2023bls,DiNoi:2024ajj,Born:2024mgz,Ibarra:2024tpt,Naterop:2024ydo,Aebischer:2025hsx,Duhr:2025zqw,Haisch:2025lvd}~\footnote{The two-loop RGEs in a general renormalizable gauge theory have been known for four decades~\cite{Machacek:1983tz,Machacek:1983fi,Machacek:1984zw,Luo:2002ti,Schienbein:2018fsw}.}. Such effort leads us to a higher precision order, i.e., the next-to-leading logarithmic (NLL) order, and   together with one-loop matching conditions, the complete NLL corrections guarantee the regularization scheme independence of predictions. In this work, we aim to calculate the complete two-loop RGEs of renormalizable couplings and Wilson coefficients of dim-5 operators in the $\nu$SMEFT with the dimension regularization (DR) and minimal subtraction (MS) schemes~\cite{Bollini:1972ui,tHooft:1972tcz,tHooft:1973mfk}. We retain both the first and second poles of $\varepsilon$ in our calculations. The former provides us with the desired two-loop RGEs, and the latter contributes to a partial cross-check on our calculations via consistency relations~\cite{tHooft:1973mfk,Machacek:1983tz,Jenkins:2023rtg}. However, it turns out that the consistency relations among the first and second poles of $\varepsilon = (4-d)/2$ hold for most renormalization constants, but those for lepton doublet and right-handed neutrino wave-function renormalization constants, as well as the renormalization constants of charged-lepton and neutrino Yukawa coupling matrices, are spoiled if one starts with only physical operators. As a result, the RG functions for lepton doublet and right-handed neutrino fields, as well as those for the charged-lepton and neutrino Yukawa coupling matrices, become divergent. To the best of our knowledge, this is the first instance of infinite RG functions for couplings appearing within the EFT framework. Such infinite RG functions begin to show up at the three-loop level in the SM~\cite{Bednyakov:2014pia,Herren:2017uxn} due to renormalization ambiguity from flavor symmetry~\cite{Herren:2021yur}, and infinite field anomalous dimensions appearing at the two-loop level in an $O(n)$ scalar EFT has been discussed in Ref.~\cite{Manohar:2024xbh} but no divergent RGEs for couplings arise there. Therefore, the other goal of this work is to understand the origin of those infinite RG functions for fields and couplings within the EFT framework and clarify how to deal with them. Moreover, we emphasize that though exploiting fields' equation of motions (EoMs) to convert redundant operators into physical ones is proved to be equivalent to field redefinitions at the leading linear order~\cite{Criado:2018sdb}, the former still has some ambiguities in determining reduction relations. One explicit instance is given in this work. Therefore, one has better directly use field redefinitions to get the reduction relations among redundant and physical operators.

The remainder of this paper is organized as follows. In Sec.~\ref{sec:bfm-re}, we introduce the quantization and renormalization of the $\nu$SMEFT Lagrangian with the background field method (BFM) and present formulae for RG functions and consistency relations in the DR and MS schemes. We discuss redundant and evanescent operators involved in two-loop calculations and show the ambiguity in reduction relations by means of EoMs in Sec.~\ref{sec:nuisance-operators}, and in Sec.~\ref{sec:infrared}, propagators are decomposed by introducing a universal spurious mass to rearrange infrared (IR) divergences, and then, only two types of two-loop massive vacuum integrals are left after reducing tensor integrals. We present two-loop RGEs of all renormalizable couplings and Wilson coefficients in the $\nu$SMEFT and check the consistency relations in Sec.~\ref{sec:two-loop-rges}. Some renormalization constants do not satisfy the consistency relations, leading to infinite RG functions, and hence we explain the origin of those infinite RG functions and show how to handle them in Sec~\ref{sec:infinite-RG}. Finally, we present our conclusions in Sec~\ref{sec:conlusions}.

\section{Background Field Method and Renormalization}\label{sec:bfm-re}

The Lagrangian of the $\nu$SMEFT up to dimension five is~\cite{Anisimov:2006hv,delAguila:2008ir,Anisimov:2008gg,Aparici:2009fh}
\begin{eqnarray}\label{eq:L-nuSMEFT}
    \mathcal{L}^{}_{\nu \rm SMEFT} &=& \mathcal{L}^{}_{\rm SM} + \mathcal{L}^{}_N + \mathcal{L}^{}_{\rm dim-5} \;
\end{eqnarray}
with
\begin{eqnarray}
    \mathcal{L}^{}_{\rm SM} &=& - \frac{1}{4} G^{A}_{\mu\nu} G^{A\mu\nu} - \frac{1}{4} W^I_{\mu\nu} W^{I\mu\nu} - \frac{1}{4} B^{}_{\mu\nu} B^{\mu\nu} + \left( D^{}_\mu H \right)^\dagger \left( D^\mu H \right) - m^2 H^\dagger H - \lambda \left( H^\dagger H \right)^2 
    \nonumber
    \\
    && + \sum^{}_f \overline{f} \rmi \slashed{D} f - \left( \overline{Q^{}_{\rm L}} Y^{}_{\rm u}  \widetilde{H} U^{}_{\rm R} +  \overline{Q^{}_{\rm L}}  Y^{}_{\rm d} H D^{}_{ \rm R}+ \overline{\ell^{}_{ \rm L}} Y^{}_l H E^{}_{ \rm R}+ {\rm h.c.} \right) \;,
    \nonumber
    \\
    \mathcal{L}^{}_{N} &=& \overline{N^{}_{\rm R}} \rmi \slashed{\partial} N^{}_{\rm R} - \left( \frac{1}{2} \overline{N^{\rm c}_{\rm R}} M^{}_N N^{}_{\rm R} + \overline{\ell^{}_{\rm L}} Y^{}_\nu \widetilde{H} N^{}_{\rm R} + {\rm h.c.} \right) \;,
    \nonumber
    \\
    \mathcal{L}^{}_{\rm dim-5} &=& \frac{1}{2} C^{\alpha\beta}_5 \Op^{\alpha\beta}_5 + C^{\alpha\beta}_{HN} \Op^{\alpha\beta}_{HN} + C^{\alpha\beta}_{BN} \Op^{\alpha\beta}_{BN} + {\rm h.c.} \;,
\end{eqnarray}
where $f= Q^{}_{\rm L}, U^{}_{\rm R}, D^{}_{\rm R}, \ell^{}_{\rm L}, E^{}_{\rm R}$ and
\begin{eqnarray}\label{eq:phy-op}
    \Op^{\alpha\beta}_5 = \overline{\ell^{}_{\alpha\rm L}} \widetilde{H} \widetilde{H}^{\rm T} \ell^{\rm c}_{\beta \rm L} \;,\quad \Op^{\alpha\beta}_{HN} = \overline{N^{\rm c}_{\alpha \rm R}} N^{}_{\beta \rm R} H^\dagger H \;,\quad \Op^{\alpha\beta}_{BN} = \overline{N^{\rm c}_{\alpha\rm R}} \sigma^{\mu\nu} N^{}_{\beta\rm R} B^{}_{\mu\nu} \;
\end{eqnarray}
are the three chosen physical dim-5 operators, and their Wilson coefficients $C^{}_5$, $C^{}_{HN}$ and $C^{}_{BN}$ are suppressed by the cut-off scale $\Lambda$ and satisfy $C^{\alpha\beta}_{5} = C^{\beta\alpha}_5$, $C^{\alpha\beta}_{HN} = C^{\beta\alpha}_{HN}$ and  $C^{\alpha\beta}_{BN} = - C^{\beta\alpha}_{BN}$, respectively. Obviously, if there is only one right-handed neutrino, the right-handed neutrino dipole operator $\Op^{}_{BN}$ will vanish. If one switches off the non-renormalizable Lagrangian $\mathcal{L}_{\rm dim-5}$ in Eq.~\eqref{eq:L-nuSMEFT}, it becomes the Lagrangian for the canonical seesaw mechanism~\cite{Minkowski:1977sc,Yanagida:1979as,Gell-Mann:1979vob,Glashow:1979nm,Mohapatra:1979ia} but here the right-handed neutrino mass scale has not been specified and could be much lower than the usual seesaw scales. Moreover, if one removes all terms involving right-handed neutrinos, it turns out to be the SMEFT with the unique Weinberg operator~\cite{Weinberg:1979sa}. On the other hand, the $\nu$SMEFT Lagrangian in Eq.~\eqref{eq:L-nuSMEFT} can describe the case among seesaw scales where one or more heavy right-handed neutrinos are integrated out in the canonical seesaw mechanism~\cite{Zhang:2024weq}. 

One has to quantize the gauge-invariant classical Lagrangian in Eq.~\eqref{eq:L-nuSMEFT} by fixing a gauge when considering physics beyond the tree level. In the conventional formulation, choosing a  gauge breaks the gauge symmetry and leads to gauge non-invariance in intermediate steps of calculations though final results for physical quantities are gauge invariant and independent of the chosen gauge. To avoid explicitly breaking the gauge symmetry, we adopt the BFM~\cite{Abbott:1980hw,Abbott:1981ke,Abbott:1983zw} to quantize the theory, where the gauge invariance remains for background fields even as a gauge is fixed for quantum fields, and Green's functions obey the naive Ward identities of the gauge symmetry instead of the complicated Slavnov-Taylor identities~\cite{Taylor:1971ff,Slavnov:1972fg}. This greatly simplifies the renormalization in the gauge sector. 

With the BFM, all fields are split into background and quantum parts, respectively appearing as external lines (or tree-level propagators) and loop propagators. The following gauge-fixing condition
\begin{eqnarray}\label{eq:gc}
    \mathcal{G}^i_V = D^{}_\mu \hat{V}^{i\mu}= \partial_\mu\hat {V}^{i\mu} - \rmi g_V T^j_{ik} V^j_\mu \hat{V}^{k\mu} 
\end{eqnarray}
preserves the gauge invariance of  the background-field effective action with respect to gauge transformations of background fields~\cite{Abbott:1980hw}. In Eq.~\eqref{eq:gc}, $V = B,W,G$ (or $\hat{V} = \hat{B},\hat{W},\hat{G}$) are background (or quantum) gauge fields, and $g_V$ and $T^j$, with $j$ being adjoint index, are respectively the gauge coupling and the adjoint representation of the gauge group. The gauge-fixing condition in Eq.~\eqref{eq:gc} results in the following gauge-fixing and ghost field terms
\begin{eqnarray}\label{eq:LGF}
    \mathcal{L}^{}_{\rm GF} &=&  - \frac{1}{2\xi_{\hat{B}}} \left( \partial^\mu \hat{B}^{}_\mu \right)^2 - \frac{1}{2\xi_{\hat{W}}}  \left( D^\mu \hat{W}^I_{\mu} \right)^2 - \frac{1}{2\xi_{\hat{G}}}  \left( D^\mu \hat{G}^A_{\mu} \right)^2 \;,
    \nonumber
    \\
    \mathcal{L}^{}_{\rm Ghost} &=& - \overline{\theta}^{}_{\hat{B}} \partial^{}_\mu \partial^\mu \theta^{}_{\hat{B}} - \overline{\theta}^{I}_{\hat{W}} D^{}_\mu \hat{D}^\mu \theta^{I}_{\hat{W}} - \overline{\theta}^{A}_{\hat{G}}D^{}_\mu \hat{D}^\mu \theta^{A}_{\hat{G}} \;,
\end{eqnarray}
where $\xi^{}_{\hat{V}}$ and $\theta^{}_{\hat{V}}$ denote the gauge-fixing parameter and ghost field for the quantum gauge field $\hat{V}$, and $\hat{D}^{}_\mu$ is the covariant derivative containing both background and quantum gauge fields and obtained by substituting $V^{}_\mu+\hat{V}^{}_\mu$ for $V^{}_\mu$ in the background covariant derivative $D^{}_\mu$. Since the ghost field $\theta^{}_{\hat{B}}$ does not couple to any other fields, one can remove its kinetic term from the ghost Lagrangian.  Moreover, the gauge condition in Eq.~\eqref{eq:gc} does not involve fermion and Higgs fields. This indicates that all interactions of the background and quantum fields for fermions and Higgs are exactly the same and indistinguishable. Therefore, there is no need to split fermion and Higgs fields, and besides the gauge-fixing and ghost Lagrangian in Eq.~\eqref{eq:LGF}, one only needs to substitute $V^{}_\mu+\hat{V}^{}_\mu$ for $V^{}_\mu$ in the classical Lagrangian given in Eq.~\eqref{eq:L-nuSMEFT}. 

So far, the Lagrangian is still the bare one and needs to be renormalized. In the BFM, quantum fields and ghost fields are always in loops, and hence, their wave-function renormalization constants associated with propagators and the adjacent vertices are canceled out. Consequently, the renormalization of quantum and ghost fields is unnecessary.  However, gauge fixing parameters for quantum gauge fields still need to be renormalized owing to the unrenormalized longitudinal part of the quantum gauge field propagator. The required renormalization constants for gauge fixing parameters, background fields, and couplings in the $d = 4 -2\varepsilon$ dimensional space-time~\cite{Bollini:1972ui,tHooft:1972tcz} are defined below:
\begin{eqnarray}\label{eq:renorm-cont}
    && \xi^{}_{\hat{V}} = Z^{}_{\xi_{\hat{V}}} \xi^{}_{\hat{V},r} \;,\quad  
    \nonumber
    \\
    && \varphi = Z^{1/2}_\varphi \varphi^{}_r \;,\quad \psi^{}_\alpha = \left( Z^{1/2}_\psi \right)^{}_{\alpha\beta} \psi^{}_{\beta, r} \;,\quad 
    \nonumber
    \\
    && \kappa = \mu^{\rho_\kappa \varepsilon} Z^{}_\kappa \kappa^{}_r \;,\quad  Y^{}_{\alpha\beta} = \mu^\varepsilon \left( Y^{}_{r} \right)^{}_{\alpha\gamma} \left( Z^{}_{Y} \right)^{}_{\gamma\beta} \;,\quad   \mathcal{M}^{\alpha\beta} = \mu^{\rho^{}_{\mathcal{M}}\varepsilon} \left( \mathcal{M}^{\alpha\beta}_{r} + \delta \mathcal{M}^{\alpha\beta} \right) \;,
\end{eqnarray}
with $\hat{V} \in \{ {\hat{B}}, {\hat{W}}, {\hat{G}} \}$, $\varphi \in \{H, B, W, G \}$,  $\psi \in \{ Q^{}_{\rm L}, U^{}_{\rm R}, D^{}_{\rm R}, \ell^{}_{\rm L} , E^{}_{\rm R}, N^{}_{\rm R} \}$, $\kappa \in \{ g^{}_1, g^{}_2, g^{}_3, m^2, \lambda \}$, $Y \in \{ Y^{}_{\rm u}, Y^{}_{\rm d}, Y^{}_l, Y^{}_\nu \}$, $\mathcal{M} \in \{ M^{}_N, C^{}_{5}, C^{}_{HN}, C^{}_{BN} \}$, and $\mu$, $\rho^{}_{\kappa ~(\text{or}~ \mathcal{M})}$ and the subscript $r$ denoting an arbitrary parameter of mass-dimension one, the tree-level anomalous dimension of $\kappa ~(\text{or}~ \mathcal{M})$ and the renormalized quantity, respectively. In Eq.~\eqref{eq:renorm-cont}, flavor indices (e.g., $\alpha,\beta,\gamma$, \dots) are shown explicitly. Thanks to the explicit gauge invariance in the BFM, the renormalization constants for background gauge fields and the corresponding gauge couplings are related by~\cite{Abbott:1980hw}
\begin{eqnarray}
    Z^{}_{g^{}_1} = Z^{-1/2}_B \;,\quad Z^{}_{g^{}_2} = Z^{-1/2}_W \;,\quad Z^{}_{g^{}_3} = Z^{-1/2}_G \;.
\end{eqnarray}
Accordingly, the renormalization of background gauge fields and gauge-fixing parameters is competent to renormalize the whole gauge sector, which is extremely simplified compared with the conventional renormalization of a gauge theory. One may introduce the abbreviation
\begin{eqnarray}
    \delta Z := Z - 1
\end{eqnarray}
for all renormalization constants given in Eq.~\eqref{eq:renorm-cont}. Then, the bare Lagrangian can be written into a renormalized Lagrangian with all counterterms. In the MS scheme~\cite{tHooft:1973mfk}, the explicit expressions of those counterterms or renormalization constants can be achieved by calculating relevant 1PI Green's functions and requiring full cancellation of all divergences at each loop level. They can be written as 
\begin{eqnarray}
    \delta Z^{}_\phi = \sum^{\infty}_{k=1} \frac{z^{(k)}_\phi}{\varepsilon^k}  \;,\quad \delta \mathtt{g}^{}_I  = \sum^{\infty}_{k=1} \frac{a^{(k)}_I}{\varepsilon^k} \qquad {\rm with}\qquad z^{(k)}_\phi = \sum^{}_{L=1} z^{(L,k)}_\phi \;,\quad a^{(k)}_I = \sum^{}_{L=1} a^{(L,k)}_I \;,
\end{eqnarray}
where $L$ is the number of loops, $\phi \in \{ \varphi, \psi \}$ and $\delta \mathtt{g}^{}_I \in \{ \delta Z^{}_\kappa \kappa^{}_r, Y^{}_r \delta Z^{}_Y, \delta \mathcal{M} \}$. Considering the fact that bare quantities are independent of the renormalization scale $\mu$, one can derive couplings' beta functions and field anomalous dimensions together with consistency relations among poles of $\varepsilon$ in the 4-dimension limit, namely~\cite{tHooft:1973mfk,Machacek:1983tz,Jenkins:2023rtg}
\begin{eqnarray}\label{eq:finite-bf}
    \beta^{(0)}_I &=& \mu \frac{{\rm d} \mathtt{g}^{}_I}{{\rm d} \mu}  = 2{\bf L} a^{(1)}_I \;,\qquad \gamma^{(0)}_\phi = \frac{1}{2} \mu \frac{{\rm d} \ln Z^{}_\phi}{{\rm d} \mu} = - {\bf L} z^{(1)}_\phi \;,
\end{eqnarray}
and
\begin{eqnarray}\label{eq:finite-consistency}
    2 {\bf L} a^{(n+1)}_I  &=&  \sum_J \beta^{(0)}_J \partial^J a^{(n)}_I \;,\quad   2{\bf L} z^{(n+1)}_\phi = \sum^{}_{I} \beta^{(0)}_I \partial^I z^{(n)}_\phi - 2 z^{(n)}_\phi \gamma^{(0)}_\phi  \;,  \quad n \geq 1
\end{eqnarray}
in which $\beta^{(0)}_I $ and $\gamma^{(0)}_\phi$ can be parametrized as a loop expansion, i.e., $\beta^{(0)}_I = \sum_L \beta^{(L,0)}_I $ and $\gamma^{(0)}_\phi = \sum_L \gamma^{(L,0)}_\phi$, and the loop operator $\bf L$ is introduced for conciseness and it acts on $x=a^{}_I, z^{}_\phi$ as ${\bf L} x^{(k)} = \sum_L L \cdot x^{(L,k)}$. As can be seen from Eqs.~\eqref{eq:finite-bf} and~\eqref{eq:finite-consistency}, only the first pole of $\varepsilon$ contributes to beta functions and field anomalous dimensions, and higher poles of $\varepsilon$ are iteratively related to the lower ones via consistency relations. However, this is not always the case. As discussed in Refs.~\cite{Bednyakov:2014pia,Herren:2017uxn,Herren:2021yur}, even in the SM, the consistency relations in Eq.~\eqref{eq:finite-consistency} are spoiled, and infinite contributions to RG functions appear at the three-loop level, resulting from the renormalization ambiguity associated with flavor symmetry. We will discuss infinite RG functions appearing in our calculations in Sec.~\ref{sec:infinite-RG}.

\section{Redundant and Evanescent Operators}\label{sec:nuisance-operators}

Besides physical operators, the renormalization of an EFT with gauge symmetry, in general, requires redundant, evanescent, and BRST-exact operators. The first ones are those in the so-called Green's basis~\cite{Jiang:2018pbd} and can be converted to the physical ones by field redefinitions. Even if they do not appear in the Lagrangian initially, they can be generated at the loop level via operator mixing in the off-shell scheme. The second ones result from the Dirac algebra in $d$ dimensional space-time and typically are of $\mathcal{O} \left( \varepsilon \right)$~\cite{Buras:1989xd,Dugan:1990df,Herrlich:1994kh}. Therefore, they are vanishing in the 4-dimension limit and potentially lead to contributions of $\mathcal{O} \left( \varepsilon^0 \right)$ and $\mathcal{O} \left( \varepsilon^{-1} \right)$ at the one- and two-loop level, respectively. As shown in Eq.~\eqref{eq:finite-bf}, the $\mathcal{O} \left( \varepsilon^{-1} \right)$ contributions at the two-loop level will enter RGEs of couplings. The last ones usually arise due to the fact that the gauge symmetry is explicitly broken by gauge-fixing and ghost field terms, whereas the theory still has the BRST symmetry~\cite{Becchi:1975nq,Tyutin:1975qk}. In the BFM, although the background-field effective action is gauge invariant under the background-gauge transformation, such gauge-variant BRST-exact operators are still needed to eliminate the possible sub-divergences~\cite{Misiak:1994zw}. These operators are the BRST variation of some other operators. In our case, the latter must be dim-4 operators with ghost number $-1$ since the BRST variation raises both mass dimension and ghost number by one unit. Fortunately, in the $\nu$SMEFT at $\mathcal{O} \left( \Lambda^{-1} \right)$, no BRST-exact operators can be constructed. Hence, we only need to consider redundant and evanescent operators in our calculations.

Apart from the three physical dim-5 operators in Eq.~\eqref{eq:phy-op}, there are three additional dim-5 operators (and their Hermitian conjugates) in the Green's basis of $\nu$SMEFT, namely~\cite{Zhang:2024weq}
\begin{eqnarray}\label{eq:redundant-opr}
    \Opr^{\alpha\beta}_{DN} = \overline{N^{\rm c}_{\alpha\rm R}} \partial^2 N^{}_{\beta\rm R} \;,\quad \Opr^{\alpha\beta}_{\ell HN1} = \overline{\ell^{}_{\alpha\rm L}} \widetilde{H} \rmi \slashed{\partial} N^{\rm c}_{\beta\rm R} \;,\quad \Opr^{\alpha\beta}_{\ell HN2} = \overline{\ell^{}_{\alpha\rm L}} \rmi \Dls  \widetilde{H} N^{\rm c}_{\beta\rm R} \;
\end{eqnarray}
with their Wilson coefficients $G^{\alpha\beta}_{DN} = G^{\beta\alpha}_{DN}$, $G^{\alpha\beta}_{\ell HN1}$ and $G^{\alpha\beta}_{\ell HN2}$, respectively. Generally, these redundant operators can be dealt with in two ways~\cite{Manohar:2024xbh}. One is to include them in the Lagrangian with their Wilson coefficients being arbitrary and then exploit field redefinitions to convert them and their counterterms to the physical ones after all calculations. The other one is to exclude them from the Lagrangian, i.e., with their Wilson coefficients vanishing, and convert their counterterms induced by operator mixing under renormalization to the physical ones via field redefinitions. These two schemes will lead us to the same results after taking remnant redundant operators' Wilson coefficients in renormalization constants to be zero in the former scheme. In this work, we mainly adopt the latter scheme since there are no redundant operators at the tree level, simplifying calculations. However, as we will see in Sec.~\ref{sec:infinite-RG}, the former scheme plays a significant role in understanding the breaking of consistency relations in Eq.~\eqref{eq:finite-consistency}, which gives rise to infinite anomalous dimensions. Therefore, we will also achieve all one-loop renormalization constants in the first scheme. 

To convert the redundant operators in Eq.~\eqref{eq:redundant-opr} into the physical ones, one can do the following field redefinitions:
\begin{eqnarray}\label{eq:field-redefinition}
    N^{}_{\rm R} &\to& N^{}_{\rm R} + \frac{1}{2} \left( G^\dagger_{DN} M^{}_N + M^\dagger_N G^{}_{DN} \right)N^{}_{\rm R} + \rmi G^\dagger_{DN} \slashed{\partial} N^{\rm c}_{\rm R} +\left(  Y^{}_\nu G^\dagger_{DN} - G^{}_{\ell HN1} \right)^{\rm T} \widetilde{H}^{\rm T} \ell^{\rm c}_{\rm L} \;,
    \nonumber
    \\
    \ell^{}_{\rm L} &\to& \ell^{}_{\rm L} + G^{}_{\ell HN2} \widetilde{H} N^{\rm c}_{\rm R} \;,
\end{eqnarray}
for right-handed neutrino and lepton doublet fields, respectively. They lead us to reduction relations as follows:
\begin{eqnarray}\label{eq:shift}
    M^{\prime}_N &=& M^{}_N + M^{}_N G^{\dagger}_{DN} M^{}_N + \frac{1}{2} \left( M^{}_N M^\dagger_N  G^{}_{DN} + G^{}_{DN} M^\dagger_N M^{}_N \right) \;,
    \nonumber
    \\
    Y^{\prime}_\nu &=& Y^{}_\nu - G^{}_{\ell HN1} M^{}_N + \frac{3}{2} Y^{}_\nu G^{\dagger}_{DN}  M^{}_N + \frac{1}{2} Y^{}_\nu M^\dagger_N  G^{}_{DN}\;,
    \nonumber
    \\
    C^{\prime}_5 &=&  C^{}_5 + \left( G^{}_{\ell HN1} Y^{\rm T}_\nu + Y^{}_\nu G^{\rm T}_{\ell HN1} \right)  -  2 Y^{}_\nu G^{\dagger}_{DN} Y^{\rm T}_\nu \;,
    \nonumber
    \\
    C^{\prime}_{HN} &=& C^{}_{HN} - \frac{1}{2} \left( G^{\dagger}_{\ell HN2} Y^{}_\nu  + Y^{\rm T}_\nu G^{\ast}_{\ell HN2}  \right) \;.
\end{eqnarray}
The above reduction relations are for the bare couplings. One can substitute bare couplings with the renormalized ones together with renormalization constants in Eq.~\eqref{eq:shift} and then obtain the reduction relations for renormalized couplings and counterterms order by order. Let us take $Y^{}_\nu$ as an example: the reduction relation for renormalized $Y^{}_\nu$ is the same as that in Eq.~\eqref{eq:shift}, and its one- and two-loop counterterms are found to be
\begin{eqnarray}\label{eq:shift-ct}
    Y^{\prime}_{\nu} \delta Z^{(L=1)}_{Y^\prime_\nu} &=& Y^{}_{\nu} \delta Z^{(L=1)}_{Y^{}_\nu} - \delta G^{(L=1)}_{\ell HN1} M^{}_N - G^{}_{\ell HN1}  \delta M^{(L=1)}_N + \frac{3}{2} \left( Y^{}_\nu \delta Z^{(L=1)}_{Y^{}_\nu} G^\dagger_{DN} M^{}_N  + Y^{}_\nu \delta G^{(L=1)\dagger}_{DN} M^{}_N \right.
    \nonumber
    \\
    && + \left. Y^{}_\nu G^\dagger_{DN} \delta M^{(L=1)}_N \right) + \frac{1}{2} \left( Y^{}_\nu \delta Z^{(L=1)}_{Y^{}_\nu} M^\dagger_N G^{}_{DN} + Y^{}_\nu \delta M^{(L=1)\dagger}_N G^{}_{DN} + Y^{}_\nu M^\dagger_N \delta G^{(L=1)}_{DN}  \right) \;,
    \nonumber
    \\
    Y^{\prime}_{\nu} \delta Z^{(L=2)}_{Y^\prime_\nu} &=& Y^{}_{\nu} \delta Z^{(L=2)}_{Y^{}_\nu} - \delta G^{(L=2)}_{\ell HN1} M^{}_N - G^{}_{\ell HN1}  \delta M^{(L=2)}_N + \frac{3}{2} \left( Y^{}_\nu \delta Z^{(L=2)}_{Y^{}_\nu} G^\dagger_{DN} M^{}_N  + Y^{}_\nu \delta G^{(L=2)\dagger}_{DN} M^{}_N \right.
    \nonumber
    \\
    && + \left. Y^{}_\nu G^\dagger_{DN} \delta M^{(L=2)}_N \right) + \frac{1}{2} \left( Y^{}_\nu \delta Z^{(L=2)}_{Y^{}_\nu} M^\dagger_N G^{}_{DN} + Y^{}_\nu \delta M^{(L=2)\dagger}_N G^{}_{DN} + Y^{}_\nu M^\dagger_N \delta G^{(L=2)}_{DN}  \right) 
    \nonumber
    \\
    && - \delta G^{(L=1)}_{\ell HN1} \delta M^{(L=1)}_N + \frac{3}{2} \left( Y^{}_\nu \delta Z^{(L=1)}_{Y^{}_\nu} \delta G^{(L=1)\dagger}_{DN} M^{}_N + Y^{}_\nu \delta Z^{(L=1)}_{Y^{}_\nu}  G^{\dagger}_{DN} \delta M^{(L=1)}_N  \right. 
    \nonumber
    \\
    && \left. + Y^{}_\nu  \delta G^{(L=1)\dagger}_{DN} \delta M^{(L=1)}_N  \right) + \frac{1}{2} \left(  Y^{}_\nu \delta Z^{(L=1)}_{Y^{}_\nu} \delta M^{(L=1)\dagger}_N G^{}_{DN} + Y^{}_\nu \delta Z^{(L=1)}_{Y^{}_\nu} M^{\dagger}_N \delta G^{(L=1)}_{DN}  \right. 
    \nonumber
    \\
    && + \left. Y^{}_\nu \delta M^{(L=1)\dagger}_N \delta G^{(L=1)}_{DN} \right) \;.
\end{eqnarray}
In the above equation, we have omitted the subscript $r$ for the renormalized couplings, and we will do so hereafter unless specified otherwise. Working with the second scheme to handle redundant operators, one takes $G^{}_{DN}$, $G^{}_{\ell HN1}$ and $G^{}_{\ell HN2}$ in Eqs.~\eqref{eq:shift} and ~\eqref{eq:shift-ct} to be zero. Then, only the reduction relations for counterterms remain.

Usually, we can alternatively reach the above reduction relations by exploiting the lowest EoMs of heavy neutrino and lepton doublet fields, i.e.,
\begin{eqnarray}\label{eq:EoMs}
    \rmi \slashed{\partial} N^{}_{\rm R} = M^\dagger_N N^{\rm c}_{\rm R} + Y^\dagger_\nu \widetilde{H}^\dagger \ell^{}_{\rm L} \;,\quad  \rmi \slashed{D} \ell^{}_{\rm L} = Y^{}_l H E^{}_{\rm R} + Y^{}_\nu \widetilde{H} N^{}_{\rm R} \;,
\end{eqnarray}
as EoMs are equivalent to the corresponding field redefinitions at the leading linear order~\cite{Criado:2018sdb}. However, when the heavy neutrino fields' EoM in Eq.~\eqref{eq:EoMs} is applied to the redundant operator $\Opr^{}_{DN}$, a flavor-structure ambiguity shows up. More specifically, $\Opr^{}_{DN}$ can be written as $\Opr^{\alpha\beta}_{DN} =\left( \overline{N^{\rm c}_{\alpha\rm R}} \rmi \dl \;\right) \left( \rmi\slashed{\partial} N^{}_{\beta\rm R} \right) $, $- \left[ \left( \overline{N^{\rm c}_{\alpha\rm R}} \rmi \dl \;\right)  \rmi\dl \;\right] N^{}_{\beta\rm R}  $, or $-\overline{N^{\rm c}_{\alpha\rm R}} \left[ \rmi \slashed{\partial} \left( \rmi\slashed{\partial} N^{}_{\beta\rm R} \right)\right] $ with the help of integration by parts, and applying the EoMs in Eq.~\eqref{eq:EoMs} to each form leads to
\begin{eqnarray}\label{eq:EoMs2}
	\left( \overline{N^{\rm c}_{\rm R}} \rmi \dl \;\right) G^{}_{DN} \left( \rmi\slashed{\partial} N^{}_{\rm R} \right) &=& - \left( \overline{N^{}_{\rm R}} M^\dagger_N + \overline{\ell^{\rm c}_{\rm L}} \widetilde{H}^\ast Y^\ast_\nu \right) G^{}_{DN} \left( M^\dagger_N N^{\rm c}_{\rm R} + Y^\dagger_\nu \widetilde{H}^\dagger \ell^{}_{\rm L} \right)
	\nonumber
	\\
	&=& - \overline{N^{}_{\rm R}} \left( M^\dagger_N G^{}_{DN} M^\dagger_N \right) N^{\rm c}_{\rm R} - 2 \overline{N^{}_{\rm R}} \left( M^\dagger_N G^{}_{DN} Y^\dagger_\nu \right) \widetilde{H}^\dagger \ell^{}_{\rm L} 
	\nonumber
	\\
	&& - \overline{\ell^{\rm c}_{\rm L}} \widetilde{H}^\ast \left( Y^\ast_\nu G^{}_{DN} Y^\dagger_\nu \right)  \widetilde{H}^\dagger \ell^{}_{\rm L} \;,
	\nonumber
	\\
	-\left[ \left( \overline{N^{\rm c}_{\rm R}} \rmi \dl \;\right)  \rmi\dl \;\right] G^{}_{DN} N^{}_{\rm R}  &=&  \left[ \left( \overline{N^{}_{\rm R}} M^\dagger_N + \overline{\ell^{\rm c}_{\rm L}} \widetilde{H}^\ast Y^\ast_\nu \right)  \rmi \dl \;\right] G^{}_{DN} N^{}_{\rm R}  
	\nonumber
	\\
	&=&  \left( \overline{N^{}_{\rm R}} \rmi \dl \right) M^\dagger_N G^{}_{DN} N^{}_{\rm R}  -  \overline{\ell^{\rm c}_{\rm L}} \widetilde{H}^\ast Y^\ast_\nu G^{}_{DN} \left( \rmi\slashed{\partial} N^{}_{\rm R} \right)
	\nonumber
	\\
	&=& - \left( \overline{N^{\rm c}_{\rm R}} M^{}_N + \overline{\ell^{}_{\rm L}} \widetilde{H} Y^{}_\nu \right) M^\dagger_N G^{}_{DN} N^{}_{\rm R} -  \overline{\ell^{\rm c}_{\rm L}} \widetilde{H}^\ast Y^\ast_\nu G^{}_{DN} \left( M^\dagger_N N^{\rm c}_{\rm R} + Y^\dagger_\nu \widetilde{H}^\dagger \ell^{}_{\rm L} \right)
	\nonumber
	\\
	&=& - \overline{N^{\rm c}_{\rm R}} \left( M^{}_N M^\dagger_N G^{}_{DN} \right)N^{}_{\rm R} -   \overline{\ell^{}_{\rm L}} \widetilde{H} \left( Y^{}_\nu M^\dagger_N G^{}_{DN} \right) N^{}_{\rm R} 
	\nonumber
	\\
	&& - \overline{N^{}_{\rm R}} \left( M^\dagger_N G^{}_{DN} Y^\dagger_\nu \right) \widetilde{H}^\dagger \ell^{}_{\rm L} - \overline{\ell^{\rm c}_{\rm L}} \widetilde{H}^\ast \left( Y^\ast_\nu G^{}_{DN} Y^\dagger_\nu \right)  \widetilde{H}^\dagger \ell^{}_{\rm L} \;,
	\nonumber
	\\
	-\overline{N^{\rm c}_{\rm R}} G^{}_{DN} \left[ \rmi \slashed{\partial} \left( \rmi\slashed{\partial} N^{}_{\rm R} \right)\right]  &=& -\overline{N^{\rm c}_{\rm R}} G^{}_{DN}\left[ \rmi \slashed{\partial}  \left( M^\dagger_N N^{\rm c}_{\rm R} + Y^\dagger_\nu \widetilde{H}^\dagger \ell^{}_{\rm L} \right) \right] 
	\nonumber
	\\
	&=& -\overline{N^{\rm c}_{\rm R}} G^{}_{DN} M^\dagger_N  \left( \rmi \slashed{\partial} N^{\rm c}_{\rm R} \right) + \left( \overline{N^{\rm c}_{\rm R}} \rmi \dl \;\right) G^{}_{DN} Y^\dagger_\nu \widetilde{H}^\dagger \ell^{}_{\rm L}
	\nonumber
	\\
	&=& -\overline{N^{\rm c}_{\rm R}} G^{}_{DN} M^\dagger_N  \left( M^{}_N N^{}_{\rm R} + Y^{\rm T}_\nu \widetilde{H}^{\rm T} \ell^{\rm c}_{\rm L} \right) - \left( \overline{N^{}_{\rm R}} M^\dagger_N + \overline{\ell^{\rm c}_{\rm L}} \widetilde{H}^\ast Y^\ast_\nu \right) G^{}_{DN} Y^\dagger_\nu \widetilde{H}^\dagger \ell^{}_{\rm L}
	\nonumber
	\\
	&=& -\overline{N^{\rm c}_{\rm R}} \left( G^{}_{DN} M^\dagger_N   M^{}_N \right) N^{}_{\rm R} - \overline{\ell^{}_{\rm L}} \widetilde{H} \left( Y^{}_\nu M^\dagger_N G^{}_{DN} \right) N^{}_{\rm R} 
	\nonumber
	\\
	&& - \overline{N^{}_{\rm R}} \left( M^\dagger_N G^{}_{DN} Y^\dagger_\nu \right) \widetilde{H}^\dagger \ell^{}_{\rm L} - \overline{\ell^{\rm c}_{\rm L}} \widetilde{H}^\ast \left( Y^\ast_\nu G^{}_{DN} Y^\dagger_\nu \right)  \widetilde{H}^\dagger \ell^{}_{\rm L} \;.
\end{eqnarray}
As shown in Eq.~\eqref{eq:EoMs2}, different forms of $\Opr^{}_{DN}$ result in different flavor structures for the right-handed neutrino mass and neutrino Yukawa interaction terms, and none of them produce full contributions from $\Opr^{}_{DN}$ to the reduction relations in Eq.~\eqref{eq:shift} but their linear combination $\Opr^{\alpha\beta}_{DN} = \left( \overline{N^{\rm c}_{\alpha\rm R}} \rmi \dl \;\right) \left( \rmi\slashed{\partial} N^{}_{\beta\rm R} \right) /2 -  \left[ \left( \overline{N^{\rm c}_{\alpha\rm R}} \rmi \dl \;\right)  \rmi\dl \;\right] N^{}_{\beta\rm R} /4 - \overline{N^{\rm c}_{\alpha\rm R}} \left[ \rmi \slashed{\partial} \left( \rmi\slashed{\partial} N^{}_{\beta\rm R} \right)\right] /4 $ does. This indicates that EoMs can not simply produce the same result as that given by field redefinitions, even at the leading linear order where the former is supposed to be equivalent to the latter. Therefore, one had better directly take advantage of field redefinitions to derive the reduction relations between redundant and physical operators, and this avoids not only missing potential contributions but also the flavor-structure ambiguity.

\begin{figure}
    \centering
    \includegraphics[width=0.4\textwidth]{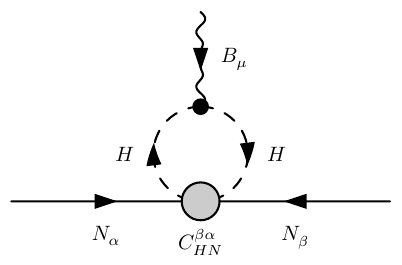}
    \caption{The 1PI diagram with a single insertion of physical dim-5 operators, potentially contributing to the evanescent operator $E^{}_{BN}$.}\label{fig:eva}
\end{figure}

As mentioned at the beginning of this section, evanescent operators are typically of $\mathcal{O}( \varepsilon)$ and may contribute to the first pole $1/\varepsilon$ at the two-loop level due to the existence of the second pole $1/\varepsilon^2$ in the two-loop integrals. Consequently, one must take into account the potential contributions from evanescent operators to two-loop anomalous dimensions. The Dirac structure of dim-5 operators in the $\nu$SMEFT is just a fermion binary and hence there are no evanescent operators resulting from Fierz and Dirac-tensor-reduction identities. The only dim-5 evanescent operator comes from the relation $\epsilon^{}_{\mu\nu\rho\sigma} \sigma^{\rho\sigma} = -2\rmi \sigma^{}_{\mu\nu} \gamma^{}_5 $ in four dimensions and can be defined as~\cite{Fuentes-Martin:2022vvu}
\begin{eqnarray} 	
    E^{\alpha\beta}_{BN} =  \Op^{\alpha\beta}_{BN} - \overline{N^{\rm c}_{\alpha\rm R}} \rmi \sigma^{\mu\nu} N^{}_{\beta\rm R} \widetilde{B}^{}_{\mu\nu}
\end{eqnarray}
with $\widetilde{B}^{}_{\mu\nu} = \epsilon^{}_{\mu\nu\rho\sigma} B^{\rho\sigma}/2$, which vanishes in four dimensions but is of $\mathcal{O} \left( \varepsilon \right) $ in $d$ dimensions. By including contributions from evanescent operators $\{E\}$ to two-loop anomalous dimension of physical operators $\{\Op\}$, the formula in Eq.~\eqref{eq:finite-bf} is generalized to~\cite{Buras:1989xd,Ciuchini:1993vr,Herrlich:1994kh,Chetyrkin:1997fm,Chetyrkin:1997gb}
\begin{eqnarray}
    \beta^{(2,0)}_{\Op^{}_i} = 4 a^{(2,1)}_{\Op^{}_i} \left( \{\Op\}  \right) - 2 \sum^{}_j a^{(1,1)}_{E^{}_j}\left(  \{\Op\}  \right)  \partial^{E^{}_j} a^{(1,0)}_{\Op^{}_i} \left( \{E\} \right) \;,
\end{eqnarray} 
where the first term on the right side is the usual contributions from physical operators, and the second one is those from evanescent operators. $a^{(1,1)}_{E^{}_i}\left(  \{\Op\}  \right) $ is extracted from the one-loop renormalization constant of evanescent operator $E^{}_i$, obtained from one-loop 1PI diagrams with a single insertion of physical operators $\{ \Op \}$. Similarly, $a^{(1,0)}_{\Op^{}_i} \left( \{E\} \right) $ is the finite term in the one-loop renormalization constant of the physical operator $\Op^{}_i$ resulting from a single insertion of evanescent operators $\{E\}$. In the $\nu$SMEFT at $\mathcal{O} \left( \Lambda^{-1} \right)$, there is only one 1PI diagram with single insertion of $\Op^{}_{HN}$ potentially contributing to $a^{(1,1)}_{E^{}_{BN}}$ and shown in Fig.~\ref{fig:eva}. However, it is easy to check that the whole amplitude for the diagram in Fig.~\ref{fig:eva} vanishes and leads to $a^{(1,1)}_{E^{}_{BN}} = 0$. This result can also be understood via the non-renormalization theorem~\cite{Cheung:2015aba}, as the antiholomorphic weight of $E^{}_{BN}$ is smaller than that of $\Op^{}_{HN}$ and then $E^{}_{BN}$ can not be renormalized by $\Op^{}_{HN}$. As a result, no contributions from the evanescent operator $E^{}_{BN}$ to the two-loop anomalous dimension of physical operators will arise in the present case.

\section{Loop Integrals and Infrared Rearrangement}\label{sec:infrared}

In a mass-independent renormalization scheme, e.g., the (modified) MS scheme, all UV divergences are polynomial in masses and momenta~\cite{Collins:1974da}. As a result, one can perform an expansion in masses and external momenta before calculating the UV divergent part of loop integrals. Even one can take all masses and external momenta to zero for calculations of some renormalization constants. In this way, the integrals to be calculated are greatly simplified. However, this leads to the appearance of spurious infrared divergences. This is not a problem for one-loop integrals as UV and IR divergences are not entangled, while it is not the case for higher-loop integrals. Usually, one has to do the IR rearrangement to avoid those IR divergences. In this work, we follow the approach put forward in Ref.~\cite{Chetyrkin:1997fm} to rearrange IR divergences by introducing a single artificial mass and decomposing each propagator~\footnote{Another popular IR rearrangement method is the  $R^{(\ast)}$-operation, see e.g., Refs.~\cite{Chetyrkin:1984xa,Smirnov:1985yck,Herzog:2017bjx}.}. In this way, there are no spurious IR divergences, and all loop integrals become massive vacuum integrals with a single spurious mass. More specifically, one can iteratively decompose a propagator as
\begin{eqnarray}\label{eq:decomposed}
    \frac{1}{\left( k + p \right)^2 - m^2} &=&  \frac{1}{k^2 - M^2} + \frac{m^2 - p^2 -2kp -M^2}{k^2-M^2} \frac{1}{\left( k + p \right)^2 - m^2}
    \nonumber
    \\
    &=&  \frac{1}{k^2 - M^2} + \frac{m^2 - p^2 -2kp}{\left( k^2-M^2 \right)^2}  + \frac{\left( m^2 - p^2 -2kp \right)^2}{\left( k^2-M^2 \right)^3} 
    \nonumber
    \\
    && - \frac{M^2}{\left( k^2-M^2 \right)^2} + \frac{M^4-2M^2\left( m^2 - p^2 - 2kp \right)}{\left( k^2 - M^2 \right)^3}
    \nonumber
    \\
    && +  \frac{\left( m^2 - p^2 -2 kp -M^2 \right)^3 }{\left( k^2- M^2 \right)^3 \left[ \left( k + p \right)^2 - m^2 \right]}
    \nonumber
    \\
    &=& \cdots
\end{eqnarray}
until the overall degree of divergence of the last term is lower than a certain value. In Eq.~\eqref{eq:decomposed}, $k$ (or $p$) is a linear combination of loop (or external) momenta, $m$ and $M$ denote the physical and artificial masses, respectively. After a sufficient number of iterations, the last term in the decomposition becomes free of UV divergences and can be discarded. Moreover, thanks to the exact decomposition, the final result is independent of the spurious mass $M$. In other words, the UV divergences polynomial in $M$ are entirely canceled out. Taking advantage of this property of the decomposition, one can discard all terms whose numerators are proportional to $M^2$ (e.g., those in the third line of Eq.~\eqref{eq:decomposed}) and instead introduce local counterterms proportional to $M^2$ to cancel the corresponding (sub)divergences in integrals without $M^2$ in numerators (e.g., those in the second line of Eq.~\eqref{eq:decomposed})~\cite{Chetyrkin:1997fm}. In our case, i.e., working in the $\nu$SMEFT at $\mathcal{O} \left( \Lambda^{-1} \right)$ with the BFM, such local counterterms turn out to be
\begin{eqnarray}\label{eq:local-cts}
    \delta^S \mathcal{L}^{}_{\rm ct} &=&  \frac{1}{2} \delta Z^{}_{M^{}_{\hat G}} M^2 \hat{G}^A_\mu \hat{G}^{A\mu} +  \frac{1}{2} \delta Z^{}_{M^{}_{\hat W}} M^2 \hat{W}^I_\mu \hat{W}^{I\mu} +  \frac{1}{2} \delta Z^{}_{M^{}_{\hat B}} M^2 \hat{B}^{}_\mu \hat{B}^{\mu} -  \delta Z^{}_{M^{}_{\theta \hat{G}}} M^2 \bar{\theta}^{A}_{\hat{G}} \theta^{A}_{\hat{G}} 
    \nonumber
    \\
    && -  \delta Z^{}_{M^{}_{\theta \hat{W}}} M^2 \bar{\theta}^{I}_{\hat{W}} \theta^{I}_{\hat{W}} - \delta Z^{}_{M^{}_H} M^2 H^\dagger H  - \left( \frac{1}{2} \delta M^{\alpha\beta}_{N^{}_s} M^2 \overline{N^{\rm c}_{\alpha R}} N^{}_{\beta R} + {\rm h.c.} \right) \;,
\end{eqnarray}
in which we do not introduce the counterterm for the $B$-ghost fields since they are decoupled from the Lagrangian. Moreover, such local counterterms for background gauge fields are unnecessary, as background fields do not appear in loops. Note that we do not renormalize quantum gauge and ghost fields, and thereby we need to subtract their wave-function contributions when extracting mass counterterms, $\delta Z_{M_{\hat V}}$ and $\delta Z_{M_{\theta \hat V}}$, from the relevant self-energy amplitudes, while it is unnecessary for $\delta Z^{}_{M_H}$ and $\delta M^{}_{N^{}_s}$. For the counterterm of right-handed neutrinos, $\delta M^{}_{N^{}_s}$ is of $\mathcal{O} \left( \Lambda^{-1} \right)$ according to the mass dimension of the counterterm and hence is supposed to contain one dim-5 Wilson coefficient. If we work at a higher order of the cut-off scale $\Lambda$, more counterterms have to be introduced, and their mass dimension must be at least two units smaller than the maximal dimension of operators~\cite{Chetyrkin:1997fm}.

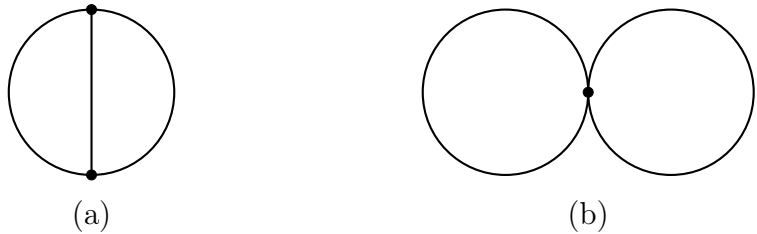
\begin{figure}
\centering
\begin{tikzpicture}[scale=1.1]
    \draw[thick] (0,0) circle (1);
    \draw[thick]  (0,-1) -- (0,1);
    \filldraw (0,-1) circle (0.06);
    \filldraw (0,1) circle (0.06);
    \draw (0,-1.5) node {(a)};
\end{tikzpicture}
\hspace{3cm}
\begin{tikzpicture}[scale=1.1]
    \draw[thick]  (-1,0) circle (1);
    \draw[thick]  (1,0) circle (1);
    \filldraw (0,0) circle (0.06);
    \draw (0,-1.5) node {(b)};
\end{tikzpicture}
\caption{Topologies of the remaining two-loop massive vacuum integrals.}\label{fig:loop}
\end{figure}

After discarding irrelevant terms in the decomposition shown in Eq.~\eqref{eq:decomposed} and reducing tensor integrals, only two types of two-loop massive vacuum integrals remain, whose topologies are shown in Fig.~\ref{fig:loop}. Diagram (a) in Fig.~\ref{fig:loop} corresponds to
\begin{eqnarray}\label{eq:integral}
    K^{(d)}_{\{1,M\},\{1,M\},\{1,M\}} &=& \frac{1}{\pi^{d}}  \int\int  \frac{ {\rm d}^d k^{}_1 {\rm d}^d k^{}_2 }{\left( k^2_1 -M^2 \right) \left( k^2_2 - M^2 \right) \left[ \left( k^{}_1 - k^{}_2 \right)^2 - M^2 \right] }
    \nonumber
    \\
    &=& -\frac{3 M^2}{2 \varepsilon^2} \left[ 1 + \varepsilon \left( 3 - 2 \gamma^{}_{\rm E} - 2 \ln M^2 \right)   \right] + {\rm ~finite~terms} \;,
\end{eqnarray}
where we follow the notations in Ref.~\cite{Mertig:1998vk} and only give the divergent part of this integral~\cite{Davydychev:1992mt,Chetyrkin:1997fm}, and Diagram (b) corresponds to a trivial two-loop integral that is the product of two identical one-loop massive integrals. The latter is given by
\begin{eqnarray}
    A^{(d)}_{\{1,M\}} &=& \frac{1}{\pi^{d/2}} \int \frac{ {\rm d}^d k }{k^2 - M^2} =  -\rmi M^{2-2\varepsilon} \Gamma \left( \varepsilon - 1 \right) \;
\end{eqnarray}
with $\Gamma(x)$ being the gamma function. There is a logarithm of $M^2$ in Eq.~\eqref{eq:integral}, but this kind of logarithm will disappear after subtracting all sub-divergences by including one-loop diagrams with one-loop counterterms.

\section{Two-loop RGEs in the $\nu$SMEFT}\label{sec:two-loop-rges}

Now, we are ready to calculate two-loop RGEs of all couplings and Wilson coefficients in the $\nu$SMEFT at $\Op (\Lambda^{-1})$. We adopt the Feynman-`t Hooft gauge ${\xi_{\hat{V}}} = 1$ (for $\hat{V} = \hat{B}, \hat{W}, \hat{G}$) and the minimal subtraction scheme to calculate all one- and two-loop counterterms in $d$-dimensional space-time, where several Mathematica packages have been exploited. More specifically, we use {\sf FeynRules}~\cite{Christensen:2008py,Alloul:2013bka} to generate Feynman rules including those for counterterms, {\sf FeynArts}~\cite{Hahn:2000kx} to generate Feynman diagrams, {\sf FeynCalc}~\cite{Shtabovenko:2016sxi,Shtabovenko:2020gxv} to calculate amplitudes, and {\sf TARCER}~\cite{Mertig:1998vk} to perform the tensor-integral reduction.

As we aim to derive the two-loop RGEs, one-loop counterterms need to be achieved first and then used to eliminate sub-divergences in two-loop calculations. Although UV divergences in one-loop integrals can be easily extracted without IR rearrangement, we have to do the propagator decomposition shown in Eq.~\eqref{eq:decomposed} in one-loop calculations as well to get the introduced local counterterms in Eq.~\eqref{eq:local-cts}. We collect all one-loop counterterms in the Green's basis in Appendix~\ref{app:one-loop-ct}, which can be used to reproduce the one-loop RGEs in the $\nu$SMEFT at $\Op (\Lambda^{-1})$. With the help of the one-loop results, we calculate all relevant two-loop diagrams, as well as one-loop diagrams with the insertion of one-loop counterterms, to extract all two-loop counterterms~\footnote{The two-loop counterterms for gauge-fixing parameters and those introduced in Eq.~\eqref{eq:local-cts} are irrelevant to two-loop RGEs, and hence we ignore them.}. The two-loop results in the Green's basis are collected in Appendix~\ref{app:two-loop-ct}. We retain both the first and second poles of $\varepsilon$ in all two-loop renormalization constants. The first pole is used to derive two-loop RGEs, while the second pole is used to verify consistency relations. Before doing so, one has to take advantage of the reduction relations induced by field redefinitions, like those in Eq.~\eqref{eq:shift-ct}, to convert counterterms of redundant operators to those of physical ones. 

The one-loop RGEs in the $\nu$SMEFT at $\Op (\Lambda^{-1})$ have been achieved in Ref.~\cite{Zhang:2024weq}~\footnote{The coefficient of the $g^2_1$ term in the RGE of $C^{}_{BN}$ in Eq.~(6) of Ref.~\cite{Zhang:2024weq} should be $41/6$ instead of $1/6$, which has been corrected in the latest arXiv version of Ref.~\cite{Zhang:2024weq}.}. Here we only list two-loop RGEs $\beta^{(2,0)}_i$ with the overall two-loop factor $1/(16\pi^2)^2$ absorbed into $\beta^{(2,0)}_i$: 
\begin{eqnarray}
    \beta^{(2,0)}_{g^{}_1} &=& \hat{\beta}^{(2,0)}_{g^{}_1} -  8 g^2_1 {\rm Tr} \left[ Y^\ast_\nu \left( M^{}_N C^\dagger_{BN} + C^{}_{BN} M^\dagger_N \right) Y^{\rm T}_\nu   \right] \;,
    \\
    \beta^{(2,0)}_{g^{}_2} &=&  \hat{\beta}^{(2,0)}_{g^{}_2} \;,
    \\
    \beta^{(2,0)}_{g^{}_3} &=&  \hat{\beta}^{(2,0)}_{g^{}_3} \;,
    \\
    \beta^{(2,0)}_{m^2} &=&	\hat{\beta}^{(2,0)}_{m^2} + {\rm Tr} \left[ \frac{9}{2}  \left( Y^\dagger_\nu C^{}_5 Y^\ast_\nu M^{}_N +  M^\dagger_N Y^{\rm T}_\nu C^\dagger_5 Y^{}_\nu \right) - \frac{8}{3} \left(  Y^\dagger_\nu Y^{}_\nu M^\dagger_N C^{}_{HN}  +  C^\dagger_{HN} M^{}_N Y^\dagger_\nu Y^{}_\nu \right) \right.
    \nonumber
    \\
    &&  + \left. 28 g^{}_1 \left(  Y^\dagger_\nu Y^{}_\nu M^\dagger_N C^{}_{BN}  + C^\dagger_{BN} M^{}_N Y^\dagger_\nu Y^{}_\nu \right) \right] m^2 - {\rm Tr} \left[ 6 \left( Y^\dagger_\nu C^{}_5 Y^\ast_\nu M^{}_N M^\dagger_N M^{}_N  \right.\right.
    \nonumber
    \\
    && + \left. M^\dagger_N M^{}_N M^\dagger_N Y^{\rm T}_\nu C^\dagger_5 Y^{}_\nu \right) + 12 \left( Y^\dagger_\nu Y^{}_\nu M^\dagger_N M^{}_N M^\dagger_N C^{}_{HN} + C^\dagger_{HN} M^{}_N M^\dagger_N M^{}_N Y^\dagger_\nu Y^{}_\nu \right.
    \nonumber
    \\
    && + \left.  2Y^\dagger_\nu Y^{}_\nu M^\dagger_N C^{}_{HN} M^\dagger_N M^{}_N + 2 M^\dagger_N M^{}_N C^\dagger_{HN} M^{}_N Y^\dagger_\nu Y^{}_\nu \right)  + 4 g^{}_1 \left( 5 Y^\dagger_\nu Y^{}_\nu M^\dagger_N M^{}_N M^\dagger_N C^{}_{BN} \right.
    \nonumber
    \\
    && + \left.\left. 5C^\dagger_{BN} M^{}_N M^\dagger_N M^{}_N Y^\dagger_\nu Y^{}_\nu  + 4 Y^\dagger_\nu Y^{}_\nu M^\dagger_N C^{}_{BN} M^\dagger_NM^{}_N + 4 M^\dagger_N M^{}_N C^\dagger_{BN} M^{}_N Y^\dagger_\nu Y^{}_\nu  \right) \right] \;,
    \\
    \beta^{(2,0)}_\lambda &=& \hat{\beta}^{(2,0)}_\lambda  - \frac{1}{2} \left( g^2_1 + g^2_2 - 10 \lambda \right) {\rm Tr} \left(  C^{}_5 Y^\ast_\nu M^{}_N Y^\dagger_\nu + Y^{}_\nu M^\dagger_N Y^{\rm T}_\nu C^\dagger_5 \right) - 32\lambda {\rm Tr} \left( C^{}_{HN} Y^\dagger_\nu Y^{}_\nu M^\dagger_N  \right.
    \nonumber
    \\
    && + \left. M^{}_N Y^\dagger_\nu Y^{}_\nu C^\dagger_{HN} \right) - 4 g^{}_1 \left( 3g^2_1 + 3g^2_2 + 10 \lambda \right) {\rm Tr} \left( C^{}_{BN} M^\dagger_N Y^{\rm T}_\nu Y^\ast_\nu + Y^{\rm T}_\nu Y^\ast_\nu M^{}_N C^\dagger_{BN}  \right)
    \nonumber
    \\
    && - 12 g^{}_1 {\rm Tr} \left( C^{}_{BN} Y^\dagger_\nu Y^{}_\nu Y^\dagger_\nu Y^{}_\nu M^\dagger_N + M^{}_N  Y^\dagger_\nu Y^{}_\nu Y^\dagger_\nu Y^{}_\nu C^\dagger_{BN} \right) + 2 {\rm Tr} \left[ 2C^{}_{HN} Y^\dagger_\nu Y^{}_l Y^\dagger_l Y^{}_\nu M^\dagger_N \right.
    \nonumber
    \\
    && + 2M^{}_N Y^\dagger_\nu Y^{}_l Y^\dagger_l Y^{}_\nu C^\dagger_{HN} - 28 \left( C^{}_{HN} Y^\dagger_\nu Y^{}_\nu Y^\dagger_\nu Y^{}_\nu M^\dagger_N  +  M^{}_N Y^\dagger_\nu Y^{}_\nu Y^\dagger_\nu Y^{}_\nu C^\dagger_{HN} \right)  
    \nonumber
    \\
    && - 14 \left( C^{}_{HN} Y^\dagger_\nu Y^{}_\nu M^\dagger_N Y^{\rm T}_\nu Y^\ast_\nu + Y^{\rm T}_\nu Y^\ast_\nu M^{}_N Y^\dagger_\nu Y^{}_\nu C^\dagger_{HN} \right) + C^{}_5 Y^\ast_\nu M^{}_\nu Y^\dagger_\nu Y^{}_l Y^\dagger_l + Y^{}_l Y^\dagger_l Y^{}_\nu M^\dagger_N Y^{\rm T}_\nu C^\dagger_5 
    \nonumber
    \\ 
    && - \left. 15  \left( C^{}_5 Y^\ast_\nu M^{}_N Y^\dagger_\nu Y^{}_\nu Y^\dagger_\nu  + Y^{}_\nu Y^\dagger_\nu Y^{}_\nu M^\dagger_N Y^{\rm T}_\nu C^\dagger_5 \right) \right] \;.
    \\
    \beta^{(2,0)}_{Y^{}_{\rm d}} &=& \hat{\beta}^{(2,0)}_{Y^{}_{\rm d}}  + {\rm Tr} \left[ \frac{9}{4} \left( Y^{}_\nu M^\dagger_N Y^{\rm T}_\nu C^\dagger_5 + C^{}_5 Y^\ast_\nu M^{}_N Y^\dagger_\nu \right) + 4Y^\dagger_\nu Y^{}_\nu M^\dagger_N C^{}_{HN} + 4C^\dagger_{HN} M^{}_N Y^\dagger_\nu Y^{}_\nu  \right.
    \nonumber
    \\
    && + \left. 10 g^{}_1 \left( Y^\dagger_\nu Y^{}_\nu M^\dagger_N C^{}_{BN} + C^\dagger_{BN} M^{}_N Y^\dagger_\nu Y^{}_\nu \right) \right] Y^{}_{\rm d} \;,
    \\
    \beta^{(2,0)}_{Y^{}_{\rm u}} &=& \hat{\beta}^{(2,0)}_{Y^{}_{\rm u}}  + {\rm Tr} \left[ \frac{9}{4} \left( Y^{}_\nu M^\dagger_N Y^{\rm T}_\nu C^\dagger_5 + C^{}_5 Y^\ast_\nu M^{}_N Y^\dagger_\nu \right) + 4 Y^\dagger_\nu Y^{}_\nu M^\dagger_N C^{}_{HN} + 4 C^\dagger_{HN} M^{}_N Y^\dagger_\nu Y^{}_\nu  \right.
    \nonumber
    \\
    && + \left. 10 g^{}_1 \left( Y^\dagger_\nu Y^{}_\nu M^\dagger_N C^{}_{BN} + C^\dagger_{BN} M^{}_N Y^\dagger_\nu Y^{}_\nu \right)  \right] Y^{}_{\rm u} \;,
    \\
    \beta^{(2,0)}_{Y^{}_l} &=& \hat{\beta}^{(2,0)}_{Y^{}_l}  + {\rm Tr} \left[ 10 g^{}_1 Y^{}_\nu  \left( M^\dagger_N C^{}_{BN} + C^\dagger_{BN} M^{}_N \right) Y^\dagger_\nu + 4Y^{}_\nu \left( M^\dagger_N C^{}_{HN} + C^\dagger_{HN} M^{}_N \right) Y^\dagger_\nu + \right.
    \nonumber
    \\
    && + \left.\frac{9}{4} \left( C^{}_5 Y^\ast_\nu M^{}_N Y^\dagger_\nu + Y^{}_\nu M^\dagger_N Y^{\rm T}_\nu C^\dagger_5 \right) \right] Y^{}_l + g^{}_1 Y^{}_\nu \left( 31 M^\dagger_N C^{}_{BN} - 19 C^\dagger_{BN} M^{}_N \right) Y^\dagger_\nu Y^{}_l 
    \nonumber
    \\
    && - Y^{}_\nu \left(2 M^\dagger_N C^{}_{HN} + 8 C^\dagger_{HN} M^{}_N \right) Y^\dagger_\nu Y^{}_l  - \frac{21}{8} \left( C^{}_5 Y^\ast_\nu M^{}_N Y^\dagger_\nu Y^{}_l + Y^{}_\nu M^{}_N Y^{\rm T}_\nu C^\dagger_5 Y^{}_l \right) \;,
    \\
    \beta^{(2,0)}_{Y^{}_\nu} &=& \hat{\beta}^{(2,0)}_{Y^{}_\nu} + {\rm Tr} \left[ 10 g^{}_1 Y^{}_\nu \left( M^\dagger_N C^{}_{BN} + C^\dagger_{BN} M^{}_N \right) Y^\dagger_\nu + 4Y^{}_\nu \left( M^\dagger_N C^{}_{HN} + C^\dagger_{HN} M^{}_N \right) Y^\dagger_\nu \right.
    \nonumber
    \\
    && + \left. \frac{9}{4} \left( C^{}_5 Y^\ast_\nu M^{}_N Y^\dagger_\nu + Y^{}_\nu M^\dagger_N Y^{\rm T}_\nu C^\dagger_5 \right) \right] Y^{}_\nu - \left( \frac{45}{4} g^2_1 + 24g^2_2 - 3 \lambda - \frac{9}{2} T \right) C^{}_5 Y^\ast_\nu M^{}_N  
    \nonumber
    \\
    && - \left( g^2_1 + 3g^2_2 - 4T - 24\lambda  \right) Y^{}_\nu M^\dagger_N C^{}_{HN}  + 12 \lambda Y^{}_\nu C^\dagger_{HN} M^{}_N 
    \nonumber
    \\
    && +  g^{}_1 \left[  - \left( \frac{91}{2} g^2_1 + \frac{27}{2} g^2_2 - 6 T \right) Y^{}_\nu C^\dagger_{BN} M^{}_N - \left( \frac{370}{3} g^2_1 + 36 g^2_2 \right) Y^{}_\nu M^\dagger_N C^{}_{BN}  \right.
    \nonumber
    \\
    && + \left( 12 Y^{}_l Y^\dagger_l - 19Y^{}_\nu Y^\dagger_\nu \right) Y^{}_\nu M^\dagger_N C^{}_{BN} +Y^{}_\nu \left( \frac{11}{2} M^\dagger_N C^{}_{BN} + 2 C^\dagger_{BN} M^{}_N \right) Y^\dagger_\nu Y^{}_\nu 
    \nonumber
    \\
    && -  \left.  \frac{69}{2} Y^{}_\nu M^\dagger_N Y^{\rm T}_\nu Y^\ast_\nu C^{}_{BN} - 9 Y^{}_l Y^\dagger_l Y^{}_\nu C^\dagger_{BN} M^{}_N + \frac{3}{2} Y^{}_\nu Y^\dagger_\nu Y^{}_\nu C^\dagger_{BN} M^{}_N - \frac{21}{2} Y^{}_\nu C^\dagger_{BN} Y^{\rm T}_\nu Y^\ast_\nu M^{}_N \vphantom{\frac{1}{6}}\right] 
    \nonumber
    \\
    && + 9 Y^{}_\nu Y^\dagger_\nu Y^{}_\nu  M^\dagger_N C^{}_{HN} + \frac{1}{2} Y^{}_\nu Y^\dagger_\nu Y^{}_\nu C^\dagger_{HN} M^{}_N  + 3 Y^{}_l Y^\dagger_l Y^{}_\nu C^\dagger_{HN} M^{}_N  - \frac{5}{2} Y^{}_\nu C^\dagger_{HN} Y^{\rm T}_\nu Y^\ast_\nu M^{}_N 
    \nonumber
    \\
    && - 3 Y^{}_\nu C^\dagger_{HN} M^{}_N Y^\dagger_\nu Y^{}_\nu - \frac{1}{2} Y^{}_\nu M^\dagger_N C^{}_{HN} Y^\dagger_\nu Y^{}_\nu + \frac{11}{2} Y^{}_\nu M^\dagger_N Y^{\rm T}_\nu Y^\ast_\nu C^{}_{HN} - \frac{45}{8} C^{}_5 Y^\ast_\nu M^{}_N Y^\dagger_\nu Y^{}_\nu 
    \nonumber
    \\
    && + \frac{3}{8} Y^{}_\nu M^\dagger_N Y^{\rm T}_\nu C^\dagger_5 Y^{}_\nu + \frac{3}{4} C^{}_5 Y^\ast_l Y^{\rm T}_l Y^\ast_\nu M^{}_N + \frac{9}{4} Y^{}_\nu Y^\dagger_\nu C^{}_5 Y^\ast_\nu M^{}_N + 3 C^{}_5 Y^\ast_\nu Y^{\rm T}_\nu Y^\ast_\nu M^{}_N  \;,
    \\
    \beta^{(2,0)}_{M_N} &=&  \hat{\beta}^{(2,0)}_{M_N} - m^2 \left[ 16 \left( g^2_1 +3g^2_2 - T \right) C^{}_{HN}  + 28 g^{}_1  \left(C^{}_{BN} Y^\dagger_\nu Y^{}_\nu - Y^{\rm T}_\nu Y^\ast_\nu C^{}_{BN} \right) - 4 C^{}_{HN} Y^\dagger_\nu Y^{}_\nu \right.
    \nonumber
    \\
    && - \left. 4Y^{\rm T}_\nu Y^\ast_\nu C^{}_{HN} \right]  + 8 C^{}_{HN} M^\dagger_N M^{}_N Y^\dagger_\nu Y^{}_\nu + 8 \left( Y^\dagger_\nu Y^{}_\nu \right)^{\rm T} M^{}_N M^\dagger_N C^{}_{HN} + C^{}_{HN} M^\dagger_N \left( Y^\dagger_\nu Y^{}_\nu \right)^{\rm T} M^{}_N 
    \nonumber
    \\
    && +  M^{}_N Y^\dagger_\nu Y^{}_\nu M^\dagger_N C^{}_{HN}  +  \left( Y^\dagger_\nu Y^{}_\nu \right)^{\rm T} M^{}_N C^\dagger_{HN} M^{}_N +  M^{}_N C^\dagger_{HN} M^{}_N Y^\dagger_\nu Y^{}_\nu - M^{}_N Y^\dagger_\nu Y^{}_\nu C^\dagger_{HN} M^{}_N
    \nonumber
    \\
    &&  - M^{}_N C^\dagger_{HN} Y^{\rm T}_\nu Y^\ast_\nu M^{}_N  - \frac{1}{2}  M^{}_N M^\dagger_N C^{}_{HN} Y^\dagger_\nu Y^{}_\nu  - \frac{1}{2}  M^{}_N M^\dagger_N Y^{\rm T}_\nu Y^\ast_\nu C^{}_{HN} - \frac{1}{2} C^{}_{HN} Y^\dagger_\nu Y^{}_\nu  M^\dagger_N M^{}_N 
    \nonumber
    \\
    && - \frac{1}{2}  Y^{\rm T}_\nu Y^\ast_\nu C^{}_{HN} M^\dagger_N M^{}_N  -  g^{}_1 \left[ 24C^{}_{BN} M^\dagger_N M^{}_N Y^\dagger_\nu Y^{}_\nu  - 24\left( Y^\dagger_\nu Y^{}_\nu \right)^{\rm T} M^{}_N M^\dagger_N C^{}_{BN} \right.
    \nonumber
    \\
    && +  C^{}_{BN} M^\dagger_N \left( Y^\dagger_\nu Y^{}_\nu \right)^{\rm T} M^{}_N - M^{}_N Y^\dagger_\nu Y^{}_\nu M^\dagger_N C^{}_{BN} - M^{}_N C^\dagger_{BN} M^{}_N Y^\dagger_\nu Y^{}_\nu + \left( Y^\dagger_\nu Y^{}_\nu \right)^{\rm T} M^{}_N C^\dagger_{BN} M^{}_N 
    \nonumber
    \\
    && + 9 M^{}_N C^\dagger_{BN} Y^{\rm T}_\nu Y^\ast_\nu M^{}_N  - 9 M^{}_N Y^\dagger_\nu Y^{}_\nu C^\dagger_{BN} M^{}_N - \frac{9}{2} M^{}_N M^\dagger_N \left( C^{}_{BN} Y^\dagger_\nu Y^{}_\nu - Y^{\rm T}_\nu Y^\ast_\nu C^{}_{BN} \right) 
    \nonumber
    \\
    && - \left.  \frac{9}{2}  \left( C^{}_{BN} Y^\dagger_\nu Y^{}_\nu - Y^{\rm T}_\nu Y^\ast_\nu C^{}_{BN} \right) M^\dagger_N M^{}_N \right] \;,
\end{eqnarray}
for renormalizable couplings, and
\begin{eqnarray}\label{eq:two-loop-rge-c5}
    \beta^{(2,0)}_{C_5} &=& \left[ - \frac{129}{8} g^4_1 - \frac{169}{24} g^4_2 - \frac{83}{4} g^2_1 g^2_2 + g^2_1 {\rm Tr} \left( \frac{5}{4} Y^{}_\nu Y^\dagger_\nu + \frac{25}{4} Y^{}_l Y^\dagger_l + \frac{25}{12} Y^{}_{\rm d} Y^\dagger_{\rm d} + \frac{85}{12} Y^{}_{\rm u} Y^\dagger_{\rm u} \right) \right.
    \nonumber
    \\
    && + \frac{15}{4} g^2_2 T + 40 g^2_3 {\rm Tr} \left( Y^{}_{\rm u} Y^\dagger_{\rm u} + Y^{}_{\rm d} Y^\dagger_{\rm d} \right) - 4\lambda g^2_1 - 28 \lambda^2  - 8\lambda T - \frac{1}{2} T^\prime +  {\rm Tr} \left( Y^{}_\nu Y^\dagger_\nu Y^{}_l Y^\dagger_l \right.
    \nonumber
    \\
    && + \left.\left. 3 Y^{}_{\rm u} Y^\dagger_{\rm u} Y^{}_{\rm d} Y^\dagger_{\rm d} \right) \right] C^{}_5 + \left( - \frac{57}{16} g^2_1 + \frac{33}{16} g^2_2 + \frac{5}{4} T  \right) \left[ Y^{}_l Y^\dagger_l C^{}_5 + C^{}_5 \left( Y^{}_l Y^\dagger_l \right)^{\rm T} \right]
    \nonumber
    \\
    && + \left( \frac{203}{16} g^2_1 + \frac{381}{16} g^2_2 - 29 \lambda - \frac{21}{4} T  \right) \left[ Y^{}_\nu Y^\dagger_\nu C^{}_5 + C^{}_5 \left( Y^{}_\nu Y^\dagger_\nu \right)^{\rm T} \right]  + 2 Y^{}_l  Y^\dagger_l C^{}_5 \left( Y^{}_l Y^\dagger_l \right)^{\rm T} 
    \nonumber
    \\
    && -\frac{5}{2} Y^{}_\nu  Y^\dagger_\nu C^{}_5 \left( Y^{}_\nu Y^\dagger_\nu \right)^{\rm T} + \frac{19}{4}   Y^{}_l Y^\dagger_l Y^{}_l Y^\dagger_l C^{}_5 + \frac{19}{4} C^{}_5 \left( Y^{}_l Y^\dagger_l Y^{}_l Y^\dagger_l  \right)^{\rm T} - Y^{}_l Y^\dagger_l Y^{}_\nu Y^\dagger_\nu C^{}_5 
    \nonumber
    \\
    && - C^{}_5 \left( Y^{}_l Y^\dagger_l Y^{}_\nu Y^\dagger_\nu  \right)^{\rm T} + \frac{3}{4} Y^{}_\nu Y^\dagger_\nu Y^{}_\nu Y^\dagger_\nu C^{}_5 + \frac{3}{4} C^{}_5 \left( Y^{}_\nu Y^\dagger_\nu Y^{}_\nu Y^\dagger_\nu  \right)^{\rm T} - \frac{3}{4} C^{}_5 Y^\ast_l Y^{\rm T}_l Y^\ast_\nu Y^{\rm T}_\nu 
    \nonumber
    \\
    &&  - \frac{3}{4} Y^{}_\nu Y^\dagger_\nu Y^{}_l  Y^\dagger_l C^{}_5 - 40 \lambda Y^{}_\nu C^\dagger_{HN} Y^{\rm T}_\nu + 5 Y^{}_\nu Y^\dagger_\nu Y^{}_\nu C^\dagger_{HN} Y^{\rm T}_\nu + 5 Y^{}_\nu C^\dagger_{HN} Y^{\rm T}_\nu Y^\ast_\nu Y^{\rm T}_\nu 
    \nonumber
    \\
    && - 3 Y^{}_l Y^\dagger_l Y^{}_\nu C^\dagger_{HN} Y^{\rm T}_\nu  - 3 Y^{}_\nu C^\dagger_{HN} Y^{\rm T}_\nu Y^\ast_l Y^{\rm T}_l - 9 g^{}_1 \left( Y^{}_\nu C^\dagger_{BN} Y^{\rm T}_\nu Y^\ast_\nu Y^{\rm T}_\nu \right.
    \nonumber
    \\
    && - \left.  Y^{}_\nu Y^\dagger_\nu Y^{}_\nu C^\dagger_{BN} Y^{\rm T}_\nu + Y^{}_\nu C^\dagger_{BN} Y^{\rm T}_\nu Y^\ast_l Y^{\rm T}_l - Y^{}_l Y^\dagger_l Y^{}_\nu C^\dagger_{BN} Y^{\rm T}_\nu  \right) \;,
    \\
    \beta^{(2,0)}_{C_{HN}} &=& \left[ \frac{557}{48} g^4_1 - \frac{145}{16} g^4_2 + \frac{15}{8} g^2_1 g^2_2 + 4 \lambda \left( 6g^2_1 + 18g^2_2 - 15\lambda - 6T \right) + \frac{5}{12} g^2_1 {\rm Tr} \left( 3Y^{}_\nu Y^\dagger_\nu + 15 Y^{}_l Y^\dagger_l \right.\right.
    \nonumber
    \\
    && + \left. 5 Y^{}_{\rm d} Y^\dagger_{\rm d} + 17 Y^{}_{\rm u} Y^\dagger_{\rm u} \right) + \frac{15}{4} g^2_2 T + 40 g^2_3 {\rm Tr} \left( Y^{}_{\rm u} Y^\dagger_{\rm u}  + Y^{}_{\rm d} Y^\dagger_{\rm d} \right) - \frac{9}{2} T^\prime - 7 {\rm Tr} \left( Y^{}_\nu Y^\dagger_\nu Y^{}_l Y^\dagger_l \right.
    \nonumber
    \\
    && + \left.\left. 3 Y^{}_{\rm u} Y^\dagger_{\rm u} Y^{}_{\rm d} Y^\dagger_{\rm d} \right)  \right] C^{}_{HN}  + \left[ \frac{41}{8} \left( g^2_1 + 3g^2_2 \right) - 45 \lambda - 5 T \right] \left[ C^{}_{HN} Y^\dagger_\nu Y^{}_\nu + \left( Y^\dagger_\nu Y^{}_\nu \right)^{\rm T} C^{}_{HN} \right] 
    \nonumber
    \\
    && + 5 \left[ C^{}_{HN} Y^\dagger_\nu Y^{}_l Y^\dagger_l Y^{}_\nu + \left( Y^\dagger_\nu Y^{}_l Y^\dagger_l Y^{}_\nu \right)^{\rm T} C^{}_{HN} \right] - \frac{25}{4} \left[ C^{}_{HN} Y^\dagger_\nu Y^{}_\nu Y^\dagger_\nu Y^{}_\nu + \left( Y^\dagger_\nu Y^{}_\nu Y^\dagger_\nu Y^{}_\nu \right)^{\rm T} C^{}_{HN} \right] 
    \nonumber
    \\
    && + \frac{5}{2} \left( Y^\dagger_\nu Y^{}_\nu \right)^{\rm T} C^{}_{HN} Y^\dagger_\nu Y^{}_\nu - \left[ \frac{9}{4} \left( g^2_1 + g^2_2 \right) + 18 \lambda \right] Y^{\rm T}_\nu C^\dagger_5 Y^{}_\nu + \frac{9}{2} Y^{\rm T}_\nu C^\dagger_5 Y^{}_\nu Y^\dagger_\nu Y^{}_\nu 
    \nonumber
    \\
    && + \frac{9}{2}Y^{\rm T}_\nu Y^\ast_\nu Y^{\rm T} C^\dagger_5 Y^{}_\nu  + g^{}_1 \left( \frac{11}{24} g^2_1 - \frac{63}{8} g^2_2 +  42 \lambda + 3 T \right) \left[ C^{}_{BN} Y^\dagger_\nu Y^{}_\nu -  \left( Y^\dagger_\nu Y^{}_\nu \right)^{\rm T} C^{}_{BN} \right] 
    \nonumber
    \\
    && - \frac{21}{4} g^{}_1 \left[ C^{}_{BN} Y^\dagger_\nu Y^{}_l Y^\dagger_l Y^{}_\nu - \left( Y^\dagger_\nu Y^{}_l Y^\dagger_l Y^{}_\nu \right)^{\rm T} C^{}_{BN} \right] + 15 g^{}_1 \left[ C^{}_{BN} Y^\dagger_\nu Y^{}_\nu Y^\dagger_\nu Y^{}_\nu \right.
    \nonumber
    \\
    && - \left. \left( Y^\dagger_\nu Y^{}_\nu Y^\dagger_\nu Y^{}_\nu \right)^{\rm T} C^{}_{BN} \right]  \;,
    \\
    \beta^{(2,0)}_{C_{BN}} &=& \left[ \frac{199}{18} g^4_1 + \frac{9}{2} g^2_1 g^2_2 + \frac{44}{3} g^2_1 g^2_3 - \frac{1}{6} g^2_1 {\rm Tr} \left( 3 Y^{}_\nu Y^\dagger_\nu + 15 Y^{}_l Y^\dagger_l + 5 Y^{}_{\rm d} Y^\dagger_{\rm d} + 17 Y^{}_{\rm u} Y^\dagger_{\rm u} \right) \right] C^{}_{BN}
    \nonumber
    \\
    && + \left[ \frac{1}{8} \left( 35g^2_1 + 51 g^2_2 \right) - \frac{3}{2} T \right] \left( C^{}_{BN} Y^\dagger_\nu Y^{}_\nu + Y^{\rm T}_\nu Y^\ast_\nu C^{}_{BN} \right) - \frac{1}{4}  g^{}_1  \left( C^{}_{HN} Y^\dagger_\nu Y^{}_\nu  \right.
    \nonumber
    \\
    && - \left. Y^{\rm T}_\nu Y^\ast_\nu C^{}_{HN} \right) - \frac{1}{4} Y^{\rm T}_\nu \left( Y^\ast_\nu Y^{\rm T}_\nu + Y^\ast_l Y^{\rm T}_l \right) Y^{}_\nu C^{}_{BN} - \frac{1}{4} C^{}_{BN} Y^\dagger_\nu \left( Y^{}_\nu Y^\dagger_\nu + Y^{}_l Y^\dagger_l \right) Y^{}_\nu 
    \nonumber
    \\
    && - 4\left( Y^\dagger_\nu Y^{}_\nu \right)^{\rm T} C^{}_{BN} Y^\dagger_\nu Y^{}_\nu \;,
\end{eqnarray}
for non-renormalizable Wilson coefficients, where the two-loop RGEs $\hat{\beta}^{(2,0)}_i$ only involve renormalizable couplings and are the very ones in the type-I seesaw model. They are found to be
\begin{eqnarray}\label{eq:renormalizable-rge-g1}
    \hat{\beta}^{(2,0)}_{g^{}_1} &=& \frac{1}{6} g^3_1 \left[ \frac{199}{3} g^2_1 + 27 g^2_2 + 88 g^2_3 - {\rm Tr} \left( 3Y^{}_\nu Y^\dagger_\nu + 15 Y^{}_l Y^\dagger_l + 17 Y^{}_{\rm u} Y^\dagger_{\rm u}  + 5 Y^{}_{\rm d} Y^\dagger_{\rm d} \right)  \right] \;,
    \\
    \hat{\beta}^{(2,0)}_{g^{}_2} &=&  \frac{1}{2} g^3_2 \left( 3g^2_1 + \frac{35}{3} g^2_2 + 24g^2_3 - T \right) \;,
    \\
    \hat{\beta}^{(2,0)}_{g^{}_3} &=&  g^3_3 \left[ \frac{11}{6} g^2_1 + \frac{9}{2}g^2_2 - 26 g^2_3 - 2 {\rm Tr} \left( Y^{}_{\rm u} Y^\dagger_{\rm u} + Y^{}_{\rm d} Y^\dagger_{\rm d} \right) \right] \;,
    \\
    \hat{\beta}^{(2,0)}_{m^2} &=& \left[ \frac{557}{48} g^4_2 + \frac{15}{8} g^2_1 g^2_2 - \frac{145}{16} g^4_2 + \frac{1}{12} g^2_1 {\rm Tr} \left( 15 Y^{}_\nu Y^\dagger_\nu + 75Y^{}_l Y^\dagger_l + 85 Y^{}_{\rm u} Y^\dagger_{\rm u} + 25 Y^{}_{\rm d} Y^\dagger_{\rm d} \right) \right.
    \nonumber
    \\
    && + \frac{15}{4} g^2_2 T + 40g^2_3 {\rm Tr} \left( Y^{}_{\rm u} Y^\dagger_{\rm u} + Y^{}_{\rm d} Y^\dagger_{\rm d} \right) + 4\lambda \left( 6g^2_1 + 18g^2_2 - 15\lambda -6T \right) - \frac{9}{2} T^\prime 
    \nonumber
    \\
    && - \left. 7 {\rm Tr} \left( Y^{}_\nu Y^\dagger_\nu Y^{}_l Y^\dagger_l + 3Y^{}_{\rm u} Y^\dagger_{\rm u} Y^{}_{\rm d} Y^\dagger_{\rm d} \right)  \right] m^2  + {\rm Tr} \left( 8Y^\dagger_\nu Y^{}_\nu M^\dagger_N Y^{\rm T}_\nu Y^\ast_\nu M^{}_N \right.
    \nonumber
    \\
    && + \left. 14Y^{}_\nu Y^\dagger_\nu Y^{}_\nu M^\dagger_N M^{}_N Y^\dagger_\nu -  2Y^{}_l Y^\dagger_l Y^{}_\nu M^\dagger_N M^{}_N Y^\dagger_\nu \right) \;,
    \\
    \hat{\beta}^{(2,0)}_\lambda &=& - \frac{379}{48} g^6_1 + \frac{305}{16} g^6_2 - \frac{289}{48} g^2_1 g^4_2 - \frac{559}{48} g^4_1 g^2_2 + \frac{629}{24} \lambda g^4_1 - \frac{73}{8} \lambda g^4_2 + \frac{39}{4} \lambda g^2_1 g^2_2 + 36\lambda^2 g^2_1 + 108 \lambda^2 g^2_2 
    \nonumber
    \\
    && - 312\lambda^3 - \frac{1}{4} g^4_1 {\rm Tr} \left( Y^{}_\nu Y^\dagger_\nu + 25 Y^{}_l Y^\dagger_l - 5Y^{}_{\rm d} Y^\dagger_{\rm d} + 19 Y^{}_{\rm u} Y^\dagger_{\rm u} \right) - \frac{3}{4} g^4_2 T + \frac{1}{2} g^2_1 g^2_2 {\rm Tr} \left( -Y^{}_\nu Y^\dagger_\nu \right.
    \nonumber
    \\
    && + \left. 11Y^{}_l Y^\dagger_l + 9 Y^{}_{\rm d} Y^\dagger_{\rm d} + 21Y^{}_{\rm u} Y^\dagger_{\rm u} \right) + \frac{5}{6} \lambda g^2_1 {\rm Tr} \left( 3Y^{}_\nu Y^\dagger_\nu + 15Y^{}_l Y^\dagger_l + 5 Y^{}_{\rm d} Y^\dagger_{\rm d} + 17 Y^{}_{\rm u} Y^\dagger_{\rm u} \right) + \frac{15}{2} \lambda g^2_2 T
    \nonumber
    \\
    && + 80\lambda g^2_3 {\rm Tr} \left( Y^{}_{\rm u} Y^\dagger_{\rm u} + Y^{}_{\rm d} Y^\dagger_{\rm d} \right) - 48\lambda^2 T -  \lambda T^\prime  - 14 \lambda  {\rm Tr} \left( Y^{}_\nu Y^\dagger_\nu Y^{}_l Y^\dagger_l + 3 Y^{}_{\rm u }Y^\dagger_{\rm u} Y^{}_{\rm d} Y^\dagger_{\rm d} \right) 
    \nonumber
    \\
    &&  + 4 g^2_1 {\rm Tr} \left( - Y^{}_l Y^\dagger_l Y^{}_l Y^\dagger_l + \frac{1}{3} Y^{}_{\rm d} Y^\dagger_{\rm d} Y^{}_{\rm d} Y^\dagger_{\rm d} - \frac{2}{3} Y^{}_{\rm u} Y^\dagger_{\rm u} Y^{}_{\rm u} Y^\dagger_{\rm u} \right) - 32 g^2_3 {\rm Tr} \left( Y^{}_{\rm d} Y^\dagger_{\rm d} Y^{}_{\rm d} Y^\dagger_{\rm d} + Y^{}_{\rm u} Y^\dagger_{\rm u} Y^{}_{\rm u} Y^\dagger_{\rm u} \right) 
    \nonumber
    \\
    && +2 {\rm Tr} \left[ 5 \left( Y^{}_\nu Y^\dagger_\nu Y^{}_\nu Y^\dagger_\nu  Y^{}_\nu Y^\dagger_\nu + Y^{}_l Y^\dagger_l Y^{}_l Y^\dagger_l  Y^{}_l Y^\dagger_l \right) - \left( Y^{}_\nu Y^\dagger_\nu Y^{}_\nu Y^\dagger_\nu  Y^{}_l Y^\dagger_l + Y^{}_l Y^\dagger_l Y^{}_l Y^\dagger_l  Y^{}_\nu Y^\dagger_\nu \right) \right.
    \nonumber
    \\
    && + \left. 15\left( Y^{}_{\rm u} Y^\dagger_{\rm u}  Y^{}_{\rm u}  Y^\dagger_{\rm u}   Y^{}_{\rm u}  Y^\dagger_{\rm u} + Y^{}_{\rm d} Y^\dagger_{\rm d}  Y^{}_{\rm d}  Y^\dagger_{\rm d}   Y^{}_{\rm d}  Y^\dagger_{\rm d}  \right) - 3 \left( Y^{}_{\rm u} Y^\dagger_{\rm u}  Y^{}_{\rm u}  Y^\dagger_{\rm u}   Y^{}_{\rm d}  Y^\dagger_{\rm d} + Y^{}_{\rm d} Y^\dagger_{\rm d}  Y^{}_{\rm d}  Y^\dagger_{\rm d}   Y^{}_{\rm u}  Y^\dagger_{\rm u}  \right)  \right] \;,
    \\
    \hat{\beta}^{(2,0)}_{Y^{}_{\rm d}} &=& \left[ - \frac{127}{216} g^4_1 - \frac{23}{4} g^4_2 - 108 g^4_3 - \frac{9}{4} g^2_1 g^2_2 + \frac{31}{9} g^2_1 g^2_3 + 9 g^2_2 g^2_3 + g^2_1 {\rm Tr} \left( \frac{5}{8} Y^{}_\nu Y^\dagger_\nu + \frac{25}{8} Y^{}_l Y^\dagger_l \right.\right.
    \nonumber
    \\
    && + \left. \frac{85}{24} Y^{}_{\rm u} Y^\dagger_{\rm u} + \frac{25}{24} Y^{}_{\rm d} Y^\dagger_{\rm d} \right) + \frac{15}{8} g^2_2 T + 20g^2_3 {\rm Tr} \left( Y^{}_{\rm u} Y^\dagger_{\rm u} + Y^{}_{\rm d} Y^\dagger_{\rm d} \right) + 6 \lambda^2 - \frac{9}{4} T^\prime 
    \nonumber
    \\
    && + \left. \frac{1}{2} {\rm Tr} \left( Y^{}_\nu Y^\dagger_\nu Y^{}_l Y^\dagger_l + 3 Y^{}_{\rm  u} Y^\dagger_{\rm u} Y^{}_{\rm d} Y^\dagger_{\rm d} \right) \right] Y^{}_{\rm d} + \left( \frac{187}{48}g^2_1 + \frac{135}{16} g^2_2 + 16g^2_3 - 12\lambda - \frac{9}{4} T \right) Y^{}_{\rm d} Y^\dagger_{\rm d} Y^{}_{\rm d}
    \nonumber
    \\
    && + \left( - \frac{79}{48} g^2_1 + \frac{9}{16} g^2_2 - 16g^2_3 + \frac{5}{4} T \right) Y^{}_{\rm u} Y^\dagger_{\rm u} Y^{}_{\rm d} - Y^{}_{\rm u} Y^\dagger_{\rm u} Y^{}_{\rm d} Y^\dagger_{\rm d} Y^{}_{\rm d} - \frac{1}{4} Y^{}_{\rm d} Y^\dagger_{\rm d} Y^{}_{\rm u} Y^\dagger_{\rm u} Y^{}_{\rm d}
    \nonumber
    \\
    && + \frac{11}{4} Y^{}_{\rm u} Y^\dagger_{\rm u} Y^{}_{\rm u} Y^\dagger_{\rm u} Y^{}_{\rm d} + \frac{3}{2} Y^{}_{\rm d} Y^\dagger_{\rm d} Y^{}_{\rm d} Y^\dagger_{\rm d} Y^{}_{\rm d}  \;,
    \\
    \hat{\beta}^{(2,0)}_{Y^{}_{\rm u}} &=& \left[ \frac{1187}{216} g^4_1 - \frac{23}{4} g^4_2 - 108 g^4_3 - \frac{3}{4} g^2_1 g^2_2 + \frac{19}{9} g^2_1 g^2_3 + 9 g^2_2 g^2_3  + g^2_1 {\rm Tr} \left( \frac{5}{8} Y^{}_\nu Y^\dagger_\nu + \frac{25}{8} Y^{}_l Y^\dagger_l \right.\right.
    \nonumber
    \\
    && + \left. \frac{85}{24} Y^{}_{\rm u} Y^\dagger_{\rm u} + \frac{25}{24} Y^{}_{\rm d} Y^\dagger_{\rm d} \right)+ \frac{15}{8} g^2_2 T + 20 g^2_3 {\rm Tr} \left( Y^{}_{\rm u} Y^\dagger_{\rm u} + Y^{}_{\rm d} Y^\dagger_{\rm d} \right) + 6 \lambda^2 - \frac{9}{4} T^\prime
    \nonumber
    \\
    && + \left. \frac{1}{2} {\rm Tr} \left( Y^{}_\nu Y^\dagger_\nu Y^{}_l Y^\dagger_l + 3 Y^{}_{\rm u} Y^\dagger_{\rm u} Y^{}_{\rm d} Y^\dagger_{\rm d} \right)  \right] Y^{}_{\rm u} + \left( \frac{223}{48} g^2_1 + \frac{135}{16} g^2_2 + 16g^2_3 - 12\lambda - \frac{9}{4} T \right) Y^{}_{\rm u} Y^\dagger_{\rm u} Y^{}_{\rm u}
    \nonumber
    \\
    && + \left( - \frac{43}{48} g^2_1 + \frac{9}{16} g^2_2 - 16g^2_3 + \frac{5}{4} T \right) Y^{}_{\rm d} Y^\dagger_{\rm d} Y^{}_{\rm u} -  Y^{}_{\rm d} Y^\dagger_{\rm d} Y^{}_{\rm u} Y^\dagger_{\rm u} Y^{}_{\rm u} - \frac{1}{4} Y^{}_{\rm u} Y^\dagger_{\rm u} Y^{}_{\rm d} Y^\dagger_{\rm d} Y^{}_{\rm u} 
    \nonumber
    \\
    && + \frac{11}{4} Y^{}_{\rm d} Y^\dagger_{\rm d} Y^{}_{\rm d} Y^\dagger_{\rm d} Y^{}_{\rm u} + \frac{3}{2} Y^{}_{\rm u} Y^\dagger_{\rm u} Y^{}_{\rm u} Y^\dagger_{\rm u} Y^{}_{\rm u} \;,
    \\
    \hat{\beta}^{(2,0)}_{Y^{}_l} &=& \left[ \frac{457}{24} g^4_1 + \frac{9}{4} g^2_1 g^2_2 - \frac{23}{4} g^4_2 + g^2_1 {\rm Tr} \left( \frac{5}{8} Y^{}_\nu Y^\dagger_\nu + \frac{25}{8} Y^{}_l Y^\dagger_l  + \frac{25}{24} Y^{}_{\rm d} Y^\dagger_{\rm d} + \frac{85}{24} Y^{}_{\rm u} Y^\dagger_{\rm u} \right) + \frac{15}{8} g^2_2 T  \right.
    \nonumber
    \\
    &&  +  \left.  20g^2_3 {\rm Tr} \left( Y^{}_{\rm d} Y^\dagger_{\rm d} + Y^{}_{\rm u} Y^\dagger_{\rm u} \right) + 6\lambda^2 + \frac{1}{2} {\rm Tr} \left( Y^{}_\nu Y^\dagger_\nu Y^{}_l Y^\dagger_l  + 3 Y^{}_{\rm u} Y^\dagger_{\rm u} Y^{}_{\rm d} Y^\dagger_{\rm d} \right) - \frac{9}{4} T^\prime \right] Y^{}_l
    \nonumber
    \\
    &&  + \left( \frac{129}{16} g^2_1 + \frac{135}{16} g^2_2 - 12\lambda - \frac{9}{4} T \right) Y^{}_l Y^\dagger_l Y^{}_l + \left( - \frac{45}{16} g^2_1 + \frac{9}{16} g^2_2 +  \frac{5}{4} T \right) Y^{}_\nu Y^\dagger_\nu Y^{}_l 
    \nonumber
    \\
    && -  Y^{}_\nu Y^\dagger_\nu Y^{}_l Y^\dagger_l Y^{}_l - \frac{1}{4} Y^{}_l Y^\dagger_l Y^{}_\nu Y^\dagger_\nu Y^{}_l + \frac{11}{4} Y^{}_\nu Y^\dagger_\nu Y^{}_\nu Y^\dagger_\nu Y^{}_l + \frac{3}{2} Y^{}_l Y^\dagger_l Y^{}_l Y^\dagger_l Y^{}_l \;,
    \\\label{eq:renormalizable-rge-ynu}
    \hat{\beta}^{(2,0)}_{Y^{}_\nu} &=& \left[ \frac{35}{24} g^4_1 - \frac{23}{4} g^4_2 - \frac{9}{4} g^2_1 g^2_2 + \frac{5}{24} g^2_1 {\rm Tr} \left( 3Y^{}_\nu Y^\dagger_\nu +15 Y^{}_l Y^\dagger_l + 5 Y^{}_{\rm d} Y^\dagger_{\rm d} + 17 Y^{}_{\rm u} Y^\dagger_{\rm u} \right) + \frac{15}{8} g^2_2 T  \right.
    \nonumber
    \\
    && + \left. 20g^2_3 {\rm Tr} \left( Y^{}_{\rm d} Y^\dagger_{\rm d} + Y^{}_{\rm u} Y^\dagger_{\rm u}  \right) + 6\lambda^2 - \frac{9}{4} T^\prime + \frac{1}{2} {\rm Tr} \left( Y^{}_\nu Y^\dagger_\nu Y^{}_l Y^\dagger_l + 3 Y^{}_{\rm d} Y^\dagger_{\rm d} Y^{}_{\rm u} Y^\dagger_{\rm u} \right) \right] Y^{}_\nu
    \nonumber
    \\
    && +  \left( \frac{93}{16} g^2_1 + \frac{135}{16} g^2_2 - 12\lambda - \frac{9}{4} T \right) Y^{}_\nu Y^\dagger_\nu Y^{}_\nu + \left( - \frac{81}{16} g^2_1 + \frac{9}{16} g^2_2 + \frac{5}{4} T \right) Y^{}_l Y^\dagger_l Y^{}_\nu 
    \nonumber
    \\
    && + \frac{11}{4} Y^{}_l Y^\dagger_l Y^{}_l Y^\dagger_l Y^{}_\nu -  Y^{}_l Y^\dagger_l Y^{}_\nu Y^\dagger_\nu Y^{}_\nu  + \frac{3}{2} Y^{}_\nu Y^\dagger_\nu Y^{}_\nu Y^\dagger_\nu Y^{}_\nu - \frac{1}{4} Y^{}_\nu Y^\dagger_\nu Y^{}_l Y^\dagger_l Y^{}_\nu \;,
    \\
    \hat{\beta}^{(2,0)}_{M_N} &=& \frac{1}{8} \left( 17g^2_1 + 51 g^2_2 - 12 T \right) \left[ M^{}_N Y^\dagger_\nu Y^{}_\nu + \left( Y^\dagger_\nu Y^{}_\nu \right)^{\rm T} M^{}_N \right] - \frac{1}{4} M^{}_N Y^\dagger_\nu \left( Y^{}_l Y^\dagger_l + Y^{}_\nu Y^\dagger_\nu \right) Y^{}_\nu 
    \nonumber
    \\\label{eq:renormalizable-rge-mn}
    && -\frac{1}{4} Y^{\rm T}_\nu \left[ \left( Y^{}_l Y^\dagger_l \right)^{\rm T} + \left( Y^{}_\nu Y^\dagger_\nu \right)^{\rm T} \right] Y^\ast_\nu M^{}_N + 4\left( Y^\dagger_\nu Y^{}_\nu \right)^{\rm T} M^{}_N Y^\dagger_\nu Y^{}_\nu  \;,
\end{eqnarray}
with $T = {\rm Tr} \left( Y^{}_\nu Y^\dagger_\nu + Y^{}_l Y^\dagger_l + 3Y^{}_{\rm u} Y^\dagger_{\rm u} + 3Y^{}_{\rm d}Y^\dagger_{\rm d}  \right)$ and $T^\prime = {\rm Tr} \left[ ( Y^{}_\nu Y^\dagger_\nu  )^2+ ( Y^{}_l Y^\dagger_l )^2 + 3( Y^{}_{\rm u}  Y^\dagger_{\rm u} )^2 + 3( Y^{}_{\rm d}  Y^\dagger_{\rm d} )^2 \right] $. The above results are subject to several cross-checks:
\begin{itemize}
    \item For each renormalization constant, the second pole of $\varepsilon$ from the sum of one-loop diagrams with insertion of one-loop counterterms is twice that from the sum of two-loop diagrams, but they have an opposite sign. This leads to the counteraction among non-local sub-divergences and then gives local results polynomial in masses and external momenta.
    
    \item Except for renormalization constants of lepton doublet and right-handed neutrino fields and those of $Y^{}_l$ and $Y^{}_\nu$, all renormalization constants satisfy the two consistency relations in Eq.~\eqref{eq:finite-consistency}. The exceptions do not point to wrong results but lead to infinite RG functions. We will show how to understand them in detail in Sec.~\ref{sec:infinite-RG}. Note that one needs to restore the gauge-fixing parameters in all one-loop wave-function renormalization constants when considering the consistency relation for field anomalous dimension. 
    
    \item Switching off non-renormalizable operators, all results are very those in the type-I seesaw mechanism. The two-loop RGEs in a general renormalizable gauge theory have been achieved for a long time~\cite{Machacek:1983tz,Machacek:1983fi,Machacek:1984zw,Luo:2002ti,Schienbein:2018fsw}, and several packages~\cite{Staub:2013tta,Litim:2020jvl,Thomsen:2021ncy,Sartore:2020gou} have been developed based on these generic results. We exploit {\sf RGBeta}~\cite{Thomsen:2021ncy} to reproduce the two-loop RGEs in the type-I seesaw mechanism, and it shows that all results are consistent with those given in Eqs.~\eqref{eq:renormalizable-rge-g1}-\eqref{eq:renormalizable-rge-mn}. The two-loop RGEs of $Y^{}_\nu$ and $M^{}_N$ have also been given in Ref.~\cite{Ibarra:2020eia} with approximations, which can be easily achieved by those in Eqs.~\eqref{eq:renormalizable-rge-ynu} and \eqref{eq:renormalizable-rge-mn} with the same approximation.
    
    \item Switching off right-handed neutrino fields, it becomes the SMEFT at $\Op (\Lambda^{-1})$. The full two-loop RGE of the Weinberg operator's Wilson coefficient has been achieved very recently~\cite{Ibarra:2024tpt}, and we can reproduce it by discarding terms involving $Y^{}_\nu$, $C^{}_{HN}$ and $C^{}_{BN}$ in Eq.~\eqref{eq:two-loop-rge-c5}.
\end{itemize}

\section{Infinite RG Functions}\label{sec:infinite-RG}
When we check consistency relations between the $1/\varepsilon$ and $1/\varepsilon^2$ poles for renormalization constants, we find that the consistency relations for the lepton doublet and right-handed neutrino wave-function renormalization constants, as well as those for the renormalization constants of charged-lepton and neutrino Yukawa coupling matrices, do not hold. This gives rise to infinite contributions to RG functions. As discussed in Ref.~\cite{Herren:2021yur}, such infinite RG functions can be calculated as~\footnote{Our definition of field wave-function renormalization constants differs from that in Ref.~\cite{Herren:2021yur}, i.e., the former is the square of the latter.}
\begin{eqnarray}\label{eq:ifad}
    \gamma^{(n)}_\phi &=&  - {\bf L} z^{(n+1)}_\phi + \sum^{n-1}_{k=0} \left( \frac{1}{2} \sum^{}_I \beta^{(k)}_I \partial^I z^{(n-k)}_\phi - z^{(n-k)}_\phi \gamma^{(k)}_\phi  \right) \;,\quad n \geq 1 \;,
\\\label{eq:ibf}
    \beta^{(n)}_I &=& 2 {\bf L} a^{(n+1)}_I - \sum^{n-1}_{k=0} \sum^{}_J \beta^{(k)}_J \partial^J a^{(n-k)}_I \;,\quad n \geq 1 \;,
\end{eqnarray}
where the field anomalous dimension and beta function are expanded as $\gamma^{}_\phi = \sum^{}_n \gamma^{(n)}_\phi/\varepsilon^n$ and $\beta^{}_I = \sum^{}_n \beta^{(n)}_I /\varepsilon^n$, respectively. Eqs.~\eqref{eq:ifad} and \eqref{eq:ibf} will regress to the consistency relations in Eq.~\eqref{eq:finite-consistency} if there are no infinite contributions to RG functions, namely $\gamma^{(n)}_\phi = \beta^{(n)}_I =0$ (for $n\geq 1$).

Substituting relevant renormalization constants into Eqs.~\eqref{eq:ifad} and~\eqref{eq:ibf}, we obtain the infinite field anomalous dimension for lepton doublet and right-handed neutrino fields, i.e.,
\begin{eqnarray}\label{eq:ifad1}
    \gamma^{(2,1)}_\ell &=& 3 g^{}_1 Y^{}_\nu \left( M^\dagger_N C^{}_{BN} + C^\dagger_{BN} M^{}_N \right) Y^\dagger_\nu - Y^{}_\nu \left( M^\dagger_N C^{}_{HN} + C^\dagger_{HN} M^{}_N \right) Y^\dagger_\nu \;,
    \nonumber
    \\
    \gamma^{(2,1)}_N &=& \frac{3}{2} M^{}_N Y^\dagger_\nu C^{}_5 Y^\ast_\nu + \frac{3}{2} Y^{\rm T}_\nu C^\dagger_5 Y^{}_\nu M^\dagger_N \;,
\end{eqnarray}
and the infinite beta functions for $Y^{}_l$ and $Y^{}_\nu$, namely
\begin{eqnarray}\label{eq:ibf1}
    \beta^{(2,1)}_{Y^{}_l}  &=&  Y^{}_\nu \left( 3g^{}_1 C^\dagger_{BN} - C^\dagger_{HN} \right) M^{}_N Y^\dagger_\nu Y^{}_l  - Y^{}_\nu M^\dagger_N \left( 3g^{}_1 C^{}_{BN} - C^{}_{HN} \right) Y^\dagger_\nu Y^{}_l \;,
    \nonumber
    \\
    \beta^{(2,1)}_{Y^{}_\nu}  &=&  Y^{}_\nu \left( 3g^{}_1 C^\dagger_{BN} - C^\dagger_{HN} \right) M^{}_N Y^\dagger_\nu Y^{}_\nu  - Y^{}_\nu M^\dagger_N \left( 3g^{}_1 C^{}_{BN} - C^{}_{HN} \right) Y^\dagger_\nu Y^{}_\nu \;.
\end{eqnarray}
Ref.~\cite{Manohar:2024xbh} shows that such infinite field anomalous dimensions in Eq.~\eqref{eq:ifad1} result from infinite field redefinitions. However, infinite beta functions like those in Eq.~\eqref{eq:ibf1}  do not and are expected not to show up in Ref.~\cite{Manohar:2024xbh}. To address these infinite RG functions, especially those of couplings, we restore contributions from redundant operators to all one-loop renormalization constants by adopting the first scheme to handle redundant operators as discussed in Sec.~\ref{sec:nuisance-operators}. It shows that infinite RG functions are attributed to the non-commutation between taking redundant operators' Wilson coefficients to be zero and taking the derivative of renormalization constants with respect to all couplings, and intrinsically, they are related to the flavor symmetry $U(3)_\ell \times U(n)_N$ embodied in the kinetic terms of lepton doublet and right-handed neutrinos. This is more or less similar to the case where one verifies the consistency relation for field anomalous dimensions in a non-$R^{}_\xi$ gauge. The gauge symmetry ensures that renormalization constants of couplings are independent of gauge-fixing parameters, but this is not the case for wave-function renormalization constants, which generally depend on those gauge-fixing parameters. This dependence is inexplicit in a non-$R^{}_\xi$ gauge, and we must restore it in order to make consistency relations for wave-function renormalization constants hold since the consistency relations involve the derivative with respect to all parameters, including gauge-fixing parameters. Otherwise, contributions from $\sum_{i} a^{(1,1)}_{\xi_i} \partial z^{(1,1)}_\phi/ \partial \xi^{}_i $  would be missed and then lead to the breaking of consistency relations and infinite field anomalous dimensions. 

Now, to understand the infinite RG functions in Eqs.~\eqref{eq:ifad1} and~\eqref{eq:ibf1}, we recalculate all one-loop renormalization constants by including redundant operators in the tree-level Lagrangian and the results in the Green's basis are collected in Appendix~\ref{app:one-loop-ct}. Then, one redefines right-handed neutrino and lepton doublet fields as Eq.~\eqref{eq:field-redefinition} to remove redundant operators in the Lagrangian and get the redefined physical couplings as Eq.~\eqref{eq:shift}. With the help of expressions of the redefined couplings and the one-loop results in the Green's basis, one can achieve all renormalization constants for the redefined couplings in terms of redefined couplings. It turns out that the obtained renormalization constants remain the same as those obtained without redundant operators in the tree-level Lagrangian, except $Z^{}_{N_{\rm L}} (\equiv Z^\ast_{N^{}_R})$, $Z^{}_\ell$, $\delta M^{}_N$, $Z^{}_{Y_\nu}$ and $Z^{}_{Y_l}$.  These excepted renormalization constants contain extra terms that involve redundant operators' Wilson coefficients and can not be entirely absorbed in the redefined couplings. Such extra terms are found to be
\begin{eqnarray}\label{eq:remanent}
    \Delta Z^{(L=1)}_{N^{}_{\rm L}} &=&  \loopf \left[ \frac{1}{2} \left( 3M^{}_N G^\dagger_{DN} + G^{}_{DN} M^\dagger_N \right)Y^{\rm T}_\nu Y^\ast_\nu + \frac{1}{2} Y^{\rm T}_\nu Y^\ast_\nu \left( 3 G^{}_{DN} M^\dagger_N + M^{}_N G^\dagger_{DN} \right) \right.
    \nonumber
    \\
    && - \left.  M^{}_N G^{\rm T}_{\ell HN1} Y^\ast_\nu - Y^{\rm T}_\nu G^\ast_{\ell HN1} M^\dagger_N  \right]  \;,
    \nonumber
    \\
    \Delta Z^{(L=1)}_\ell &=& - \loopf \left( G^{}_{\ell HN2} M^{}_N Y^\dagger_\nu + Y^{}_\nu M^\dagger_N G^\dagger_{\ell HN2} \right) \;,
    \nonumber
    \\
    \Delta \delta M^{(L=1)}_N &=& \loopf \left[ -\frac{1}{4} M^{}_N M^\dagger_N \left( Y^{\rm T}_\nu G^\ast_{\ell HN1} - G^\dagger_{\ell HN1} Y^{}_\nu \right)  + \frac{1}{4}  \left( Y^{\rm T}_\nu G^\ast_{\ell HN1} - G^\dagger_{\ell HN1} Y^{}_\nu \right)  M^\dagger_N M^{}_N \right.
    \nonumber
    \\
    && + M^{}_N M^\dagger_N \left( \frac{1}{2} Y^{\rm T}_\nu Y^\ast_\nu G^{}_{DN} - \frac{3}{4} G^{}_{DN} Y^\dagger_\nu Y^{}_\nu \right) - \left( \frac{3}{4} Y^{\rm T}_\nu Y^\ast_\nu G^{}_{DN} - \frac{1}{2} G^{}_{DN} Y^\dagger_\nu Y^{}_\nu \right) M^\dagger_N M^{}_N   
    \nonumber
    \\
    && - \frac{1}{4} M^{}_N G^\dagger_{DN} M^{}_N Y^\dagger_\nu Y^{}_\nu - \frac{1}{4} Y^{\rm T}_\nu Y^\ast_\nu M^{}_N G^\dagger_{DN} M^{}_N  + \frac{1}{4} M^{}_N Y^\dagger_\nu Y^{}_\nu M^\dagger_N G^{}_{DN} 
    \nonumber
    \\
    && + \left. \frac{1}{4} G^{}_{DN} M^\dagger_N Y^{\rm T}_\nu Y^\ast_\nu M^{}_N + \frac{1}{4} M^{}_N G^\dagger_{DN} Y^{\rm T}_\nu Y^\ast_\nu M^{}_N + \frac{1}{4} M^{}_N Y^\dagger_\nu Y^{}_\nu G^\dagger_{DN} M^{}_N \right] \;,
    \nonumber
    \\
    Y^{}_\nu \Delta Z^{(L=1)}_{Y^{}_\nu} &=& \loopf \left( \frac{3}{4} Y^{}_\nu Y^\dagger_\nu Y^{}_\nu G^\dagger_{DN} M^{}_N - \frac{3}{4} Y^{}_\nu M^\dagger_N G^{}_{DN} Y^\dagger_\nu Y^{}_\nu - \frac{1}{2} Y^{}_\nu G^\dagger_{DN} Y^{\rm T}_\nu Y^\ast_\nu M^{}_N \right.
    \nonumber
    \\
    && + \frac{1}{2} Y^{}_\nu M^\dagger_N Y^{\rm T}_\nu Y^\ast_\nu G^{}_{DN} + \frac{1}{4} Y^{}_\nu Y^\dagger_\nu Y^{}_\nu M^\dagger_N G^{}_{DN} - \frac{1}{4} Y^{}_\nu G^\dagger_{DN} M^{}_N Y^\dagger_\nu Y^{}_\nu - \frac{1}{4}  Y^{}_\nu Y^\dagger_\nu G^{}_{\ell HN1} M^{}_N 
    \nonumber
    \\
    && + \frac{1}{4} Y^{}_\nu M^\dagger_N G^\dagger_{\ell HN1} Y^{}_\nu  + \frac{1}{4} Y^{}_\nu G^{\rm T}_{\ell HN1} Y^\ast_\nu M^{}_N - \frac{1}{4} Y^{}_\nu M^\dagger_N Y^{\rm T}_\nu G^\ast_{\ell HN1} + \frac{1}{2} G^{}_{\ell HN2} M^{}_N Y^\dagger_\nu Y^{}_\nu 
    \nonumber
    \\
    && -  \left. \frac{1}{2} Y^{}_\nu M^\dagger_N G^\dagger_{\ell HN2} Y^{}_\nu   \right)  \;,
    \nonumber
    \\
    Y^{}_l \Delta Z^{(L=1)}_{Y^{}_l } &=& \loopf \left( \frac{1}{2} G^{}_{\ell HN2} M^{}_N Y^\dagger_\nu - \frac{1}{2} Y^{}_\nu M^\dagger_N G^\dagger_{\ell HN2} \right) Y^{}_l \;,
\end{eqnarray}
where we use the same symbols without the prime superscript for the redefined couplings for simplicity. As expected, these extra terms go vanishing if we take redundant operators' Wilson coefficients to be zero, and all results go back to the previous ones. However, doing so before taking the derivative of them with respect to these redundant operators' Wilson coefficients spoils consistency relations in Eq.~\eqref{eq:finite-consistency} and leads to infinite RG functions in Eqs.~\eqref{eq:ifad1} and~\eqref{eq:ibf1} via Eqs.~\eqref{eq:ifad} and \eqref{eq:ibf}. On the contrary, if one takes redundant operators' Wilson coefficients to be zero after taking the derivative with respect to them, the consistency relations hold for all renormalization constants, and thereby, no infinite RG functions arise. This is because contributions from $\delta G^{(L=1)}_{DN}$, $\delta G^{(L=1)}_{\ell HN1}$ and $\delta G^{(L=1)}_{\ell HN2}$ have been restored and they exactly eliminate the infinite terms in Eqs.~\eqref{eq:ifad1} and~\eqref{eq:ibf1}. In fact, contributions from $\delta G^{(L=1)}_{DN}$ are vanishing and those from $\delta G^{(L=1)}_{\ell HN1}$ in $\delta M^{}_N$ or $Z^{}_{Y_\nu}$ cancel out. This is the reason why there is no infinite RG function for $M^{}_N$, i.e., $\beta^{(2,1)}_{M^{}_N}$, even if Wilson coefficients of the redundant operators are taken to be zero in the very beginning. 

Now, it comes to the conclusion that no infinite contributions to field anomalous dimensions and beta functions show up, and the consistency relations hold both for them if one tracks contributions from the redundant operators. One may worry about the dependence of renormalization constants on the redundant operators' Wilson coefficients after field redefinitions like those in Eq.~\eqref{eq:remanent}, which indicates that such renormalization constants are not invariant under field redefinitions. This is not a problem for wave-function renormalization constants as they are unphysical, and moreover, their dependence on redundant operators' Wilson coefficients can be removed by infinite field redefinitions leading to infinite field anomalous dimensions~\cite{Manohar:2024xbh}. For the renormalization constants of physical couplings, such a dependence, or equivalently, such non-invariance under field redefinitions, would indicate the redundant operators' involvement in the RG running of physical couplings even if the Wilson coefficients of redundant operators are vanishing at one energy scale. Fortunately, this is not the case. After doing field redefinitions, all redundant operators and their counterterms are removed, while their Wilson coefficients are still involved in some renormalization constants. This dependence on the remanent Wilson coefficients is unphysical and related to flavor transformations, and hence those remanent Wilson coefficients can be simply taken to be zero so as to get the ``physical'' results. In fact, the Yukawa couplings themselves are not physical quantities since they are covariant under flavor symmetries. As flavor transformations are kind of linear field redefinitions, by no means the renormalization constants relying on a flavor basis are supposed to be invariant under field redefinitions unless one considers flavor invariants (see, e.g., Refs.~\cite{Jarlskog:1985ht,Jarlskog:1985cw,Bernabeu:1986fc,Branco:1986gr} for early studies on flavor invariants in quark and lepton sectors).

A closer examination of the contributions from the Wilson coefficients of redundant operators to $\delta M^{}_N$, $Z^{}_{Y_\nu}$ and $Z^{}_{Y_l}$ in Eq.~\eqref{eq:remanent} reveals that these contributions can be fully transferred to the anti-Hermitian part of the wave-function renormalization constants for lepton doublet and right-handed neutrino fields, namely~\footnote{We assume Hermitian wave-function renormalization constants in previous calculations for simplicity. Here, we include an anti-Hermitian part in $Z^{}_\ell$ and $Z^{}_{N_{\rm L}}$ by redefining the corresponding renormalized fields to get rid of the dependence of $\delta M^{}_N$, $Z^{}_{Y^{}_\nu}$, and $\delta Z^{}_{Y^{}_l}$ on redundant operators, i.e., those in Eq.~\eqref{eq:remanent}. Then, the renormalization constants transform accordingly as $\delta Z^{}_{\ell, N_{\rm L}} \to  \delta Z^{}_{\ell, N_{\rm L} } + \delta Z^{\rm anti}_{\ell, N_{\rm L} }$, $  \delta M^{}_N \to  \delta M^{ }_N +  \delta Z^{\rm anti}_{N_{\rm L}}  M^{}_N /2 + M^{}_N ( \delta Z^{\rm anti}_{N_{\rm L}} )^{\rm T}$/2, $Y^{}_\nu \delta Z^{}_{Y_\nu} \to Y^{}_\nu \delta Z^{}_{Y_\nu} + \delta Z^{\rm anti}_\ell Y^{}_\nu /2 + Y^{}_\nu ( \delta Z^{\rm anti}_{N_{\rm L}})^{\rm T} /2$, and $Y^{}_l \delta Z^{}_{Y_l} \to Y^{}_l \delta Z^{}_{Y_l} + \delta Z^{\rm anti}_\ell Y^{}_l/2$ at the one-loop level. The changes in renormalization constants of dim-5 operators' Wilson coefficients are beyond $\Op \left( \Lambda^{-1} \right)$ and hence omitted.}
\begin{eqnarray}\label{eq:anti-hermitian}
    \delta Z^{(L=1)\rm anti}_\ell &=&  \loopf \left( Y^{}_\nu M^\dagger_N G^\dagger_{\ell HN2} -  G^{}_{\ell HN2} M^{}_N Y^\dagger_\nu \right) \;,
    \nonumber
    \\
    \delta Z^{(L=1)\rm anti}_{N_{\rm L}} &=& \loopf \left(  - \frac{1}{2} Y^{\rm T}_\nu G^\ast_{\ell HN1} M^\dagger_N + \frac{1}{2} M^{}_N G^{\rm T}_{\ell HN1} Y^\ast_\nu + \frac{1}{2} G^\dagger_{\ell HN1} Y^{}_\nu M^\dagger_N - \frac{1}{2} M^{}_N Y^\dagger_\nu G^{}_{\ell HN1} \right.
    \nonumber
    \\
    && + \frac{3}{2} Y^{\rm T}_\nu Y^\ast_\nu G^{}_{DN} M^\dagger_N  - \frac{3}{2} M^{}_N G^\dagger_{DN} Y^{\rm T}_\nu Y^\ast_\nu -  G^{}_{DN} Y^\dagger_\nu Y^{}_\nu M^\dagger_N + M^{}_N Y^\dagger_\nu Y^{}_\nu G^\dagger_{DN}
    \nonumber
    \\
    && + \left. \frac{1}{2} Y^{\rm T}_\nu Y^\ast_\nu M^{}_N G^\dagger_{DN} - \frac{1}{2} G^{}_{DN} M^\dagger_N Y^{\rm T}_\nu Y^\ast_\nu \right) \;.
\end{eqnarray}
Then, $\delta M^{}_N$, $Z^{}_{Y_\nu}$ and $Z^{}_{Y_l}$ become independent of the redundant operators' Wilson coefficients; that is, $\Delta \delta M^{}_N$, $Y^{}_\nu \Delta Z_{Y_\nu}$, and $Y^{}_l \Delta Z_{Y_l}$ in Eq.~\eqref{eq:remanent} become vanishing after including those in Eq.~\eqref{eq:anti-hermitian}. In this case, the infinite RG functions rearise and will not disappear even if the derivative with respect to the redundant operator's Wilson coefficients is considered. The infinite beta functions for $Y^{}_l$ and $Y^{}_\nu$ are still given by Eq.~\eqref{eq:ibf1}, and the infinite field anomalous dimension for lepton doublet resulting from the anti-Hermitian part of its wave-function renormalization constant given in Eq.~\eqref{eq:anti-hermitian} is 
\begin{eqnarray}\label{eq:ifad2}
    \gamma^{(2,1)\rm anti}_{\ell^{}} &=&  Y^{}_\nu M^\dagger_N \left( 3g^{}_1 C^{}_{BN} - C^{}_{HN} \right) Y^\dagger_\nu - Y^{}_\nu \left( 3g^{}_1 C^\dagger_{BN} - C^\dagger_{HN} \right) M^{}_N Y^\dagger_\nu  \;.
\end{eqnarray}
The field anomalous dimension for right-handed neutrinos is vanishing due to $\delta G^{(L=1)}_{DN}=0$ and cancellation among terms involving $\delta G^{(L=1)}_{\ell HN1}$. Note that the infinite beta functions in Eq.~\eqref{eq:ibf1} and field anomalous dimension in Eq.~\eqref{eq:ifad2} satisfy the so-called RG-finiteness~\cite{Herren:2021yur} arising from the flavor Ward identity:
\begin{eqnarray}
    \beta^{(2,1)}_{Y^{}_l } &=& - \gamma^{(2,1)\rm anti}_{\ell}  Y^{}_l  \;,\qquad 	\beta^{(2,1)}_{Y^{}_\nu } = - \gamma^{(2,1)\rm anti}_{\ell}  Y^{}_\nu  \;,
\end{eqnarray}
where the flavor-covariant $\gamma^{(2,1)\rm anti}_{\ell}$ is an element of the flavor symmetry. This indicates that this infiniteness in RG functions is consistent with the finite running of Green's functions. Further, one may take advantage of the renormalization ambiguity in the form of a flavor rotation $U$ to get rid of the infiniteness, i.e.,~\cite{Herren:2021yur}
\begin{eqnarray}
    U = \exp\left( - \sum^{}_{n=1} \frac{1}{\varepsilon^n} u^{(n)} \right) \;,\quad \Delta \gamma = \widetilde{\gamma} - \gamma = \beta^{}_I U \partial^I U^\dagger \;,\quad \Delta \beta^{}_I = \widetilde{\beta}^{}_I - \beta^{}_I = - \left( \Delta \gamma \mathtt{g} \right)^{}_I \;,
\end{eqnarray}
where anti-Hermitian $u^{(n)}$ belongs to the flavor symmetry. In our case, the flavor rotation can be chosen as 
\begin{eqnarray}
    u^{(1)} = \frac{1}{16\pi^2} \frac{1}{2} \left( Y^{}_\nu M^\dagger_N G^\dagger_{\ell HN2} - G^{}_{\ell HN2} M^{}_N Y^\dagger_\nu \right) \;,
\end{eqnarray}
and then the changes of RG functions under this flavor transformation are given by
\begin{eqnarray}
    \Delta \gamma^{(2,1)}_{\ell}  &=& \left. - \beta^{(1,0)}_I \partial^I u^{(1)} \right|^{}_{G_i = 0}  =  Y^{}_\nu \left( 3g^{}_1 C^\dagger_{BN} - C^\dagger_{HN} \right) M^{}_N Y^\dagger_\nu -  Y^{}_\nu M^\dagger_N \left( 3g^{}_1 C^{}_{BN} - C^{}_{HN} \right) Y^\dagger_\nu \;,
    \nonumber
    \\
    \Delta \beta^{(2,1)}_{Y^{}_l}  &=&  - \left. \Delta \gamma^{(2,1)}_{\ell} Y^{}_l \right|^{}_{G_i = 0} = - Y^{}_\nu \left( 3g^{}_1 C^\dagger_{BN} - C^\dagger_{HN} \right) M^{}_N Y^\dagger_\nu Y^{}_l + Y^{}_\nu M^\dagger_N \left( 3g^{}_1 C^{}_{BN} - C^{}_{HN} \right) Y^\dagger_\nu Y^{}_l \;,
    \nonumber
    \\
    \Delta \beta^{(2,1)}_{Y^{}_\nu}  &=&  - \left. \Delta \gamma^{(2,1)}_{\ell} Y^{}_\nu \right|^{}_{G_i = 0} =  - Y^{}_\nu \left( 3g^{}_1 C^\dagger_{BN} - C^\dagger_{HN} \right) M^{}_N Y^\dagger_\nu Y^{}_\nu +  Y^{}_\nu M^\dagger_N \left( 3g^{}_1 C^{}_{BN} - C^{}_{HN} \right) Y^\dagger_\nu Y^{}_\nu \qquad
\end{eqnarray}
up to the two-loop level and $\mathcal{O}\left( \Lambda^{-1} \right)$. These changes eliminate the infinite field anomalous dimension in Eq.~\eqref{eq:ifad2} and infinite beta functions in Eq.~\eqref{eq:ibf1} and leave RG functions finite.

As discussed above, the breaking of consistency relations or the infiniteness of RG functions is induced by the non-invariance of some renormalization constants under flavor transformations, a kind of field redefinitions. To understand this, one had better restore the dependence of renormalization constants on the Wilson coefficients of redundant operators, even though it is unphysical. We can simply take the remnant redundant operators' Wilson coefficients to be vanishing after taking the derivative with respect to them so as to make the flavor-dependent couplings get rid of the infinite contributions to their RGEs. This is, in principle, unnecessary thanks to the flavor symmetry, but doing so simplifies the RG running of relevant couplings in a specific flavor basis. Alternatively, one can consider the RGEs of flavor invariants instead of those of Yukawa couplings. In this way, all RGEs not only get rid of infinite contributions without restoring redundant operators in the Lagrangian but also more generally, are free from the renormalization ambiguity induced by the flavor symmetry because flavor invariants, as they are called, are invariant under flavor transformations. One can partially check this by following the results and discussions with right-handed neutrinos in Refs.~\cite{Jenkins:2009dy,Hanany:2010vu,Wang:2021wdq,Yu:2021cco,Yu:2022ttm,Grojean:2024qdm}. However, the other side of the coin is that constructing a complete set of independent flavor invariants and establishing their relationships with observables are highly intricate. Therefore, it remains intriguing to explore the generic connection between infinite RG functions and the renormalization ambiguity in the EFT framework, which would be similar to that in the renormalizable gauge-Yukawa theories~\cite{Herren:2021yur,Fortin:2012hn,Jack:2013sha,Poole:2019kcm,Davies:2021mnc}, but further efforts are required to incorporate irrelevant operators and field redefinitions into this formalism.

\section{Conclusions}\label{sec:conlusions}

The EFT approach is a very promising way to explore new physics beyond the SM. As a simple extension of the SMEFT with right-handed neutrinos, the $\nu$SMEFT has gained increasing attention, and its phenomena, one-loop RGEs, and the matching of UV models onto it have been widely studied. In this work, we push the RGEs in the $\nu$SMEFT towards the next-to-leading logarithmic  accuracy. Adopting dimension regularization and the minimal subtraction scheme, we derive the two-loop RGEs for renormalizable couplings and Wilson coefficients of dim-5 operators in the $\nu$SMEFT. We use the background field method to simplify the renormalization of the gauge sector and introduce a universal spurious mass to decompose each propagator and rearrange infrared divergences in the two-loop integrals. We work out both the first and second poles of $\varepsilon$ in all renormalization constants. The first pole $1/\varepsilon$ is utilized to derive the two-loop RGEs, and the second one $1/\varepsilon^2$ provides a cross-check on calculations since it is related to the first pole by consistency relations. However, we find that the consistency relations for lepton doublet and right-handed neutrino wave-function renormalization constants, as well as for the renormalization constants of charged-lepton and neutrino Yukawa coupling matrices, do not hold. This gives rise to infinite RG functions for the relevant fields and couplings. We show that such infinite RG functions result from the non-invariance of fields and Yukawa coupling matrices under flavor transformations---a kind of linear field redefinitions. One can get rid of those unphysical infinite terms by restoring contributions from the derivative of renormalization constants with respect to Wilson coefficients of redundant operators or, alternatively, by considering the RGEs for flavor invariants instead of those for coupling matrices. Additionally, we illustrate an instance of the flavor-structure ambiguity by naively using EoMs to convert redundant operators into physical ones, even though this procedure is equivalent to exploiting field redefinitions at leading linear order. Therefore, employing field redefinitions is always recommended to avoid potential ambiguities and ensure the inclusion of higher-order contributions when reduction relations between redundant and physical operators play a significant role in calculations.

\section*{Acknowledgements}

This work was partially supported by the Alexander von Humboldt Foundation, the Collaborative Research Center SFB1258, and the Deutsche Forschungsgemeinschaft (DFG, German Research Foundation) under Germany's Excellence Strategy - EXC-2094 - 390783311.

\appendix

\section{One-loop Counterterms}\label{app:one-loop-ct}

One-loop counterterms in the Green's basis are collected below:
\begin{eqnarray}\label{eq:one-loop-ct}
    \delta Z^{(L=1)}_G &=& \frac{7g^2_3}{16\pi^2 \varepsilon} \;,
    \nonumber
    \\
    \delta Z^{(L=1)}_W &=& \frac{19g^2_2}{16\pi^2 \varepsilon \cdot 6} \;,
    \nonumber
    \\
    \delta Z^{(L=1)}_B &=& -\frac{41g^2_1}{16\pi^2 \varepsilon \cdot 6} \;,
    \nonumber
    \\
    \delta Z^{(L=1)}_{\xi_{\hat{G}}} &=& \frac{g^2_3}{16\pi^2 \varepsilon} \;,
    \nonumber
    \\
    \delta Z^{(L=1)}_{\xi_{\hat{W}}} &=& -\frac{5g^2_2}{16\pi^2 \varepsilon \cdot 6} \;,
    \nonumber
    \\
    \delta Z^{(L=1)}_{\xi_{\hat{B}}} &=& - \frac{41g^2_1}{16\pi^2 \varepsilon \cdot 6} \;,
    \nonumber
    \\
    \delta Z^{(L=1)}_H &=&  \frac{1}{16\pi^2 \varepsilon \cdot 2} \left( g^2_1 + 3g^2_2 - 2T \right)  \color{gray} + \frac{1}{16\pi^2\varepsilon} {\rm Tr} \left(  G^{}_{\ell HN1} M^{}_N Y^\dagger_\nu + Y^{}_\nu M^\dagger_N G^\dagger_{\ell HN1} \right.
    \nonumber
    \\
    && \color{gray} - \left. 2 G^\dagger_{DN} M^{}_N Y^\dagger_\nu Y^{}_\nu - 2 M^\dagger_N G^{}_{DN} Y^\dagger_\nu Y^{}_\nu \right) \;,
    \nonumber
    \\
    m^2 \delta Z^{(L=1)}_{m^2} &=& \loopf \left[ \left( - \frac{3}{4} g^2_1 - \frac{9}{4} g^2_2 + 6\lambda + T \right) m^2 + 2 {\rm Tr} \left( M^\dagger_N M^{}_N M^\dagger_N C^{}_{HN} + M^{}_N M^\dagger_N M^{}_N C^\dagger_{HN} \right.\right.
    \nonumber
    \\
    && - \left.\left. M^\dagger_N M^{}_N Y^\dagger_\nu Y^{}_\nu \right) \right]  \color{gray} + \frac{1}{16\pi^2\varepsilon} {\rm Tr} \left[ \left( -G^{}_{\ell HN1} M^{}_N Y^\dagger_\nu - Y^{}_\nu M^\dagger_N G^\dagger_{\ell HN1} + 2 G^\dagger_{DN} M^{}_N Y^\dagger_\nu Y^{}_\nu  \right. \right.
    \nonumber
    \\
    &&  \color{gray}  + \left. 2 M^\dagger_N G^{}_{DN} Y^\dagger_\nu Y^{}_\nu \right) m^2  -  2 \left( G^{}_{\ell HN2} - G^{}_{\ell HN1} \right) M^{}_N M^\dagger_N M^{}_N Y^\dagger_\nu 
    \nonumber
    \\
    && \color{gray} - 2 Y^{}_\nu M^\dagger_N M^{}_N M^\dagger_N \left( G^\dagger_{\ell HN2} - G^\dagger_{\ell HN1} \right) - 4 \left( G^\dagger_{DN} M^{}_N M^\dagger_N M^{}_N + M^\dagger_N M^{}_N M^\dagger_N G^{}_{DN} \right.
    \nonumber
    \\
    && \color{gray} + \left.\left. M^\dagger_N G^{}_{DN}  M^\dagger_N M^{}_N + M^\dagger_N M^{}_N G^\dagger_{DN} M^{}_N  \right) Y^\dagger_\nu Y^{}_\nu \right] \;,
    \nonumber
    \\
    \lambda \delta Z^{(L=1)}_\lambda &=& \frac{1}{16\pi^2 \varepsilon \cdot 16} \left[ 32\lambda T + 3 (g^2_1+g^2_2)^2 + 6g^4_2 - 24\lambda(g^2_1 + 3g^2_2) + 192 \lambda^2 - 16 T^\prime \right.
    \nonumber
    \\
    && + \left.  16{\rm Tr} \left( 4 C^\dagger_{HN} M^{}_N Y^\dagger_\nu Y^{}_\nu + 4Y^\dagger_\nu Y^{}_\nu M^\dagger_N C^{}_{HN} + C^{}_5 Y^\ast_\nu M^{}_N Y^\dagger_\nu + Y^{}_\nu M^\dagger_N Y^{\rm T}_\nu C^\dagger_5 \right)  \right] 
    \nonumber
    \\
    && \color{gray}  + \frac{2}{16\pi^2 \varepsilon} {\rm Tr} \left[ \lambda \left( -G^{}_{\ell HN1} M^{}_N Y^\dagger_\nu - Y^{}_\nu M^\dagger_N G^\dagger_{\ell HN1} + 2 G^\dagger_{DN} M^{}_N Y^\dagger_\nu Y^{}_\nu + 2 M^\dagger_N G^{}_{DN} Y^\dagger_\nu Y^{}_\nu \right)  \right.
    \nonumber
    \\
    && \color{gray} - \left( G^{}_{\ell HN2} - G^{}_{\ell HN1} \right) \left( M^{}_N Y^\dagger_\nu Y^{}_\nu Y^\dagger_\nu  + Y^{\rm T}_\nu Y^\ast_\nu M^{}_N Y^\dagger_\nu \right)
    \nonumber
    \\
    && \color{gray}   -  \left( Y^{}_\nu Y^\dagger_\nu Y^{}_\nu M^\dagger_N + Y^{}_\nu M^\dagger_N Y^{\rm T}_\nu Y^\ast_\nu \right) \left( G^\dagger_{\ell HN2} - G^\dagger_{\ell HN1} \right) 
    \nonumber
    \\
    && \color{gray} - \left. 2 Y^\dagger_\nu Y^{}_\nu Y^\dagger_\nu Y^{}_\nu \left( G^\dagger_{DN} M^{}_N + M^\dagger_N G^{}_{DN} \right)  - Y^\dagger_\nu Y^{}_\nu \left( G^\dagger_{DN} Y^{\rm T}_\nu Y^\ast_\nu M^{}_N + M^\dagger_N Y^{\rm T}_\nu Y^\ast_\nu G^{}_{DN} \right)   \right] \;,
    \nonumber
    \\ 
    \delta Z^{(L=1)}_\ell &=& -\frac{1}{16\pi^2 \varepsilon \cdot 4} \left( g^2_1 + 3g^2_2 + 2 Y^{}_l Y^\dagger_l + 2Y^{}_\nu Y^\dagger_\nu \right)  \color{gray} - \frac{1}{16\pi^2\varepsilon} \left[  -\frac{1}{2} \left( G^{}_{\ell HN1} M^{}_N Y^\dagger_\nu + Y^{}_\nu M^\dagger_N G^\dagger_{\ell HN1} \right) \right.
    \nonumber
    \\
    && \color{gray} + \left. G^{}_{\ell HN2} M^{}_N Y^\dagger_\nu + Y^{}_\nu M^\dagger_N G^\dagger_{\ell HN2} + Y^{}_\nu \left( G^\dagger_{DN} M^{}_N + M^\dagger_N G^{}_{DN} \right) Y^\dagger_\nu \right] \;,
    \nonumber
    \\
    \delta Z^{(L=1)}_e &=& -\frac{1}{16\pi^2 \varepsilon} \left( g^2_1 + Y^\dagger_l Y^{}_l  \right) \;,
    \nonumber
    \\
    \delta Z^{(L=1)}_q &=& -\frac{1}{16\pi^2 \varepsilon \cdot 36} \left( g^2_1 + 27g^2_2 + 48g^2_3 + 18 Y^{}_{\rm d} Y^\dagger_{\rm d} + 18 Y^{}_{\rm u} Y^\dagger_{\rm u} \right) \;,
    \nonumber
    \\
    \delta Z^{(L=1)}_u &=& -\frac{1}{16\pi^2 \varepsilon \cdot 9} \left( 4g^2_1 + 12g^2_3 + 9 Y^\dagger_{\rm u} Y^{}_{\rm u} \right) \;,
    \nonumber
    \\
    \delta Z^{(L=1)}_d &=& -\frac{1}{16\pi^2 \varepsilon \cdot 9} \left( g^2_1 + 12g^2_3 + 9 Y^\dagger_{\rm d} Y^{}_{\rm d} \right) \;,
    \nonumber
    \\
    \delta Z^{(L=1)}_{N^{}_{\rm L}} &=& -\loopf \left( Y^\dagger_\nu Y^{}_\nu \right)^{\rm T}  \;,
    \nonumber
    \\
    \delta M^{(L=1)}_N &=& \frac{1}{16\pi^2 \varepsilon \cdot 2} \left[ M^{}_N Y^\dagger_\nu Y^{}_\nu + \left( Y^\dagger_\nu Y^{}_\nu \right)^{\rm T} M^{}_N - 8 m^2 C^{}_{HN}\right]  \color{gray} + \frac{2 m^2}{16\pi^2 \varepsilon} \left( G^\dagger_{\ell HN2} Y^{}_\nu + Y^{\rm T}_\nu G^\ast_{\ell HN2} \right) \;,
    \nonumber
    \\
    Y^{}_l \delta Z^{(L=1)}_{Y_l} &=& \frac{1}{16\pi^2 \varepsilon \cdot 8} \left[ Y^{}_l \left( 4T - 15g^2_1 - 9g^2_2 \right)  + 6 \left( Y^{}_l Y^\dagger_l  -  Y^{}_\nu Y^\dagger_\nu \right) Y^{}_l\right]   \color{gray} +   \frac{1}{16\pi^2\varepsilon} \left[ - \frac{1}{2} {\rm Tr} \left( G^{}_{\ell HN1} M^{}_N Y^\dagger_\nu   \right. \right. 
    \nonumber
    \\
    &&  \color{gray} + \left. Y^{}_\nu M^\dagger_N G^\dagger_{\ell HN1} \right) + \left( \frac{3}{4}  G^{}_{\ell HN1} + \frac{1}{2} G^{}_{\ell HN2} \right) M^{}_N Y^\dagger_\nu + Y^{}_\nu M^\dagger_N \left( \frac{3}{4} G^\dagger_{\ell HN1}  - \frac{1}{2} G^\dagger_{\ell HN2} \right)  
    \nonumber
    \\
    && \color{gray} + \left. \tr{G^\dagger_{DN} M^{}_N Y^\dagger_\nu Y^{}_\nu +   M^\dagger_N G^{}_{DN} Y^\dagger_\nu Y^{}_\nu } - \frac{3}{2} Y^{}_\nu \left( G^\dagger_{DN} M^{}_N + M^\dagger_N G^{}_{DN} \right) Y^\dagger_\nu \right] Y^{}_l \;,
    \nonumber
    \\
    Y^{}_\nu \delta Z^{(L=1)}_{Y_\nu} &=&  \frac{1}{16\pi^2 \varepsilon \cdot 8}  \left[ Y^{}_\nu \left( 4T - 3g^2_1 -9 g^2_2 \right)  + 6 \left( Y^{}_\nu Y^\dagger_\nu - Y^{}_l Y^\dagger_l   \right) Y^{}_\nu + 48 g^{}_1 Y^{}_\nu M^\dagger_N C^{}_{BN} \right.
    \nonumber
    \\
    && - \left. 16 Y^{}_\nu M^\dagger_N C^{}_{HN} \right] 
    \nonumber 
    \\
    && \color{gray}  + \frac{1}{16\pi^2\varepsilon} \left[ \frac{1}{2} {\rm Tr} \left( -G^{}_{\ell HN1} M^{}_N Y^\dagger_\nu  - Y^{}_\nu M^\dagger_N G^\dagger_{\ell HN1}  + 2 G^\dagger_{DN} M^{}_N Y^\dagger_\nu Y^{}_\nu + 2 M^\dagger_N G^{}_{DN} Y^\dagger_\nu Y^{}_\nu \right) Y^{}_\nu \right.
    \nonumber
    \\
    && \color{gray}  -\frac{1}{4}  \left( G^{}_{\ell HN1} - 2 G^{}_{\ell HN2} \right) M^{}_N Y^\dagger_\nu Y^{}_\nu - \frac{1}{4} Y^{}_\nu M^\dagger_N \left( G^\dagger_{\ell HN1}  - 2 G^\dagger_{\ell HN2} \right) Y^{}_\nu + Y^{}_\nu M^\dagger_N Y^{\rm T}_\nu G^\ast_{\ell HN2} 
    \nonumber
    \\
    && \color{gray} + \left. \frac{1}{2} Y^{}_\nu \left(G^\dagger_{DN} M^{}_N +  M^\dagger_N G^{}_{DN} \right) Y^\dagger_\nu Y^{}_\nu \right] \;,
    \nonumber
    \\
    Y^{}_{\rm u} \delta Z^{(L=1)}_{Y_{\rm u}} &=& \frac{1}{16\pi^2 \varepsilon \cdot 24}  \left[ Y^{}_{\rm u} \left( 12T - 17g^2_1 - 27g^2_2 - 96 g^2_3 \right)  + 18 Y^{}_{\rm u} Y^\dagger_{\rm u} Y^{}_{\rm u} - 18 Y^{}_{\rm d} Y^\dagger_{\rm d} Y^{}_{\rm u}  \right] 
    \nonumber
    \\
    && \color{gray}  + \frac{1}{16\pi^2\varepsilon \cdot 2} {\rm Tr} \left( -G^{}_{\ell HN1} M^{}_N Y^\dagger_\nu  - Y^{}_\nu M^\dagger_N G^\dagger_{\ell HN1}  + 2 G^\dagger_{DN} M^{}_N Y^\dagger_\nu Y^{}_\nu +2 M^\dagger_N G^{}_{DN} Y^\dagger_\nu Y^{}_\nu \right) Y^{}_{\rm u}  \;,
    \nonumber
    \\
    Y^{}_{\rm d} \delta Z^{(L=1)}_{Y_{\rm d}} &=& \frac{1}{16\pi^2 \varepsilon \cdot 24} \left[ Y^{}_{\rm d} \left( 12T - 5g^2_1 - 27g^2_2 - 96 g^2_3 \right)  - 18 Y^{}_{\rm u} Y^\dagger_{\rm u} Y^{}_{\rm d} + 18 Y^{}_{\rm d} Y^\dagger_{\rm d} Y^{}_{\rm d}  \right] 
    \nonumber
    \\
    &&  \color{gray} + \frac{1}{16\pi^2\varepsilon \cdot 2} {\rm Tr} \left( -G^{}_{\ell HN1} M^{}_N Y^\dagger_\nu  - Y^{}_\nu M^\dagger_N G^\dagger_{\ell HN1} + 2 G^\dagger_{DN} M^{}_N Y^\dagger_\nu Y^{}_\nu + 2 M^\dagger_N G^{}_{DN} Y^\dagger_\nu Y^{}_\nu \right) Y^{}_{\rm d} \;,
    \nonumber
    \\
    \delta C^{(L=1)}_5 &=&  \frac{1}{16\pi^2 \varepsilon \cdot 4}  \left[ 2\left( 4\lambda -3g^2_2 + 2T \right)C^{}_5  +  \left( Y^{}_\nu Y^\dagger_\nu - 3Y^{}_l Y^\dagger_l \right) C^{}_5 +  C^{}_5 \left( Y^{}_\nu Y^\dagger_\nu - 3 Y^{}_lY^\dagger_l  \right)^{\rm T} \right] 
    \nonumber
    \\
    &&  \color{gray} - \frac{1}{16\pi^2\varepsilon} \left[ \frac{3}{4} \left( g^2_1 + g^2_2 \right) + 2\lambda \right] \left( -Y^{}_\nu G^{\rm T}_{\ell HN1} - G^{}_{\ell HN1} Y^{\rm T}_\nu + 2 Y^{}_\nu G^\dagger_{DN} Y^{\rm T}_\nu  \right) \;,
    \nonumber
    \\
    \delta C^{(L=1)}_{BN} &=&  \frac{1}{16\pi^2 \varepsilon \cdot 12}  \left[ 41 g^2_1 C^{}_{BN} + 6 C^{}_{BN}Y^\dagger_\nu Y^{}_\nu + 6\left( Y^\dagger_\nu Y^{}_\nu \right)^{\rm T} C^{}_{BN} \right] \;,
    \nonumber
    \\
    \delta C^{(L=1)}_{HN} &=&  \frac{1}{16\pi^2 \varepsilon \cdot 4}  \left[  \left( 4T - 3 g^2_1 - 9g^2_2  +24\lambda \right) C^{}_{HN} + 6 C^{}_{HN} Y^\dagger_\nu Y^{}_\nu + 6 \left( Y^\dagger_\nu Y^{}_\nu \right)^{\rm T} C^{}_{HN} \right] 
    \nonumber
    \\
    &&   \color{gray} - \frac{1}{16\pi^2\varepsilon} \left[ 3\lambda \left( G^\dagger_{\ell HN2} Y^{}_\nu + Y^{\rm T}_\nu G^\ast_{\ell HN2} \right) + \frac{1}{2} G^\dagger_{\ell HN2} \left( Y^{}_\nu Y^\dagger_\nu + Y^{}_l Y^\dagger_l  \right) Y^{}_\nu \right.
    \nonumber
    \\
    &&  \color{gray} + \left. \frac{1}{2} Y^{\rm T}_\nu \left( Y^{}_\nu Y^\dagger_\nu + Y^{}_l Y^\dagger_l \right)^{\rm T} G^\ast_{\ell HN2}  \right] \;,
    \nonumber
    \\
    \delta G^{(L=1)}_{\ell HN1} &=& \frac{3 C^{}_5 Y^\ast_\nu}{16\pi^2 \varepsilon \cdot 2}  \color{gray} + \loopf \left[ \left( - \frac{3}{8} g^2_1 - \frac{9}{8} g^2_2 + \frac{1}{2} T \right) G^{}_{\ell HN1} + \frac{1}{4} \left( Y^{}_\nu Y^\dagger_\nu - 3Y^{}_l Y^\dagger_l \right) G^{}_{\ell HN1} \right.
    \nonumber
    \\
    &&  \color{gray} + \left. G^{}_{\ell HN1} Y^{\rm T}_\nu Y^\ast_\nu + \frac{1}{2} Y^{}_\nu G^{\rm T}_{\ell HN1} Y^\ast_\nu  - Y^{}_\nu G^\dagger_{DN} Y^{\rm T}_\nu Y^\ast_\nu \right]\;,
    \nonumber
    \\
    \delta G^{(L=1)}_{\ell HN2} &=& \frac{1}{16\pi^2 \varepsilon} Y^{}_\nu \left( 3g^{}_1 C^\dagger_{BN} - C^\dagger_{HN}  \right) \color{gray} + \loopf \left[ \left( - \frac{3}{8} g^2_1 - \frac{9}{8} g^2_2 + \frac{1}{2} T \right) G^{}_{\ell HN2} \right.
    \nonumber
    \\
    &&  \color{gray} + \left. \frac{1}{4} \left( Y^{}_\nu Y^\dagger_\nu - Y^{}_l Y^\dagger_l \right) G^{}_{\ell HN2} + \frac{3}{2} G^{}_{\ell HN2} Y^{\rm T}_\nu Y^\ast_\nu + \frac{1}{2} Y^{}_\nu G^{\rm T}_{\ell HN2} Y^\ast_\nu \right] \;,
    \nonumber
    \\
    \delta G^{(L=1)}_{DN} &=& \color{gray} \frac{1}{16\pi^2 \varepsilon \cdot 2} \left( G^{}_{DN} Y^\dagger_\nu Y^{}_\nu + Y^{\rm T}_\nu Y^\ast_\nu G^{}_{DN} - G^\dagger_{\ell HN1} Y^{}_\nu - Y^{\rm T}_\nu G^\ast_{\ell HN1} \right) \;,
\end{eqnarray}
with $T = {\rm Tr} \left( Y^{}_\nu Y^\dagger_\nu + Y^{}_l Y^\dagger_l + 3Y^{}_{\rm u} Y^\dagger_{\rm u} + 3Y^{}_{\rm d}Y^\dagger_{\rm d}  \right)$ and $T^\prime = {\rm Tr} \left[ \left( Y^{}_\nu Y^\dagger_\nu  \right)^2+ \left( Y^{}_l Y^\dagger_l \right)^2 + 3\left( Y^{}_{\rm u}  Y^\dagger_{\rm u}  \right)^2 + 3 \left( Y^{}_{\rm d}  Y^\dagger_{\rm d}  \right)^2 \right] $, and the local counterterms introduced in Eq.~\eqref{eq:local-cts} are found to be
\begin{eqnarray}\label{eq:spurious-mass-ct-exp}
    \delta Z^{(L=1)}_{M_H} &=& \frac{1}{16\pi^2\varepsilon} \left[ \frac{1}{2} \left( g^2_1 + 3g^2_2 \right) + 6\lambda - 4T + 2 {\rm Tr} \left( M^{}_N C^\dagger_{HN} + M^\dagger_N C^{}_{HN}  \right) \right] \;,
    \nonumber
    \\
    \delta Z^{(L=1)}_{M_{N_s}} &=& -  \frac{1}{16\pi^2 \varepsilon} 4C^{}_{HN} \;,
    \nonumber
    \\	
    \delta Z^{(L=1)}_{M_{\hat{B}}} &=& - \frac{1}{16\pi^2\varepsilon} \frac{79}{6} g^2_1 \;,
    \nonumber
    \\
    \delta Z^{(L=1)}_{M_{\hat{W}}} &=& -\frac{1}{16\pi^2\varepsilon} \frac{79}{6} g^2_2 \;,
    \nonumber
    \\
    \delta Z^{(L=1)}_{M_{\hat{G}}} &=& -\frac{1}{16\pi^2\varepsilon} 16 g^2_3 \;,
    \nonumber
    \\
    \delta Z^{(L=1)}_{M_{\theta{\hat{W}}}} &=& - \frac{1}{16\pi^2\varepsilon} g^2_2 \;,
    \nonumber
    \\
    \delta Z^{(L=1)}_{M_{\theta{\hat{G}}}} &=& - \frac{1}{16\pi^2\varepsilon} \frac{3}{2} g^2_3 \;.
\end{eqnarray}
In Eq.~\eqref{eq:one-loop-ct}, we restore the contributions (i.e., those in grey) from redundant operators, which will play an important role in understanding the appearance of infinite RG functions at the two-loop level. However, we do not do so for the local counterterms in Eq.~\eqref{eq:spurious-mass-ct-exp} because these local counterterms are only used to eliminate sub-divergences proportional to $M^2$ in two-loop calculations, where we will not include contributions from tree-level redundant operators. Note that all the above results are those still in the Green's basis, and one has to convert them to the physical ones by field redefinitions before deriving the one-loop RGEs in the $\nu$SMEFT up to $\Op \left( \Lambda^{-1} \right)$, which have been achieved in Ref.~\cite{Zhang:2024weq}.

\section{Two-loop Counterterms} \label{app:two-loop-ct}

We collect all two-loop counterterms in the Green's basis in this appendix. They are obtained by calculating relevant two-loop diagrams and one-loop diagrams with the insertion of one-loop counterterms listed in Appendix~\ref{app:one-loop-ct}. The latter eliminates sub-divergences in the former, guaranteeing the locality of two-loop counterterms. Since the two-loop renormalization of gauge-fixing parameters and the two-loop contributions to the local counterterms in Eq.~\eqref{eq:local-cts} are irrelevant for our purpose, we will not calculate and show them here. Both the first and second poles of $\varepsilon$ in the two-loop counterterms are given, which can be exploited to derive RGEs and verify consistency relations, respectively. However, contributions involving one-loop wave-function renormalization constants to the second pole of $\varepsilon$ are not explicitly shown for conciseness. One can easily get the explicit results by substituting relevant one-loop counterterms listed in Appendix~\ref{app:one-loop-ct} into the expressions.

The two-loop wave-function renormalization constants are given by
\begin{eqnarray}
    \delta Z^{(L=2)}_B &=& \looptf \left\{ \frac{1}{12} g^2_1 \left[ -\frac{199}{3} g^2_1 - 27 g^2_2 - 88 g^2_3 + {\rm Tr} \left( 3Y^{}_\nu Y^\dagger_\nu + 15 Y^{}_l Y^\dagger_l + 17 Y^{}_{\rm u} Y^\dagger_{\rm u}  + 5 Y^{}_{\rm d} Y^\dagger_{\rm d} \right)  \right] \right.
    \nonumber
    \\
    && + \left. 4 g^{}_1 {\rm Tr} \left[ Y^\ast_\nu \left( M^{}_N C^\dagger_{BN} + C^{}_{BN} M^\dagger_N \right) Y^{\rm T}_\nu   \right] \right\} \;,
    \nonumber
    \\
    \delta Z^{(L=2)}_W &=& - \looptf \frac{1}{4} g^2_2 \left( 3g^2_1 + \frac{35}{3} g^2_2 + 24g^2_3 - T \right) \;,
    \nonumber
    \\
    \delta Z^{(L=2)}_G &=& - \looptf g^2_3 \left[ \frac{11}{12} g^2_1 + \frac{9}{4}g^2_2 - 13 g^2_3 - {\rm Tr} \left( Y^{}_{\rm u} Y^\dagger_{\rm u} + Y^{}_{\rm d} Y^\dagger_{\rm d} \right) \right] \;,
    \nonumber
    \\
    \delta Z^{(L=2)}_H &=& \looptfs \left\{ \frac{43}{16} g^4_1 - \frac{15}{16} g^4_2 + \frac{3}{4} g^2_1 g^2_2 + \frac{1}{8} g^2_1 \tr{ - Y^{}_\nu Y^\dagger_\nu + 11 Y^{}_l Y^\dagger_l + 5 Y^{}_{\rm u} Y^\dagger_{\rm u} - 7 Y^{}_{\rm d} Y^\dagger_{\rm d} }   \right.
    \nonumber
    \\
    && - \frac{3}{8} g^2_2 T + 12 g^2_3 \tr{Y^{}_{\rm u} Y^\dagger_{\rm u} + Y^{}_{\rm d} Y^\dagger_{\rm d}} - \frac{3}{4} T^\prime + \frac{3}{2} \tr{Y^{}_\nu Y^\dagger_\nu Y^{}_l Y^\dagger_l + 3 Y^{}_{\rm u} Y^\dagger_{\rm u} Y^{}_{\rm d} Y^\dagger_{\rm d} }  
    \nonumber
    \\
    && + \frac{3}{4} \tr{ Y^{}_\nu M^\dagger_N Y^{\rm T}_\nu C^\dagger_5 + C^{}_5 Y^\ast_\nu M^{}_N Y^\dagger_\nu } + \tr{ Y^\dagger_\nu Y^{}_\nu M^\dagger_N C^{}_{HN}  + C^\dagger_{HN} M^{}_N Y^\dagger_\nu Y^{}_\nu }
    \nonumber
    \\
    && - \left. 3 g^{}_1 \tr{Y^\dagger_\nu Y^{}_\nu M^\dagger_N C^{}_{BN} + C^\dagger_{BN} M^{}_N Y^\dagger_\nu Y^{}_\nu }  \vphantom{\frac{1}{1}}\right\}
    \nonumber
    \\
    && + \looptf \left[ -\frac{431}{192} g^4_1 - \frac{9}{32} g^2_1 g^2_2 + \frac{163}{64} g^4_2 - 3\lambda^2 - \frac{5}{48} g^2_1 {\rm Tr} \left( 3Y^{}_\nu Y^\dagger_\nu + 15 Y^{}_l Y^\dagger_l  \right.\right.
    \nonumber
    \\
    && + \left.17 Y^{}_{\rm u} Y^\dagger_{\rm u} + 5 Y^{}_{\rm d} Y^\dagger_{\rm d} \right)- \frac{15}{16} g^2_2 T - 10 g^2_3 \tr{ Y^{}_{\rm u} Y^\dagger_{\rm u} + Y^{}_{\rm d} Y^\dagger_{\rm d} } + \frac{9}{8} T^\prime 
    \nonumber
    \\
    && - \frac{1}{4} \tr{ Y^{}_\nu Y^\dagger_\nu Y^{}_l Y^\dagger_l + 3 Y^{}_{\rm u} Y^\dagger_{\rm u} Y^{}_{\rm d} Y^\dagger_{\rm d} } - \frac{9}{8} {\rm Tr} \left( Y^{}_\nu M^\dagger_N Y^{\rm T}_\nu C^\dagger_5 + C^{}_5 Y^\ast_\nu M^{}_N Y^\dagger_\nu \right) 
    \nonumber
    \\
    && - \left. 2 {\rm Tr} \left( Y^\dagger_\nu Y^{}_\nu M^\dagger_N C^{}_{HN}  + C^\dagger_{HN} M^{}_N Y^\dagger_\nu Y^{}_\nu \right) - 5  g^{}_1 {\rm Tr} \left( Y^\dagger_\nu Y^{}_\nu M^\dagger_N C^{}_{BN} + C^\dagger_{BN} M^{}_N Y^\dagger_\nu Y^{}_\nu \right) \right] \;,
    \nonumber
    \\
    \delta Z^{(L=2)}_q &=& \looptfs \left[ \frac{1}{2592} g^4_1 + \frac{57}{32} g^4_2 + \frac{44}{9} g^4_3 + \frac{1}{48} g^2_1 g^2_2 + \frac{1}{27} g^2_1 g^2_3 + g^2_2 g^2_3 \right.
    \nonumber
    \\
    && + \left( \frac{17}{144} g^2_1 + \frac{15}{16} g^2_2 + \frac{8}{3} g^2_3 - \frac{1}{4} T \right) Y^{}_{\rm d} Y^\dagger_{\rm d} +  \left( \frac{53}{144} g^2_1 + \frac{15}{16} g^2_2 + \frac{8}{3} g^2_3 - \frac{1}{4} T \right) Y^{}_{\rm u} Y^\dagger_{\rm u}  
    \nonumber
    \\
    && + \left. \frac{1}{2}Y^{}_{\rm u} Y^\dagger_{\rm u} Y^{}_{\rm d} Y^\dagger_{\rm d} + \frac{1}{2} Y^{}_{\rm d} Y^\dagger_{\rm d} Y^{}_{\rm u} Y^\dagger_{\rm u} - \frac{1}{4} Y^{}_{\rm u} Y^\dagger_{\rm u} Y^{}_{\rm u} Y^\dagger_{\rm u}  - \frac{1}{4} Y^{}_{\rm d} Y^\dagger_{\rm d} Y^{}_{\rm d} Y^\dagger_{\rm d} \right]
    \nonumber
    \\
    && + \looptf \left[ \frac{247}{1728} g^4_1 - \frac{231}{64} g^4_2 - \frac{35}{3} g^4_3 + \frac{1}{32} g^2_1 g^2_2 + \frac{1}{18} g^2_1 g^2_3 + \frac{3}{2} g^2_2 g^2_3 \right. 
    \nonumber
    \\
    && + \left( - \frac{151}{288} g^2_1 - \frac{33}{32} g^2_2 + \frac{4}{3} g^2_3 + \frac{3}{8} T  \right) Y^{}_{\rm d} Y^\dagger_{\rm d} + \left( - \frac{139}{288} g^2_1 - \frac{33}{32} g^2_2 + \frac{4}{3} g^2_3 + \frac{3}{8} T  \right) Y^{}_{\rm u} Y^\dagger_{\rm u}  
    \nonumber
    \\
    && + \left.\frac{1}{8} Y^{}_{\rm d} Y^\dagger_{\rm d} Y^{}_{\rm d} Y^\dagger_{\rm d} + \frac{1}{8} Y^{}_{\rm u} Y^\dagger_{\rm u} Y^{}_{\rm u} Y^\dagger_{\rm u}  \right] \;,
    \nonumber
    \\
    \delta Z^{(L=2)}_u &=& \looptfs \left[ \frac{8}{81} g^4_1  +  \frac{44}{9} g^4_3 + \frac{16}{27} g^2_1 g^2_3 + \left( \frac{83}{72} g^2_1 + \frac{9}{8} g^2_2 + \frac{16}{3} g^2_3 - \frac{1}{2} T  \right)  Y^\dagger_{\rm u} Y^{}_{\rm u}  \right.
    \nonumber
    \\
    && -  \left.\frac{1}{4} Y^\dagger_{\rm u} Y^{}_{\rm u} Y^\dagger_{\rm u} Y^{}_{\rm u} + \frac{3}{4} Y^\dagger_{\rm u} Y^{}_{\rm d} Y^\dagger_{\rm d} Y^{}_{\rm u} \right] 
    \nonumber
    \\
    &&+  \looptf \left[ \frac{131}{54} g^4_1  - \frac{35}{3} g^4_3 + \frac{8}{9} g^2_1 g^2_3 + \left( - \frac{49}{144} g^2_1 - \frac{51}{16} g^2_2 + \frac{8}{3} g^2_3 + \frac{3}{4} T  \right)  Y^\dagger_{\rm u} Y^{}_{\rm u}  \right.
    \nonumber
    \\
    && + \left. \frac{1}{8} Y^\dagger_{\rm u} Y^{}_{\rm u} Y^\dagger_{\rm u} Y^{}_{\rm u} +  \frac{1}{8} Y^\dagger_{\rm u} Y^{}_{\rm d} Y^\dagger_{\rm d} Y^{}_{\rm u} \right] \;,
    \nonumber
    \\
    \delta Z^{(L=2)}_d &=& \looptfs \left[ \frac{1}{162} g^4_1  + \frac{44}{9} g^4_3 + \frac{4}{27} g^2_1 g^2_3 + \left( \frac{23}{72} g^2_1 + \frac{9}{8} g^2_2 + \frac{16}{3} g^2_3 - \frac{1}{2} T \right)  Y^\dagger_{\rm d} Y^{}_{\rm d}  \right.
    \nonumber
    \\
    && - \left. \frac{1}{4} Y^\dagger_{\rm d} Y^{}_{\rm d} Y^\dagger_{\rm d} Y^{}_{\rm d} +  \frac{3}{4} Y^\dagger_{\rm d} Y^{}_{\rm u} Y^\dagger_{\rm u} Y^{}_{\rm d} \right] 
    \nonumber
    \\
    && + \looptf \left[ \frac{125}{216} g^4_1  - \frac{35}{3} g^4_3 + \frac{2}{9} g^2_1 g^2_3 + \left( - \frac{133}{144} g^2_1 - \frac{51}{16} g^2_2 + \frac{8}{3} g^2_3 + \frac{3}{4} T  \right)  Y^\dagger_{\rm d} Y^{}_{\rm d}  \right.
    \nonumber
    \\
    && + \left. \frac{1}{8} Y^\dagger_{\rm d} Y^{}_{\rm d} Y^\dagger_{\rm d} Y^{}_{\rm d} + \frac{1}{8} Y^\dagger_{\rm d} Y^{}_{\rm u} Y^\dagger_{\rm u} Y^{}_{\rm d} \right] \;,
    \nonumber
    \\
    \delta Z^{(L=2)}_\ell &=& \looptfs \left[ \frac{1}{32} g^4_1 + \frac{57}{32} g^4_2 + \frac{3}{16} g^2_1 g^2_2 + \frac{1}{16} g^2_1 \left( 17 Y^{}_l Y^\dagger_l + 5 Y^{}_\nu Y^\dagger_\nu \right) \right.
    \nonumber
    \\
    && + \left(  \frac{15}{16} g^2_2  -  \frac{1}{4} T \right) \left( Y^{}_l Y^\dagger_l + Y^{}_\nu Y^\dagger_\nu \right)  - \frac{1}{4} \left( Y^{}_l Y^\dagger_l Y^{}_l Y^\dagger_l + Y^{}_\nu Y^\dagger_\nu Y^{}_\nu Y^\dagger_\nu \right) + \frac{1}{2} \left(  Y^{}_\nu Y^\dagger_\nu Y^{}_l Y^\dagger_l \right.
    \nonumber
    \\
    && + \left. Y^{}_l Y^\dagger_l Y^{}_\nu Y^\dagger_\nu  \right) + \frac{3}{8} C^{}_5 Y^\ast_\nu M^{}_N Y^\dagger_\nu + \frac{3}{8} Y^{}_\nu M^\dagger_N Y^{\rm T}_\nu C^\dagger_5 + Y^{}_\nu \left( M^\dagger_N C^{}_{HN} + C^\dagger_{HN} M^{}_N \right) Y^\dagger_\nu 
    \nonumber
    \\
    && - \left.  3  g^{}_1 Y^{}_\nu \left( M^\dagger_N C^{}_{BN} + C^\dagger_{BN} M^{}_N \right) Y^\dagger_\nu \right] 
    \nonumber
    \\
    && + \looptf \left[ \frac{85}{64} g^4_1 - \frac{231}{64} g^4_2 + \frac{9}{32} g^2_1 g^2_2 - \frac{1}{32} g^2_1 \left( 7 Y^{}_l Y^\dagger_l + 11 Y^{}_\nu Y^\dagger_\nu \right) \right.
    \nonumber
    \\
    && - \frac{33}{32} g^2_2 \left( Y^{}_l Y^\dagger_l + Y^{}_\nu Y^\dagger_\nu \right) + \frac{3}{8} T \left( Y^{}_l Y^\dagger_l + Y^{}_\nu Y^\dagger_\nu \right) + \frac{1}{8} Y^{}_l Y^\dagger_l Y^{}_l Y^\dagger_l + \frac{1}{8} Y^{}_\nu Y^\dagger_\nu Y^{}_\nu Y^\dagger_\nu 
    \nonumber
    \\
    && - \left. \frac{3}{16} C^{}_5 Y^\ast_\nu M^{}_N Y^\dagger_\nu - \frac{3}{16} Y^{}_\nu M^\dagger_N Y^{\rm T}_\nu C^\dagger_5 -  \frac{1}{2}  g^{}_1 Y^{}_\nu \left( M^\dagger_N C^{}_{BN} + C^\dagger_{BN} M^{}_N \right) Y^\dagger_\nu \right] \;,
    \nonumber
    \\
    \delta Z^{(L=2)}_e &=& \looptfs \left[ \frac{1}{2} g^4_1 + \left( \frac{23}{8} g^2_1 + \frac{9}{8} g^2_2 - \frac{1}{2} T \right) Y^\dagger_l Y^{}_l - \frac{1}{4} Y^\dagger_l \left( Y^{}_l Y^\dagger_l - 3 Y^{}_\nu Y^\dagger_\nu \right) Y^{}_l  \right]
    \nonumber
    \\
    && + \looptf \left[ \frac{47}{8} g^4_1 + \left( \frac{11}{16} g^2_1 - \frac{51}{16} g^2_2 + \frac{3}{4} T \right) Y^\dagger_l Y^{}_l + \frac{1}{8} Y^\dagger_l \left( Y^{}_l Y^\dagger_l + Y^{}_\nu Y^\dagger_\nu \right) Y^{}_l  \right] \;,
    \nonumber
    \\
    \delta Z^{(L=2)}_{N_L} &=& \looptfs \left\{  \left[ \frac{3}{8} \left( g^2_1 + 3g^2_2 \right) - \frac{1}{2} T \right] \left( Y^\dagger_\nu Y^{}_\nu \right)^{\rm T} + \frac{1}{4} Y^{\rm T}_\nu \left[ 3 \left( Y^{}_l Y^\dagger_l \right)^{\rm T} - \left( Y^{}_\nu Y^\dagger_\nu \right)^{\rm T} \right] Y^\ast_\nu  \right.
    \nonumber
    \\
    && + 3 g^{}_1 \left[ C^{}_{BN} M^\dagger_N \left( Y^\dagger_\nu Y^{}_\nu \right)^{\rm T} + \left( Y^\dagger_\nu Y^{}_\nu \right)^{\rm T} M^{}_N C^\dagger_{BN} \right] + C^{}_{HN} M^\dagger_N \left( Y^\dagger_\nu Y^{}_\nu \right)^{\rm T} 
    \nonumber
    \\
    && + \left.  \left( Y^\dagger_\nu Y^{}_\nu \right)^{\rm T} M^{}_N C^\dagger_{HN}  \right\}
    \nonumber
    \\
    && + \looptf \left\{ \left( - \frac{17}{16} g^2_1 - \frac{51}{16} g^2_2 + \frac{3}{4} T \right) \left( Y^\dagger_\nu Y^{}_\nu \right)^{\rm T} + \frac{1}{8} Y^{\rm T}_\nu \left[ \left( Y^{}_l Y^\dagger_l \right)^{\rm T} + \left( Y^{}_\nu Y^\dagger_\nu \right)^{\rm T} \right] Y^\ast_\nu  \right.
    \nonumber
    \\
    && + \frac{1}{2}  g^{}_1 \left[ C^{}_{BN} M^\dagger_N \left( Y^\dagger_\nu Y^{}_\nu \right)^{\rm T} + \left( Y^\dagger_\nu Y^{}_\nu \right)^{\rm T} M^{}_N C^\dagger_{BN} \right]  -  \frac{1}{2} C^{}_{HN} M^\dagger_N \left( Y^\dagger_\nu Y^{}_\nu \right)^{\rm T} 
    \nonumber
    \\
    && - \left.\frac{1}{2} \left( Y^\dagger_\nu Y^{}_\nu \right)^{\rm T} M^{}_N C^\dagger_{HN}  \right\}\;.
\end{eqnarray}
The renormalization constants for the renormalizable couplings are
\begin{eqnarray}
    \delta Z^{(L=2)}_{m^2} &=&  \looptfs \left\{ -\frac{67}{32} g^4_1 + \frac{9}{16} g^2_1 g^2_2 +\frac{141}{32} g^4_2 + \frac{1}{8} g^2_1 {\rm Tr} \left(  Y^{}_\nu Y^\dagger_\nu - 11 Y^{}_l Y^\dagger_l - 5 Y^{}_{\rm u} Y^\dagger_{\rm u} + 7 Y^{}_{\rm d} Y^\dagger_{\rm d} \right)    \right.
    \nonumber
    \\
    && + \frac{3}{8} g^2_2 T - 12 g^2_3 {\rm Tr} \left( Y^{}_{\rm u} Y^\dagger_{\rm u} + Y^{}_{\rm d} Y^\dagger_{\rm d} \right) - \lambda \left( 6g^2_1 + 18g^2_2 - 54\lambda - 6T \right) - \frac{3}{2} {\rm Tr} \left( Y^{}_\nu Y^\dagger_\nu Y^{}_l Y^\dagger_l \right.
    \nonumber
    \\
    && + \left. 3Y^{}_{\rm u} Y^\dagger_{\rm u} Y^{}_{\rm d} Y^\dagger_{\rm d} \right) - \frac{9}{4} T^\prime - \frac{1}{m^2} {\rm Tr} \left[ Y^\dagger_\nu Y^{}_\nu M^\dagger_N \left( Y^\dagger_\nu Y^{}_\nu \right)^{\rm T} M^{}_N + \frac{5}{2} Y^{}_\nu Y^\dagger_\nu Y^{}_\nu M^\dagger_N M^{}_N Y^\dagger_\nu \right.
    \nonumber
    \\
    && - \left. \frac{3}{2} Y^{}_l Y^\dagger_l Y^{}_\nu M^\dagger_N M^{}_N Y^\dagger_\nu   \right] + \frac{1}{m^2}{\rm Tr} \left[ \left( \frac{1}{2} g^2_1 + \frac{3}{2} g^2_2 - 6\lambda \right) Y^\dagger_\nu Y^{}_\nu M^\dagger_N M^{}_N \right.
    \nonumber
    \\
    && -  \left.  \left( \frac{1}{2} g^2_1 + \frac{3}{2} g^2_2 - 12\lambda \right) \left( C^{}_{HN}   M^\dagger_N M^{}_N M^\dagger_N  + M^{}_N M^\dagger_N M^{}_N C^\dagger_{HN} \right)  \right] 
    \nonumber
    \\
    && + g^{}_1 {\rm Tr} \left[ 3 Y^\dagger_\nu Y^{}_\nu M^\dagger_N C^{}_{BN}  + 3 C^\dagger_{BN} M^{}_N Y^\dagger_\nu Y^{}_\nu  - \frac{1}{m^2} \left( 3 Y^\dagger_\nu Y^{}_\nu M^\dagger_N M^{}_N M^\dagger_N C^{}_{BN} \right. \right.
    \nonumber
    \\
    && + \left.\left. 3C^\dagger_{BN} M^{}_N M^\dagger_N M^{}_N Y^\dagger_\nu Y^{}_\nu + 6 Y^\dagger_\nu Y^{}_\nu M^\dagger_N C^{}_{BN} M^\dagger_N M^{}_N +  6 M^\dagger_N M^{}_N C^\dagger_{BN} M^{}_N Y^\dagger_\nu Y^{}_\nu  \right) \right] 
    \nonumber
    \\
    && + {\rm Tr} \left[ \frac{9}{4} \left( Y^\dagger_\nu C^{}_5 Y^\ast_\nu M^{}_N + M^\dagger_N Y^{\rm T}_\nu C^\dagger_5 Y^{}_\nu  \right) + \frac{3}{2m^2} \left( Y^\dagger_\nu C^{}_5 Y^\ast_\nu M^{}_N M^\dagger_N M^{}_N \right.\right.
    \nonumber
    \\
    && + \left.\left. M^\dagger_N M^{}_N M^\dagger_N Y^{\rm T}_\nu C^\dagger_5 Y^{}_\nu \right)  \right] + {\rm Tr} \left[ 15 \left( Y^\dagger_\nu Y^{}_\nu M^\dagger_N C^{}_{HN} + C^\dagger_{HN} M^{}_N Y^\dagger_\nu Y^{}_\nu \right) \right.
    \nonumber
    \\
    && + \frac{1}{m^2} \left( 5Y^\dagger_\nu Y^{}_\nu M^\dagger_N M^{}_N M^\dagger_N C^{}_{HN}  + 5 C^\dagger_{HN} M^{}_N M^\dagger_N M^{}_N Y^\dagger_\nu Y^{}_\nu \right.
    \nonumber
    \\
    && + \left.\left.\left.   4Y^\dagger_\nu Y^{}_\nu M^\dagger_N C^{}_{HN} M^\dagger_N M^{}_N + 4 M^\dagger_N M^{}_N C^\dagger_{HN} M^{}_N Y^\dagger_\nu Y^{}_\nu \right) \right] \vphantom{\frac{1}{1}} \right\}   - \delta Z^{(L=1)}_H \delta Z^{(L=1)}_{m^2}  
    \nonumber
    \\
    && +  \looptf \left\{ \frac{557}{192} g^4_1 + \frac{15}{32} g^2_1 g^2_2 - \frac{145}{64} g^4_2 + \frac{1}{48} g^2_1 {\rm Tr} \left( 15 Y^{}_\nu Y^\dagger_\nu + 75Y^{}_l Y^\dagger_l + 85 Y^{}_{\rm u} Y^\dagger_{\rm u}  \right. \right.
    \nonumber
    \\
    && + \left. 25 Y^{}_{\rm d} Y^\dagger_{\rm d} \right) + \frac{15}{16} g^2_2 T + 10g^2_3 {\rm Tr} \left( Y^{}_{\rm u} Y^\dagger_{\rm u} + Y^{}_{\rm d} Y^\dagger_{\rm d} \right) + \lambda \left( 6g^2_1 + 18g^2_2 - 15\lambda -6T \right) 
    \nonumber
    \\
    && - \frac{7}{4} {\rm Tr} \left( Y^{}_\nu Y^\dagger_\nu Y^{}_l Y^\dagger_l + 3Y^{}_{\rm u} Y^\dagger_{\rm u} Y^{}_{\rm d} Y^\dagger_{\rm d} \right) - \frac{9}{8} T^\prime + \frac{1}{m^2} {\rm Tr} \left[ 2Y^\dagger_\nu Y^{}_\nu M^\dagger_N \left( Y^\dagger_\nu Y^{}_\nu \right)^{\rm T} M^{}_N \right.
    \nonumber
    \\
    && + \left. \frac{7}{2} Y^{}_\nu Y^\dagger_\nu Y^{}_\nu M^\dagger_N M^{}_N Y^\dagger_\nu - \frac{1}{2} Y^{}_l Y^\dagger_l Y^{}_\nu M^\dagger_N M^{}_N Y^\dagger_\nu  \right] + {\rm Tr} \left[ \frac{9}{8} \left( Y^\dagger_\nu C^{}_5 Y^\ast_\nu M^{}_N + M^\dagger_N Y^{\rm T}_\nu C^\dagger_5 Y^{}_\nu  \right) \right.
    \nonumber
    \\
    && - \left. \frac{3}{2m^2} \left( Y^\dagger_\nu C^{}_5 Y^\ast_\nu M^{}_N M^\dagger_N M^{}_N + M^\dagger_N M^{}_N M^\dagger_N Y^{\rm T}_\nu C^\dagger_5 Y^{}_\nu \right)  \right] - {\rm Tr} \left[ \frac{2}{3} \left( Y^\dagger_\nu Y^{}_\nu M^\dagger_N C^{}_{HN} \right.\right.
    \nonumber
    \\
    && + \left.\left. C^\dagger_{HN} M^{}_N Y^\dagger_\nu Y^{}_\nu \right) + \frac{3}{m^2} \left( Y^\dagger_\nu Y^{}_\nu M^\dagger_N M^{}_N M^\dagger_N C^{}_{HN} + C^\dagger_{HN} M^{}_N M^\dagger_N M^{}_N Y^\dagger_\nu Y^{}_\nu \right.\right.
    \nonumber
    \\
    && + \left.\left. 2Y^\dagger_\nu Y^{}_\nu M^\dagger_N C^{}_{HN} M^\dagger_N M^{}_N + 2 M^\dagger_N M^{}_N C^\dagger_{HN} M^{}_N Y^\dagger_\nu Y^{}_\nu \right) \right] + g^{}_1 {\rm Tr} \left[ 7 Y^\dagger_\nu Y^{}_\nu M^\dagger_N C^{}_{BN} \right.
    \nonumber
    \\
    && + 7 C^\dagger_{BN} M^{}_N Y^\dagger_\nu Y^{}_\nu - \frac{1}{m^2} \left( 5 Y^\dagger_\nu Y^{}_\nu M^\dagger_N M^{}_N M^\dagger_N C^{}_{BN} + 5C^\dagger_{BN} M^{}_N M^\dagger_N M^{}_N Y^\dagger_\nu Y^{}_\nu \right.
    \nonumber
    \\
    && + \left.\left.\left. 4 Y^\dagger_\nu Y^{}_\nu M^\dagger_N C^{}_{BN} M^\dagger_NM^{}_N + 4 M^\dagger_N M^{}_N C^\dagger_{BN} M^{}_N Y^\dagger_\nu Y^{}_\nu  \right) \right] \vphantom{\frac{1}{1}}\right\} \;,
    \nonumber
    \\
    \lambda \delta Z^{(L=2)}_\lambda &=& \looptfs \left\{ \frac{85}{64} g^6_1 - \frac{87}{64} g^6_2 - \frac{11}{64} g^2_1 g^4_2 + \frac{97}{64} g^4_1 g^2_2 - 3\lambda g^4_1 + \frac{51}{4} \lambda g^4_2 + \frac{15}{4} \lambda g^2_1 g^2_2 - 15\lambda^2 g^2_1   \right.
    \nonumber
    \\
    && - 45 \lambda^2 g^2_2 + 144 \lambda^3 - \frac{3}{16} g^4_1 T - \frac{9}{16} g^4_2 T -  \frac{3}{8} g^2_1 g^2_2 T + \frac{1}{4} \lambda g^2_1 {\rm Tr} \left( Y^{}_\nu Y^\dagger_\nu - 11Y^{}_l Y^\dagger_l + 7 Y^{}_{\rm d} Y^\dagger_{\rm d} \right.
    \nonumber
    \\
    && - \left. 5 Y^{}_{\rm u} Y^\dagger_{\rm u} \right) + \frac{3}{4} \lambda g^2_2 T + \frac{1}{2} g^2_1 \left( Y^{}_\nu Y^\dagger_\nu Y^{}_\nu Y^\dagger_\nu + 7 Y^{}_l Y^\dagger_l Y^{}_l Y^\dagger_l + Y^{}_{\rm d} Y^\dagger_{\rm d} Y^{}_{\rm d} Y^\dagger_{\rm d} + 7 Y^{}_{\rm u} Y^\dagger_{\rm u} Y^{}_{\rm u} Y^\dagger_{\rm u} \right)
    \nonumber
    \\
    &&  + \frac{3}{2} g^2_2 T^\prime  - 24\lambda g^2_3 {\rm Tr} \left( Y^{}_{\rm u} Y^\dagger_{\rm u} + Y^{}_{\rm d} Y^\dagger_{\rm d} \right) + 12\lambda^2 T - \frac{21}{2} \lambda T^\prime 
    \nonumber
    \\
    &&  - 3\lambda  {\rm Tr} \left( Y^{}_\nu Y^\dagger_\nu Y^{}_l Y^\dagger_l + 3 Y^{}_{\rm u }Y^\dagger_{\rm u} Y^{}_{\rm d} Y^\dagger_{\rm d} \right) + 24 g^2_3 {\rm Tr} \left( Y^{}_{\rm d} Y^\dagger_{\rm d} Y^{}_{\rm d} Y^\dagger_{\rm d} + Y^{}_{\rm u} Y^\dagger_{\rm u} Y^{}_{\rm u} Y^\dagger_{\rm u} \right) 
    \nonumber
    \\
    && + \frac{1}{2} {\rm Tr} \left[ -3 \left( Y^{}_\nu Y^\dagger_\nu Y^{}_\nu Y^\dagger_\nu  Y^{}_\nu Y^\dagger_\nu + Y^{}_l Y^\dagger_l Y^{}_l Y^\dagger_l  Y^{}_l Y^\dagger_l \right) +3 \left( Y^{}_\nu Y^\dagger_\nu Y^{}_\nu Y^\dagger_\nu  Y^{}_l Y^\dagger_l + Y^{}_l Y^\dagger_l Y^{}_l Y^\dagger_l  Y^{}_\nu Y^\dagger_\nu \right) \right. 
    \nonumber
    \\
    && - \left. 9\left( Y^{}_{\rm u} Y^\dagger_{\rm u}  Y^{}_{\rm u}  Y^\dagger_{\rm u}   Y^{}_{\rm u}  Y^\dagger_{\rm u} + Y^{}_{\rm d} Y^\dagger_{\rm d}  Y^{}_{\rm d}  Y^\dagger_{\rm d}   Y^{}_{\rm d}  Y^\dagger_{\rm d}  \right)  + 9 \left( Y^{}_{\rm u} Y^\dagger_{\rm u}  Y^{}_{\rm u}  Y^\dagger_{\rm u}   Y^{}_{\rm d}  Y^\dagger_{\rm d} + Y^{}_{\rm d} Y^\dagger_{\rm d}  Y^{}_{\rm d}  Y^\dagger_{\rm d}   Y^{}_{\rm u}  Y^\dagger_{\rm u}  \right)  \right] 
    \nonumber
    \\
    && - \frac{1}{8} \left( g^2_1 + 9g^2_2 - 92 \lambda \right) {\rm Tr} \left(  C^{}_5 Y^\ast_\nu M^{}_N Y^\dagger_\nu + Y^{}_\nu M^\dagger_N Y^{\rm T}_\nu C^\dagger_5 \right)  
    \nonumber
    \\
    &&  - \left( 2g^2_1 + 6 g^2_2 - 58 \lambda \right) {\rm Tr} \left( C^{}_{HN} Y^\dagger_\nu Y^{}_\nu M^\dagger_N +  M^{}_N Y^\dagger_\nu Y^{}_\nu C^\dagger_{HN} \right) - 6  g^{}_1 \lambda {\rm Tr} \left( C^{}_{BN} M^\dagger_N Y^{\rm T}_\nu Y^\ast_\nu  \right.
    \nonumber
    \\
    && + \left. Y^{\rm T}_\nu Y^\ast_\nu M^{}_N C^\dagger_{BN}  \right) - 9 g^{}_1 {\rm Tr} \left( C^{}_{BN} Y^\dagger_\nu Y^{}_\nu Y^\dagger_\nu Y^{}_\nu M^\dagger_N + M^{}_N  Y^\dagger_\nu Y^{}_\nu Y^\dagger_\nu Y^{}_\nu C^\dagger_{BN} \right) 
    \nonumber
    \\
    && + {\rm Tr} \left[ -3C^{}_{HN} Y^\dagger_\nu Y^{}_l Y^\dagger_l Y^{}_\nu M^\dagger_N -3 M^{}_N Y^\dagger_\nu Y^{}_l Y^\dagger_l Y^{}_\nu C^\dagger_{HN} + 10  C^{}_{HN} Y^\dagger_\nu Y^{}_\nu Y^\dagger_\nu Y^{}_\nu M^\dagger_N \right.
    \nonumber
    \\
    &&  +  10 M^{}_N Y^\dagger_\nu Y^{}_\nu Y^\dagger_\nu Y^{}_\nu C^\dagger_{HN} + 5 \left( C^{}_{HN} Y^\dagger_\nu Y^{}_\nu M^\dagger_N Y^{\rm T}_\nu Y^\ast_\nu + Y^{\rm T}_\nu Y^\ast_\nu M^{}_N Y^\dagger_\nu Y^{}_\nu C^\dagger_{HN} \right) 
    \nonumber
    \\
    && -  \frac{3}{2} \left( C^{}_5 Y^\ast_\nu M^{}_\nu Y^\dagger_\nu Y^{}_l Y^\dagger_l + Y^{}_l Y^\dagger_l Y^{}_\nu M^\dagger_N Y^{\rm T}_\nu C^\dagger_5  \right) + \frac{9}{2} \left( C^{}_5 Y^\ast_\nu M^{}_N Y^\dagger_\nu Y^{}_\nu Y^\dagger_\nu  \right.
    \nonumber
    \\
    && +  \left.\left.\left. Y^{}_\nu Y^\dagger_\nu Y^{}_\nu M^\dagger_N Y^{\rm T}_\nu C^\dagger_5 \right) \right] \right\} - \lambda \left( \delta Z^{(L=1)}_H \right)^2 -2\lambda \delta Z^{(L=1)}_\lambda \delta Z^{(L=1)}_H
    \nonumber
    \\
    && + \looptf \left\{ - \frac{379}{192} g^6_1 + \frac{305}{64} g^6_2 - \frac{289}{192} g^2_1 g^4_2 - \frac{559}{192} g^4_1 g^2_2 + \frac{629}{96} \lambda g^4_1 - \frac{73}{32} \lambda g^4_2 + \frac{39}{16} \lambda g^2_1 g^2_2  \right.
    \nonumber
    \\
    && + 9\lambda^2 g^2_1 + 27 \lambda^2 g^2_2 - 78\lambda^3 - \frac{1}{16} g^4_1 {\rm Tr} \left( Y^{}_\nu Y^\dagger_\nu + 25 Y^{}_l Y^\dagger_l - 5Y^{}_{\rm d} Y^\dagger_{\rm d} + 19 Y^{}_{\rm u} Y^\dagger_{\rm u} \right) - \frac{3}{16} g^4_2 T 
    \nonumber
    \\
    && + \frac{1}{8} g^2_1 g^2_2 {\rm Tr} \left( -Y^{}_\nu Y^\dagger_\nu + 11Y^{}_l Y^\dagger_l + 9 Y^{}_{\rm d} Y^\dagger_{\rm d} + 21Y^{}_{\rm u} Y^\dagger_{\rm u} \right) + \frac{5}{24} \lambda g^2_1 {\rm Tr} \left( 3Y^{}_\nu Y^\dagger_\nu + 15Y^{}_l Y^\dagger_l \right.
    \nonumber
    \\
    && + \left. 5 Y^{}_{\rm d} Y^\dagger_{\rm d} + 17 Y^{}_{\rm u} Y^\dagger_{\rm u} \right) + \frac{15}{8} \lambda g^2_2 T + 20\lambda g^2_3 {\rm Tr} \left( Y^{}_{\rm u} Y^\dagger_{\rm u} + Y^{}_{\rm d} Y^\dagger_{\rm d} \right) - 12\lambda^2 T - \frac{1}{4} \lambda T^\prime 
    \nonumber
    \\
    &&  - \frac{7}{2} \lambda  {\rm Tr} \left( Y^{}_\nu Y^\dagger_\nu Y^{}_l Y^\dagger_l + 3 Y^{}_{\rm u }Y^\dagger_{\rm u} Y^{}_{\rm d} Y^\dagger_{\rm d} \right) + g^2_1 {\rm Tr} \left( - Y^{}_l Y^\dagger_l Y^{}_l Y^\dagger_l + \frac{1}{3} Y^{}_{\rm d} Y^\dagger_{\rm d} Y^{}_{\rm d} Y^\dagger_{\rm d} \right.
    \nonumber
    \\
    && - \left. \frac{2}{3} Y^{}_{\rm u} Y^\dagger_{\rm u} Y^{}_{\rm u} Y^\dagger_{\rm u} \right) - 8 g^2_3 {\rm Tr} \left( Y^{}_{\rm d} Y^\dagger_{\rm d} Y^{}_{\rm d} Y^\dagger_{\rm d} + Y^{}_{\rm u} Y^\dagger_{\rm u} Y^{}_{\rm u} Y^\dagger_{\rm u} \right) 
    \nonumber
    \\
    && + \frac{1}{2} {\rm Tr} \left[ 5 \left( Y^{}_\nu Y^\dagger_\nu Y^{}_\nu Y^\dagger_\nu  Y^{}_\nu Y^\dagger_\nu + Y^{}_l Y^\dagger_l Y^{}_l Y^\dagger_l  Y^{}_l Y^\dagger_l \right) - \left( Y^{}_\nu Y^\dagger_\nu Y^{}_\nu Y^\dagger_\nu  Y^{}_l Y^\dagger_l + Y^{}_l Y^\dagger_l Y^{}_l Y^\dagger_l  Y^{}_\nu Y^\dagger_\nu \right) \right.
    \nonumber
    \\
    && + \left. 15\left( Y^{}_{\rm u} Y^\dagger_{\rm u}  Y^{}_{\rm u}  Y^\dagger_{\rm u}   Y^{}_{\rm u}  Y^\dagger_{\rm u} + Y^{}_{\rm d} Y^\dagger_{\rm d}  Y^{}_{\rm d}  Y^\dagger_{\rm d}   Y^{}_{\rm d}  Y^\dagger_{\rm d}  \right) - 3 \left( Y^{}_{\rm u} Y^\dagger_{\rm u}  Y^{}_{\rm u}  Y^\dagger_{\rm u}   Y^{}_{\rm d}  Y^\dagger_{\rm d} + Y^{}_{\rm d} Y^\dagger_{\rm d}  Y^{}_{\rm d}  Y^\dagger_{\rm d}   Y^{}_{\rm u}  Y^\dagger_{\rm u}  \right)  \right]
    \nonumber
    \\
    &&  - \frac{1}{8} \left( g^2_1 + g^2_2 - 10 \lambda \right) {\rm Tr} \left(  C^{}_5 Y^\ast_\nu M^{}_N Y^\dagger_\nu + Y^{}_\nu M^\dagger_N Y^{\rm T}_\nu C^\dagger_5 \right) - 8 \lambda \left( C^{}_{HN} Y^\dagger_\nu Y^{}_\nu M^\dagger_N  \right.
    \nonumber
    \\
    && + \left. M^{}_N Y^\dagger_\nu Y^{}_\nu C^\dagger_{HN} \right)  -  g^{}_1 \left( 3g^2_1 + 3g^2_2 + 10 \lambda \right) {\rm Tr} \left( C^{}_{BN} M^\dagger_N Y^{\rm T}_\nu Y^\ast_\nu + Y^{\rm T}_\nu Y^\ast_\nu M^{}_N C^\dagger_{BN}  \right)
    \nonumber
    \\
    && - 3 g^{}_1 {\rm Tr} \left( C^{}_{BN} Y^\dagger_\nu Y^{}_\nu Y^\dagger_\nu Y^{}_\nu M^\dagger_N +  M^{}_N  Y^\dagger_\nu Y^{}_\nu Y^\dagger_\nu Y^{}_\nu C^\dagger_{BN} \right) + {\rm Tr} \left[ C^{}_{HN} Y^\dagger_\nu Y^{}_l Y^\dagger_l Y^{}_\nu M^\dagger_N \right.
    \nonumber
    \\
    && + M^{}_N Y^\dagger_\nu Y^{}_l Y^\dagger_l Y^{}_\nu C^\dagger_{HN} - 14 \left( C^{}_{HN} Y^\dagger_\nu Y^{}_\nu Y^\dagger_\nu Y^{}_\nu M^\dagger_N  +  M^{}_N Y^\dagger_\nu Y^{}_\nu Y^\dagger_\nu Y^{}_\nu C^\dagger_{HN} \right)  
    \nonumber
    \\
    && - 7 \left( C^{}_{HN} Y^\dagger_\nu Y^{}_\nu M^\dagger_N Y^{\rm T}_\nu Y^\ast_\nu + Y^{\rm T}_\nu Y^\ast_\nu M^{}_N Y^\dagger_\nu Y^{}_\nu C^\dagger_{HN} \right) + \frac{1}{2} \left( C^{}_5 Y^\ast_\nu M^{}_\nu Y^\dagger_\nu Y^{}_l Y^\dagger_l \right.
    \nonumber
    \\ 
    && + \left.\left.\left.Y^{}_l Y^\dagger_l Y^{}_\nu M^\dagger_N Y^{\rm T}_\nu C^\dagger_5  \right) -  \frac{15}{2} \left( C^{}_5 Y^\ast_\nu M^{}_N Y^\dagger_\nu Y^{}_\nu Y^\dagger_\nu  + Y^{}_\nu Y^\dagger_\nu Y^{}_\nu M^\dagger_N Y^{\rm T}_\nu C^\dagger_5 \right) \right] \right\}  \;,
    \nonumber
    \\
    \delta M^{(L=2)}_N &=& \looptfs \left\{ - \left[ \frac{3}{16} \left( g^2_1 + 3 g^2_2 \right) - \frac{1}{4} T \right] \left[ M^{}_N Y^\dagger_\nu Y^{}_\nu + \left( Y^\dagger_\nu Y^{}_\nu \right)^{\rm T} M^{}_N \right] \right.
    \nonumber
    \\
    && - \frac{1}{8} M^{}_N Y^\dagger_\nu \left( 3 Y^{}_l Y^\dagger_l - Y^{}_\nu Y^\dagger_\nu \right) Y^{}_\nu  -\frac{1}{8} Y^{\rm T}_\nu \left[ 3 \left( Y^{}_l Y^\dagger_l \right)^{\rm T} - \left( Y^{}_\nu Y^\dagger_\nu \right)^{\rm T} \right] Y^\ast_\nu M^{}_N 
    \nonumber
    \\
    && - \frac{1}{2} C^{}_{HN} M^\dagger_N \left( Y^\dagger_\nu Y^{}_\nu \right)^{\rm T} M^{}_N - \frac{1}{2} M^{}_N Y^\dagger_\nu Y^{}_\nu M^\dagger_N C^{}_{HN}  -  \frac{1}{2} \left( Y^\dagger_\nu Y^{}_\nu \right)^{\rm T} M^{}_N C^\dagger_{HN} M^{}_N
    \nonumber
    \\
    &&   - \frac{1}{2}M^{}_N C^\dagger_{HN} M^{}_N Y^\dagger_\nu Y^{}_\nu - \frac{3}{2}  g^{}_1 \left[  C^{}_{BN} M^\dagger_N \left( Y^\dagger_\nu Y^{}_\nu \right)^{\rm T} M^{}_N -  M^{}_N Y^\dagger_\nu Y^{}_\nu M^\dagger_N C^{}_{BN} \right.
    \nonumber
    \\
    && - \left. M^{}_N C^\dagger_{BN} M^{}_N Y^\dagger_\nu Y^{}_\nu +  \left( Y^\dagger_\nu Y^{}_\nu \right)^{\rm T} M^{}_N C^\dagger_{BN} M^{}_N \right]  + 4 \tr{M^{}_N Y^\dagger_\nu Y^{}_\nu M^\dagger_N} C^{}_{HN} 
    \nonumber
    \\
    && + m^2\left( 3g^2_1 + 9g^2_2 - 24\lambda - 4T \right) C^{}_{HN}- 3m^2 \left[ C^{}_{HN} Y^\dagger_\nu Y^{}_\nu + \left( Y^\dagger_\nu Y^{}_\nu \right)^{\rm T} C^{}_{HN} \right] 
    \nonumber
    \\
    && + \left. 3 m^2 g^{}_1  \left[ C^{}_{BN} Y^\dagger_\nu Y^{}_\nu -  \left( Y^\dagger_\nu Y^{}_\nu \right)^{\rm T} C^{}_{BN} \right] \right\} - \frac{1}{4} \delta Z^{(L=1)}_{N_L} M^{}_N \delta Z^{(L=1)\rm T}_{N_L} 
    \nonumber
    \\
    && - \frac{1}{2} \delta Z^{(L=1)}_{N_L} \delta M^{(L=1)}_N  - \frac{1}{2} \left( \delta Z^{(L=1)}_{N_L} \delta M^{(L=1)}_N  \right)^{\rm T} + \frac{1}{8} \delta Z^{(L=1)}_{N_L} \delta Z^{(L=1)}_{N_L} M^{}_N 
    \nonumber
    \\
    && + \frac{1}{8}  \left( \delta Z^{(L=1)}_{N_L} \delta Z^{(L=1)}_{N_L} M^{}_N \right)^{\rm T}
    \nonumber
    \\
    && + \looptf \left\{ \frac{1}{32} \left( 17g^2_1 + 51 g^2_2 - 12 T \right) \left[ M^{}_N Y^\dagger_\nu Y^{}_\nu + \left( Y^\dagger_\nu Y^{}_\nu \right)^{\rm T} M^{}_N \right] \right.
    \nonumber
    \\
    && - \frac{1}{16} M^{}_N Y^\dagger_\nu \left( Y^{}_l Y^\dagger_l + Y^{}_\nu Y^\dagger_\nu \right) Y^{}_\nu - \frac{1}{16} Y^{\rm T}_\nu \left[ \left( Y^{}_l Y^\dagger_l \right)^{\rm T} + \left( Y^{}_\nu Y^\dagger_\nu \right)^{\rm T} \right] Y^\ast_\nu M^{}_N 
    \nonumber
    \\
    && + \left( Y^\dagger_\nu Y^{}_\nu \right)^{\rm T} M^{}_N Y^\dagger_\nu Y^{}_\nu - 4m^2\left( g^2_1 +3g^2_2 - T \right) C^{}_{HN}  + m^2 \left[ C^{}_{HN} Y^\dagger_\nu Y^{}_\nu \right.
    \nonumber
    \\
    && + \left. \left( Y^\dagger_\nu Y^{}_\nu \right)^{\rm T} C^{}_{HN} \right] -7 m^2 g^{}_1  \left[ C^{}_{BN} Y^\dagger_\nu Y^{}_\nu - \left( Y^\dagger_\nu Y^{}_\nu \right)^{\rm T} C^{}_{BN} \right] +2 C^{}_{HN} M^\dagger_N M^{}_N Y^\dagger_\nu Y^{}_\nu 
    \nonumber
    \\
    &&  + 2\left( Y^\dagger_\nu Y^{}_\nu \right)^{\rm T} M^{}_N M^\dagger_N C^{}_{HN} + \frac{1}{4} C^{}_{HN} M^\dagger_N \left( Y^\dagger_\nu Y^{}_\nu \right)^{\rm T} M^{}_N + \frac{1}{4} M^{}_N Y^\dagger_\nu Y^{}_\nu M^\dagger_N C^{}_{HN}  
    \nonumber
    \\
    && + \frac{1}{4} \left( Y^\dagger_\nu Y^{}_\nu \right)^{\rm T} M^{}_N C^\dagger_{HN} M^{}_N  + \frac{1}{4}M^{}_N C^\dagger_{HN} M^{}_N Y^\dagger_\nu Y^{}_\nu -  g^{}_1 \left[ 6C^{}_{BN} M^\dagger_N M^{}_N Y^\dagger_\nu Y^{}_\nu  \right.
    \nonumber
    \\
    && - 6\left( Y^\dagger_\nu Y^{}_\nu \right)^{\rm T} M^{}_N M^\dagger_N C^{}_{BN} + \frac{1}{4} C^{}_{BN} M^\dagger_N \left( Y^\dagger_\nu Y^{}_\nu \right)^{\rm T} M^{}_N - \frac{1}{4} M^{}_N Y^\dagger_\nu Y^{}_\nu M^\dagger_N C^{}_{BN} 
    \nonumber
    \\
    && - \left.\left. \frac{1}{4}  M^{}_N C^\dagger_{BN} M^{}_N Y^\dagger_\nu Y^{}_\nu + \frac{1}{4} \left( Y^\dagger_\nu Y^{}_\nu \right)^{\rm T} M^{}_N C^\dagger_{BN} M^{}_N \right]  \right\} \;,
    \nonumber
    \\
    Y^{}_l \delta Z^{(L=2)}_{Y^{}_l} &=& \looptfs \left\{ \left[ -\frac{269}{64} g^4_1 + \frac{39}{32} g^2_1 g^2_2 + \frac{87}{64} g^4_2 + g^2_1 {\rm Tr} \left( \frac{1}{16} Y^{}_\nu Y^\dagger_\nu - \frac{11}{16} Y^{}_l Y^\dagger_l  + \frac{7}{16} Y^{}_{\rm d} Y^\dagger_{\rm d} \right. \right.\right.
    \nonumber
    \\
    &&  - \left. \frac{5}{16} Y^{}_{\rm u} Y^\dagger_{\rm u} \right) + \frac{3}{16} g^2_2 T - 6 g^2_3 {\rm Tr} \left( Y^{}_{\rm d} Y^\dagger_{\rm d} + Y^{}_{\rm u} Y^\dagger_{\rm u} \right) - \frac{3}{4} {\rm Tr} \left( Y^{}_\nu Y^\dagger_\nu Y^{}_l Y^\dagger_l  + 3 Y^{}_{\rm u} Y^\dagger_{\rm u} Y^{}_{\rm d} Y^\dagger_{\rm d} \right) 
    \nonumber
    \\
    && + \left. \frac{3}{8} T^\prime \right] Y^{}_l - \left(  \frac{63}{32} g^2_1 + \frac{33}{32} g^2_2  - \frac{3}{8} T \right) Y^{}_l Y^\dagger_l Y^{}_l + \left(  \frac{79}{32} g^2_1 + \frac{45}{32} g^2_2 -  \frac{3}{8} T \right) Y^{}_\nu Y^\dagger_\nu Y^{}_l 
    \nonumber
    \\
    &&  - \frac{1}{8} Y^{}_l Y^\dagger_l Y^{}_\nu Y^\dagger_\nu Y^{}_l - \frac{1}{8} Y^{}_\nu Y^\dagger_\nu Y^{}_\nu Y^\dagger_\nu Y^{}_l + \frac{1}{4} Y^{}_l Y^\dagger_l Y^{}_l Y^\dagger_l Y^{}_l  \nonumber
    \\
    && + {\rm Tr} \left[ \frac{3}{2} g^{}_1 Y^{}_\nu  \left( M^\dagger_N C^{}_{BN} + C^\dagger_{BN} M^{}_N \right) Y^\dagger_\nu  - \frac{1}{2} Y^{}_\nu \left( M^\dagger_N C^{}_{HN} + C^\dagger_{HN} M^{}_N \right) Y^\dagger_\nu \right. 
    \nonumber
    \\
    && - \left. \frac{3}{8} \left( C^{}_5 Y^\ast_\nu M^{}_N Y^\dagger_\nu + Y^{}_\nu M^\dagger_N Y^{\rm T}_\nu C^\dagger_5 \right) \right] Y^{}_l - g^{}_1 Y^{}_\nu \left( 3 M^\dagger_N C^{}_{BN} + \frac{3}{2} C^\dagger_{BN} M^{}_N \right) Y^\dagger_\nu Y^{}_l  
    \nonumber
    \\
    &&  + \left. Y^{}_\nu \left( M^\dagger_N C^{}_{HN} + \frac{1}{2} C^\dagger_{HN} M^{}_N \right) Y^\dagger_\nu Y^{}_l  +  \frac{9}{16} \left( C^{}_5 Y^\ast_\nu M^{}_N Y^\dagger_\nu Y^{}_l + Y^{}_\nu M^{}_N Y^{\rm T}_\nu C^\dagger_5 Y^{}_l \right) \right\} 
    \nonumber
    \\
    && + \frac{1}{8} \left( \delta Z^{(L=1)}_H \right)^2 Y^{}_l - \frac{1}{4} \delta Z^{(L=1)}_H \delta Z^{(L=1)}_\ell Y^{}_l - \frac{1}{4} \delta Z^{(L=1)}_H Y^{}_l \delta Z^{(L=1)}_e 
    \nonumber
    \\
    && - \frac{1}{2} \delta Z^{(L=1)}_H Y^{}_l \delta Z^{(L=1)}_{Y^{}_l} - \frac{1}{4} \delta Z^{(L=1)}_\ell Y^{}_l \delta Z^{(L=1)}_e + \frac{1}{8} \delta Z^{(L=1)}_\ell  \delta Z^{(L=1)}_\ell  Y^{}_l 
    \nonumber
    \\
    && + \frac{1}{8} Y^{}_l \delta Z^{(L=1)}_e \delta Z^{(L=1)}_e - \frac{1}{2} \delta Z^{(L=1)}_\ell  Y^{}_l \delta Z^{(L=1)}_{Y^{}_l} - \frac{1}{2} Y^{}_l \delta Z^{(L=1)}_{Y^{}_l} \delta Z^{(L=1)}_e
    \nonumber
    \\
    && + \looptf \left\{ \left[ \frac{457}{96} g^4_1 + \frac{9}{16} g^2_1 g^2_2 - \frac{23}{16} g^4_2 + g^2_1 {\rm Tr} \left( \frac{5}{32} Y^{}_\nu Y^\dagger_\nu + \frac{25}{32} Y^{}_l Y^\dagger_l  + \frac{25}{96} Y^{}_{\rm d} Y^\dagger_{\rm d} \right.\right.\right.
    \nonumber
    \\
    &&  + \left. \frac{85}{96} Y^{}_{\rm u} Y^\dagger_{\rm u} \right) + \frac{15}{32} g^2_2 T + 5g^2_3 {\rm Tr} \left( Y^{}_{\rm d} Y^\dagger_{\rm d} + Y^{}_{\rm u} Y^\dagger_{\rm u} \right) + \frac{3}{2}\lambda^2 + \frac{1}{8} {\rm Tr} \left( Y^{}_\nu Y^\dagger_\nu Y^{}_l Y^\dagger_l  \right.
    \nonumber
    \\
    && + \left. \left. 3 Y^{}_{\rm u} Y^\dagger_{\rm u} Y^{}_{\rm d} Y^\dagger_{\rm d} \right) - \frac{9}{16} T^\prime \right] Y^{}_l + \left( \frac{129}{64} g^2_1 + \frac{135}{64} g^2_2 - 3\lambda - \frac{9}{16} T \right) Y^{}_l Y^\dagger_l Y^{}_l 
    \nonumber
    \\
    && + \left( - \frac{45}{64} g^2_1 + \frac{9}{64} g^2_2 + \frac{5}{16} T \right) Y^{}_\nu Y^\dagger_\nu Y^{}_l - \frac{1}{4} Y^{}_\nu Y^\dagger_\nu Y^{}_l Y^\dagger_l Y^{}_l - \frac{1}{16} Y^{}_l Y^\dagger_l Y^{}_\nu Y^\dagger_\nu Y^{}_l 
    \nonumber
    \\
    && + \frac{11}{16} Y^{}_\nu Y^\dagger_\nu Y^{}_\nu Y^\dagger_\nu Y^{}_l + \frac{3}{8} Y^{}_l Y^\dagger_l Y^{}_l Y^\dagger_l Y^{}_l  + {\rm Tr} \left[ \frac{5}{2} g^{}_1 Y^{}_\nu  \left( M^\dagger_N C^{}_{BN} + C^\dagger_{BN} M^{}_N \right) Y^\dagger_\nu \right.
    \nonumber
    \\
    && + \left. Y^{}_\nu \left( M^\dagger_N C^{}_{HN} + C^\dagger_{HN} M^{}_N \right) Y^\dagger_\nu + \frac{9}{16} \left( C^{}_5 Y^\ast_\nu M^{}_N Y^\dagger_\nu + Y^{}_\nu M^\dagger_N Y^{\rm T}_\nu C^\dagger_5 \right) \right] Y^{}_l 
    \nonumber
    \\
    && + g^{}_1 Y^{}_\nu \left( \frac{31}{4} M^\dagger_N C^{}_{BN} - \frac{19}{4} C^\dagger_{BN} M^{}_N \right) Y^\dagger_\nu Y^{}_l - Y^{}_\nu \left( \frac{1}{2} M^\dagger_N C^{}_{HN} + 2 C^\dagger_{HN} M^{}_N \right) Y^\dagger_\nu Y^{}_l 
    \nonumber
    \\
    && - \left. \frac{21}{32} \left( C^{}_5 Y^\ast_\nu M^{}_N Y^\dagger_\nu Y^{}_l + Y^{}_\nu M^{}_N Y^{\rm T}_\nu C^\dagger_5 Y^{}_l \right) \right\} \;,
    \nonumber
    \\
    Y^{}_\nu \delta Z^{(L=2)}_{Y^{}_\nu} &=& \looptfs \left\{ \left[ -\frac{85}{64} g^4_1 + \frac{87}{64} g^4_2 - \frac{9}{32} g^2_1 g^2_2 + \frac{1}{16} g^2_1 {\rm Tr} \left( Y^{}_\nu Y^\dagger_\nu - 11 Y^{}_l Y^\dagger_l + 7 Y^{}_{\rm d} Y^\dagger_{\rm d} \right.\right.\right.
    \nonumber
    \\
    && - \left. 5 Y^{}_{\rm u} Y^\dagger_{\rm u} \right)  + \frac{3}{16} g^2_2 T -  6g^2_3 {\rm Tr} \left( Y^{}_{\rm d} Y^\dagger_{\rm d} + Y^{}_{\rm u} Y^\dagger_{\rm u}  \right) + \frac{3}{8} T^\prime - \frac{3}{4} {\rm Tr} \left( Y^{}_\nu Y^\dagger_\nu Y^{}_l Y^\dagger_l \right.
    \nonumber
    \\
    && +  \left.\left. 3 Y^{}_{\rm d} Y^\dagger_{\rm d} Y^{}_{\rm u} Y^\dagger_{\rm u} \right) \right] Y^{}_\nu  -  \left( \frac{11}{32} g^2_1 + \frac{33}{32} g^2_2 - \frac{3}{8} T \right) Y^{}_\nu Y^\dagger_\nu Y^{}_\nu
    \nonumber
    \\
    && + \left( \frac{51}{32} g^2_1 + \frac{45}{32} g^2_2 - \frac{3}{8} T \right) Y^{}_l Y^\dagger_l Y^{}_\nu - \frac{1}{8} Y^{}_l Y^\dagger_l Y^{}_l Y^\dagger_l Y^{}_\nu  + \frac{1}{4} Y^{}_\nu Y^\dagger_\nu Y^{}_\nu Y^\dagger_\nu Y^{}_\nu 
    \nonumber
    \\
    && - \frac{1}{8} Y^{}_\nu Y^\dagger_\nu Y^{}_l Y^\dagger_l Y^{}_\nu + \left( \frac{5}{4} g^2_1 + \frac{15}{4} g^2_2 - T - 6\lambda  \vphantom{\frac{3}{4}} \right) Y^{}_\nu M^\dagger_N C^{}_{HN}  
    \nonumber
    \\
    && + g^{}_1 \left(19 g^2_1 - \frac{9}{2} g^2_2 \right) Y^{}_\nu M^\dagger_N C^{}_{BN} + {\rm Tr} \left[ \frac{3}{2}  g^{}_1 Y^{}_\nu \left( M^\dagger_N C^{}_{BN} + C^\dagger_{BN} M^{}_N \right) Y^\dagger_\nu \right.
    \nonumber
    \\
    && - \left. \frac{1}{2} Y^{}_\nu \left( M^\dagger_N C^{}_{HN} + C^\dagger_{HN} M^{}_N \right) Y^\dagger_\nu -  \frac{3}{8} \left( C^{}_5 Y^\ast_\nu M^{}_N Y^\dagger_\nu + Y^{}_\nu M^\dagger_N Y^{\rm T}_\nu C^\dagger_5 \right) \right] Y^{}_\nu 
    \nonumber
    \\
    && + g^{}_1 \left[ \left( -6Y^{}_l Y^\dagger_l + \frac{9}{2} Y^{}_\nu Y^\dagger_\nu \right) Y^{}_\nu M^\dagger_N C^{}_{BN} + \frac{3}{2} Y^{}_\nu \left( M^\dagger_N C^{}_{BN} + 2 C^\dagger_{BN} M^{}_N \right) Y^\dagger_\nu Y^{}_\nu \right.
    \nonumber
    \\
    && + \left. \frac{3}{2} Y^{}_\nu M^\dagger_N Y^{\rm T}_\nu Y^\ast_\nu C^{}_{BN} \right] - \frac{3}{2} Y^{}_\nu Y^\dagger_\nu Y^{}_\nu  M^\dagger_N C^{}_{HN} +  2Y^{}_l Y^\dagger_l Y^{}_\nu  M^\dagger_N C^{}_{HN} 
    \nonumber
    \\
    && -  Y^{}_\nu C^\dagger_{HN} M^{}_N Y^\dagger_\nu Y^{}_\nu  - \frac{3}{2} Y^{}_\nu M^\dagger_N C^{}_{HN} Y^\dagger_\nu Y^{}_\nu  - \frac{5}{2} Y^{}_\nu M^\dagger_N Y^{\rm T}_\nu Y^\ast_\nu C^{}_{HN} 
    \nonumber
    \\
    && -  \left. \frac{3}{16} C^{}_5 Y^\ast_\nu M^{}_N Y^\dagger_\nu Y^{}_\nu - \frac{3}{16} Y^{}_\nu M^\dagger_N Y^{\rm T}_\nu C^\dagger_5 Y^{}_\nu \right\} + \frac{1}{8} \left( \delta Z^{(L=1)}_H \right)^2 Y^{}_\nu
    \nonumber
    \\
    &&  - \frac{1}{4} \delta Z^{(L=1)}_H \delta Z^{(L=1)}_\ell Y^{}_\nu - \frac{1}{4} \delta Z^{(L=1)}_H Y^{}_\nu \delta Z^{(L=1)\rm T}_{N_L} - \frac{1}{2} \delta Z^{(L=1)}_H Y^{}_\nu \delta Z^{(L=1)}_{Y^{}_\nu} 
    \nonumber
    \\
    && - \frac{1}{4} \delta Z^{(L=1)}_\ell Y^{}_\nu \delta Z^{(L=1)\rm T}_{N_L} + \frac{1}{8} \delta Z^{(L=1)}_\ell  \delta Z^{(L=1)}_\ell  Y^{}_\nu + \frac{1}{8} Y^{}_\nu \delta Z^{(L=1)\rm T}_{N_L} \delta Z^{(L=1)\rm T}_{N_L} 
    \nonumber
    \\
    && - \frac{1}{2} \delta Z^{(L=1)}_\ell  Y^{}_\nu \delta Z^{(L=1)}_{Y^{}_\nu} - \frac{1}{2} Y^{}_\nu \delta Z^{(L=1)}_{Y^{}_\nu} \delta Z^{(L=1)\rm T}_{N_L}
    \nonumber
    \\
    && + \looptf \left\{ \left[ \frac{35}{96} g^4_1 - \frac{23}{16} g^4_2 - \frac{9}{16} g^2_1 g^2_2 + \frac{5}{96} g^2_1 {\rm Tr} \left( 3Y^{}_\nu Y^\dagger_\nu +15 Y^{}_l Y^\dagger_l + 5 Y^{}_{\rm d} Y^\dagger_{\rm d} \right.\right.\right.
    \nonumber
    \\
    && + \left. 17 Y^{}_{\rm u} Y^\dagger_{\rm u} \right) + \frac{15}{32} g^2_2 T +  5g^2_3 {\rm Tr} \left( Y^{}_{\rm d} Y^\dagger_{\rm d} + Y^{}_{\rm u} Y^\dagger_{\rm u}  \right) + \frac{3}{2} \lambda^2 - \frac{9}{16} T^\prime + \frac{1}{8} {\rm Tr} \left( Y^{}_\nu Y^\dagger_\nu Y^{}_l Y^\dagger_l \right.
    \nonumber
    \\
    && + \left. \left.  3 Y^{}_{\rm d} Y^\dagger_{\rm d} Y^{}_{\rm u} Y^\dagger_{\rm u} \right) \vphantom{\frac{1}{1}} \right] Y^{}_\nu +  \left( \frac{93}{64} g^2_1 + \frac{135}{64} g^2_2 - 3\lambda - \frac{9}{16} T \right) Y^{}_\nu Y^\dagger_\nu Y^{}_\nu 
    \nonumber
    \\
    && + \left( - \frac{81}{64} g^2_1 + \frac{9}{64} g^2_2 + \frac{5}{16} T \right) Y^{}_l Y^\dagger_l Y^{}_\nu + \frac{11}{16} Y^{}_l Y^\dagger_l Y^{}_l Y^\dagger_l Y^{}_\nu - \frac{1}{4} Y^{}_l Y^\dagger_l Y^{}_\nu Y^\dagger_\nu Y^{}_\nu 
    \nonumber
    \\
    &&  + \frac{3}{8} Y^{}_\nu Y^\dagger_\nu Y^{}_\nu Y^\dagger_\nu Y^{}_\nu - \frac{1}{16} Y^{}_\nu Y^\dagger_\nu Y^{}_l Y^\dagger_l Y^{}_\nu - \left( \frac{1}{4} g^2_1 + \frac{3}{4} g^2_2 - T - 6\lambda  \vphantom{\frac{3}{4}} \right) Y^{}_\nu M^\dagger_N C^{}_{HN}  
    \nonumber
    \\
    && -  g^{}_1 \left( \frac{185}{6} g^2_1 + 9 g^2_2 \right) Y^{}_\nu M^\dagger_N C^{}_{BN}  + {\rm Tr} \left[ \frac{5}{2} g^{}_1 Y^{}_\nu \left( M^\dagger_N C^{}_{BN} + C^\dagger_{BN} M^{}_N \right) Y^\dagger_\nu  \right.
    \nonumber
    \\
    && + \left. Y^{}_\nu \left( M^\dagger_N C^{}_{HN} + C^\dagger_{HN} M^{}_N \right) Y^\dagger_\nu + \frac{9}{16} \left( C^{}_5 Y^\ast_\nu M^{}_N Y^\dagger_\nu + Y^{}_\nu M^\dagger_N Y^{\rm T}_\nu C^\dagger_5 \right) \right] Y^{}_\nu 
    \nonumber
    \\
    && + g^{}_1 \left[ \left( 3Y^{}_l Y^\dagger_l - \frac{19}{4} Y^{}_\nu Y^\dagger_\nu \right) Y^{}_\nu M^\dagger_N C^{}_{BN} + \frac{1}{4} Y^{}_\nu \left( M^\dagger_N C^{}_{BN} + 2 C^\dagger_{BN} M^{}_N \right) Y^\dagger_\nu Y^{}_\nu \right.
    \nonumber
    \\
    && - \left. \frac{15}{2} Y^{}_\nu M^\dagger_N Y^{\rm T}_\nu Y^\ast_\nu C^{}_{BN} \right] + \frac{9}{4} Y^{}_\nu Y^\dagger_\nu Y^{}_\nu  M^\dagger_N C^{}_{HN} - \frac{3}{4} Y^{}_\nu C^\dagger_{HN} M^{}_N Y^\dagger_\nu Y^{}_\nu 
    \nonumber
    \\
    && + \left. \frac{3}{2} Y^{}_\nu M^\dagger_N Y^{\rm T}_\nu Y^\ast_\nu C^{}_{HN} - \frac{45}{32} C^{}_5 Y^\ast_\nu M^{}_N Y^\dagger_\nu Y^{}_\nu + \frac{3}{32} Y^{}_\nu M^\dagger_N Y^{\rm T}_\nu C^\dagger_5 Y^{}_\nu \right\} \;,
    \nonumber
    \\
    Y^{}_{\rm d} \delta Z^{(L=2)}_{Y^{}_{\rm d}} &=& \looptfs \left\{ \left[ - \frac{1343}{1728} g^4_1 + \frac{87}{64} g^4_2 + \frac{82}{3} g^4_3 - \frac{35}{96} g^2_1 g^2_2 + \frac{1}{18} g^2_1 g^2_3 + \frac{7}{2} g^2_2 g^2_3 \right. \right.
    \nonumber
    \\
    &&  + g^2_1 {\rm Tr} \left( \frac{1}{16} Y^{}_\nu Y^\dagger_\nu - \frac{11}{16} Y^{}_l Y^\dagger_l - \frac{5}{16} Y^{}_{\rm u} Y^\dagger_{\rm u} + \frac{7}{16} Y^{}_{\rm d} Y^\dagger_{\rm d} \right) + \frac{3}{16} g^2_2 T   
    \nonumber
    \\
    && - \left. 6 g^2_3 {\rm Tr} \left( Y^{}_{\rm u} Y^\dagger_{\rm u} + Y^{}_{\rm d} Y^\dagger_{\rm d} \right) - \frac{3}{4} {\rm Tr} \left( Y^{}_\nu Y^\dagger_\nu Y^{}_l Y^\dagger_l + 3 Y^{}_{\rm  u} Y^\dagger_{\rm u} Y^{}_{\rm d} Y^\dagger_{\rm d} \right)  +  \frac{3}{8} T^\prime \right] Y^{}_{\rm d} 
    \nonumber
    \\
    && - \left( \frac{7}{32}g^2_1 + \frac{33}{32} g^2_2 + 4g^2_3  - \frac{3}{8} T \right) Y^{}_{\rm d} Y^\dagger_{\rm d} Y^{}_{\rm d} + \left( \frac{53}{96} g^2_1 + \frac{45}{32} g^2_2 + 8g^2_3 - \frac{3}{8} T \right) Y^{}_{\rm u} Y^\dagger_{\rm u} Y^{}_{\rm d}
    \nonumber
    \\
    &&  - \frac{1}{8} Y^{}_{\rm d} Y^\dagger_{\rm d} Y^{}_{\rm u} Y^\dagger_{\rm u} Y^{}_{\rm d} - \frac{1}{8} Y^{}_{\rm u} Y^\dagger_{\rm u} Y^{}_{\rm u} Y^\dagger_{\rm u} Y^{}_{\rm d} + \frac{1}{4} Y^{}_{\rm d} Y^\dagger_{\rm d} Y^{}_{\rm d} Y^\dagger_{\rm d} Y^{}_{\rm d}  - {\rm Tr} \left[  \frac{3}{8}  Y^{}_\nu M^\dagger_N Y^{\rm T}_\nu C^\dagger_5 \right.
    \nonumber
    \\
    && + \frac{3}{8}  C^{}_5 Y^\ast_\nu M^{}_N Y^\dagger_\nu  +\frac{1}{2}Y^\dagger_\nu Y^{}_\nu M^\dagger_N C^{}_{HN} + \frac{1}{2}C^\dagger_{HN} M^{}_N Y^\dagger_\nu Y^{}_\nu - \frac{3}{2} g^{}_1 \left( Y^\dagger_\nu Y^{}_\nu M^\dagger_N C^{}_{BN} \right.
    \nonumber
    \\
    && + \left.\left.\left. C^\dagger_{BN} M^{}_N Y^\dagger_\nu Y^{}_\nu \right) \right] Y^{}_{\rm d} \right\} + \frac{1}{8} \left( \delta Z^{(L=1)}_H \right)^2 Y^{}_{\rm d} - \frac{1}{4} \delta Z^{(L=1)}_H \delta Z^{(L=1)}_q Y^{}_{\rm d}
    \nonumber
    \\
    &&  - \frac{1}{4} \delta Z^{(L=1)}_H Y^{}_{\rm d} \delta Z^{(L=1)}_d - \frac{1}{2} \delta Z^{(L=1)}_H Y^{}_{\rm d} \delta Z^{(L=1)}_{Y^{}_{\rm d}} - \frac{1}{4} \delta Z^{(L=1)}_q Y^{}_{\rm d} \delta Z^{(L=1)}_d
    \nonumber
    \\
    && + \frac{1}{8} \delta Z^{(L=1)}_q  \delta Z^{(L=1)}_q  Y^{}_{\rm d} + \frac{1}{8} Y^{}_{\rm d} \delta Z^{(L=1)}_d \delta Z^{(L=1)}_d - \frac{1}{2} \delta Z^{(L=1)}_q Y^{}_{\rm d} \delta Z^{(L=1)}_{Y^{}_{\rm d}} 
    \nonumber
    \\
    && - \frac{1}{2} Y^{}_{\rm d} \delta Z^{(L=1)}_{Y^{}_{\rm d}} \delta Z^{(L=1)}_d
    \nonumber
    \\
    && + \looptf \left\{ \left[ - \frac{127}{864} g^4_1 - \frac{23}{16} g^4_2 - 27 g^4_3 - \frac{9}{16} g^2_1 g^2_2 + \frac{31}{36} g^2_1 g^2_3 + \frac{9}{4} g^2_2 g^2_3 \right. \right.
    \nonumber
    \\
    && + g^2_1 {\rm Tr} \left( \frac{5}{32} Y^{}_\nu Y^\dagger_\nu + \frac{25}{32} Y^{}_l Y^\dagger_l + \frac{85}{96} Y^{}_{\rm u} Y^\dagger_{\rm u} + \frac{25}{96} Y^{}_{\rm d} Y^\dagger_{\rm d} \right) + \frac{15}{32} g^2_2 T 
    \nonumber
    \\
    && + 5g^2_3 {\rm Tr} \left( Y^{}_{\rm u} Y^\dagger_{\rm u} + Y^{}_{\rm d} Y^\dagger_{\rm d} \right) + \frac{3}{2} \lambda^2 - \frac{9}{16} T^\prime + \left. \frac{1}{8} {\rm Tr} \left( Y^{}_\nu Y^\dagger_\nu Y^{}_l Y^\dagger_l + 3 Y^{}_{\rm  u} Y^\dagger_{\rm u} Y^{}_{\rm d} Y^\dagger_{\rm d} \right) \right] Y^{}_{\rm d} 
    \nonumber
    \\
    && + \left( \frac{187}{192}g^2_1 + \frac{135}{64} g^2_2 + 4g^2_3 - 3\lambda - \frac{9}{16} T \right) Y^{}_{\rm d} Y^\dagger_{\rm d} Y^{}_{\rm d} + \left( - \frac{79}{192} g^2_1 + \frac{9}{64} g^2_2 - 4g^2_3 \right.
    \nonumber
    \\
    &&  + \left. \frac{5}{16} T \right) Y^{}_{\rm u} Y^\dagger_{\rm u} Y^{}_{\rm d} - \frac{1}{4} Y^{}_{\rm u} Y^\dagger_{\rm u} Y^{}_{\rm d} Y^\dagger_{\rm d} Y^{}_{\rm d} - \frac{1}{16} Y^{}_{\rm d} Y^\dagger_{\rm d} Y^{}_{\rm u} Y^\dagger_{\rm u} Y^{}_{\rm d} + \frac{11}{16} Y^{}_{\rm u} Y^\dagger_{\rm u} Y^{}_{\rm u} Y^\dagger_{\rm u} Y^{}_{\rm d}
    \nonumber
    \\
    &&  + \frac{3}{8} Y^{}_{\rm d} Y^\dagger_{\rm d} Y^{}_{\rm d} Y^\dagger_{\rm d} Y^{}_{\rm d} + {\rm Tr} \left[ \frac{9}{16} \left( Y^{}_\nu M^\dagger_N Y^{\rm T}_\nu C^\dagger_5 + C^{}_5 Y^\ast_\nu M^{}_N Y^\dagger_\nu \right) + Y^\dagger_\nu Y^{}_\nu M^\dagger_N C^{}_{HN} \right.
    \nonumber
    \\
    && + \left.\left. C^\dagger_{HN} M^{}_N Y^\dagger_\nu Y^{}_\nu + \frac{5}{2} g^{}_1 \left( Y^\dagger_\nu Y^{}_\nu M^\dagger_N C^{}_{BN} + C^\dagger_{BN} M^{}_N Y^\dagger_\nu Y^{}_\nu \right) \right] Y^{}_{\rm d} \right\} \;,
    \nonumber
    \\
    Y^{}_{\rm u} \delta Z^{(L=2)}_{Y^{}_{\rm u}} &=& \looptfs \left\{ \left[ - \frac{3959}{1728} g^4_1 + \frac{87}{64} g^4_2 + \frac{82}{3} g^4_3 + \frac{13}{96} g^2_1 g^2_2 + \frac{61}{18} g^2_1 g^2_3 + \frac{7}{2} g^2_2 g^2_3  \right.\right.
    \nonumber
    \\
    && + g^2_1 {\rm Tr} \left( \frac{1}{16} Y^{}_\nu Y^\dagger_\nu - \frac{11}{16} Y^{}_l Y^\dagger_l - \frac{5}{16} Y^{}_{\rm u} Y^\dagger_{\rm u} + \frac{7}{16} Y^{}_{\rm d} Y^\dagger_{\rm d} \right)+ \frac{3}{16} g^2_2 T 
    \nonumber
    \\
    && - \left. 6 g^2_3 {\rm Tr} \left( Y^{}_{\rm u} Y^\dagger_{\rm u} + Y^{}_{\rm d} Y^\dagger_{\rm d} \right) - \frac{3}{4} {\rm Tr} \left( Y^{}_\nu Y^\dagger_\nu Y^{}_l Y^\dagger_l + 3 Y^{}_{\rm u} Y^\dagger_{\rm u} Y^{}_{\rm d} Y^\dagger_{\rm d} \right) + \frac{3}{8} T^\prime \right] Y^{}_{\rm u} 
    \nonumber
    \\
    && - \left( \frac{73}{96} g^2_1 + \frac{33}{32} g^2_2 + 4g^2_3 - \frac{3}{8} T \right) Y^{}_{\rm u} Y^\dagger_{\rm u} Y^{}_{\rm u} + \left(  \frac{27}{32} g^2_1 + \frac{45}{32} g^2_2 + 8 g^2_3 - \frac{3}{8} T \right) Y^{}_{\rm d} Y^\dagger_{\rm d} Y^{}_{\rm u} 
    \nonumber
    \\
    && - \frac{1}{8} Y^{}_{\rm u} Y^\dagger_{\rm u} Y^{}_{\rm d} Y^\dagger_{\rm d} Y^{}_{\rm u}  - \frac{1}{8} Y^{}_{\rm d} Y^\dagger_{\rm d} Y^{}_{\rm d} Y^\dagger_{\rm d} Y^{}_{\rm u} + \frac{1}{4} Y^{}_{\rm u} Y^\dagger_{\rm u} Y^{}_{\rm u} Y^\dagger_{\rm u} Y^{}_{\rm u} - {\rm Tr} \left[ \frac{3}{8} Y^{}_\nu M^\dagger_N Y^{\rm T}_\nu C^\dagger_5  \right.
    \nonumber
    \\
    && + \frac{3}{8} C^{}_5 Y^\ast_\nu M^{}_N Y^\dagger_\nu  + \frac{1}{2} Y^\dagger_\nu Y^{}_\nu M^\dagger_N C^{}_{HN} + \frac{1}{2} C^\dagger_{HN} M^{}_N Y^\dagger_\nu Y^{}_\nu - \frac{3}{2}  g^{}_1 \left( Y^\dagger_\nu Y^{}_\nu M^\dagger_N C^{}_{BN} \right.
    \nonumber
    \\
    && +  \left.\left.\left. C^\dagger_{BN} M^{}_N Y^\dagger_\nu Y^{}_\nu \right)  \right] Y^{}_{\rm u}  \right\} + \frac{1}{8} \left( \delta Z^{(L=1)}_H \right)^2 Y^{}_{\rm u} - \frac{1}{4} \delta Z^{(L=1)}_H \delta Z^{(L=1)}_q Y^{}_{\rm u} 
    \nonumber
    \\
    && - \frac{1}{4} \delta Z^{(L=1)}_H Y^{}_{\rm u} \delta Z^{(L=1)}_u - \frac{1}{2} \delta Z^{(L=1)}_H Y^{}_{\rm u} \delta Z^{(L=1)}_{Y^{}_{\rm u}} - \frac{1}{4} \delta Z^{(L=1)}_q Y^{}_{\rm u} \delta Z^{(L=1)}_u
    \nonumber
    \\
    && + \frac{1}{8} \delta Z^{(L=1)}_q  \delta Z^{(L=1)}_q  Y^{}_{\rm u} + \frac{1}{8} Y^{}_{\rm u} \delta Z^{(L=1)}_u \delta Z^{(L=1)}_u - \frac{1}{2} \delta Z^{(L=1)}_q Y^{}_{\rm u} \delta Z^{(L=1)}_{Y^{}_{\rm u}} 
    \nonumber
    \\
    && - \frac{1}{2} Y^{}_{\rm u} \delta Z^{(L=1)}_{Y^{}_{\rm u}} \delta Z^{(L=1)}_u
    \nonumber
    \\
    && + \looptf \left\{ \left[ \frac{1187}{864} g^4_1 - \frac{23}{16} g^4_2 - 27 g^4_3 - \frac{3}{16} g^2_1 g^2_2 + \frac{19}{36} g^2_1 g^2_3 + \frac{9}{4} g^2_2 g^2_3  \right.\right.
    \nonumber
    \\
    && + g^2_1 {\rm Tr} \left( \frac{5}{32} Y^{}_\nu Y^\dagger_\nu + \frac{25}{32} Y^{}_l Y^\dagger_l + \frac{85}{96} Y^{}_{\rm u} Y^\dagger_{\rm u} + \frac{25}{96} Y^{}_{\rm d} Y^\dagger_{\rm d} \right)+ \frac{15}{32} g^2_2 T 
    \nonumber
    \\
    && +  \left. 5 g^2_3 {\rm Tr} \left( Y^{}_{\rm u} Y^\dagger_{\rm u} + Y^{}_{\rm d} Y^\dagger_{\rm d} \right) + \frac{3}{2} \lambda^2 - \frac{9}{16} T^\prime + \frac{1}{8} {\rm Tr} \left( Y^{}_\nu Y^\dagger_\nu Y^{}_l Y^\dagger_l + 3 Y^{}_{\rm u} Y^\dagger_{\rm u} Y^{}_{\rm d} Y^\dagger_{\rm d} \right)  \right] Y^{}_{\rm u}
    \nonumber
    \\
    &&  + \left( \frac{223}{192} g^2_1 + \frac{135}{64} g^2_2 + 4g^2_3 - 3\lambda - \frac{9}{16} T \right) Y^{}_{\rm u} Y^\dagger_{\rm u} Y^{}_{\rm u} + \left( - \frac{43}{192} g^2_1 + \frac{9}{64} g^2_2 - 4g^2_3 \right.
    \nonumber
    \\
    && + \left. \frac{5}{16} T \right) Y^{}_{\rm d} Y^\dagger_{\rm d} Y^{}_{\rm u} - \frac{1}{4} Y^{}_{\rm d} Y^\dagger_{\rm d} Y^{}_{\rm u} Y^\dagger_{\rm u} Y^{}_{\rm u} - \frac{1}{16} Y^{}_{\rm u} Y^\dagger_{\rm u} Y^{}_{\rm d} Y^\dagger_{\rm d} Y^{}_{\rm u} + \frac{11}{16} Y^{}_{\rm d} Y^\dagger_{\rm d} Y^{}_{\rm d} Y^\dagger_{\rm d} Y^{}_{\rm u} 
    \nonumber
    \\
    && + \frac{3}{8} Y^{}_{\rm u} Y^\dagger_{\rm u} Y^{}_{\rm u} Y^\dagger_{\rm u} Y^{}_{\rm u} + {\rm Tr} \left[ \frac{9}{16} \left( Y^{}_\nu M^\dagger_N Y^{\rm T}_\nu C^\dagger_5 + C^{}_5 Y^\ast_\nu M^{}_N Y^\dagger_\nu \right) + Y^\dagger_\nu Y^{}_\nu M^\dagger_N C^{}_{HN} \right.
    \nonumber
    \\
    && + \left.\left.  C^\dagger_{HN} M^{}_N Y^\dagger_\nu Y^{}_\nu + \frac{5}{2}  g^{}_1 \left( Y^\dagger_\nu Y^{}_\nu M^\dagger_N C^{}_{BN} + C^\dagger_{BN} M^{}_N Y^\dagger_\nu Y^{}_\nu \right)  \right] Y^{}_{\rm u}  \right\} \;,
\end{eqnarray}
and those for the non-renormalizable Wilson coefficients are found to be 
\begin{eqnarray}
    \delta C^{(L=2)}_5 &=& \looptfs  \left\{ \left[ \frac{1}{16} g^4_1 + \frac{29}{16} g^4_2 - \frac{3}{4} g^2_1 g^2_2 + g^2_1 {\rm Tr} \left( \frac{1}{8} Y^{}_\nu Y^\dagger_\nu - \frac{11}{8} Y^{}_l Y^\dagger_l + \frac{7}{8} Y^{}_{\rm d} Y^\dagger_{\rm d} - \frac{5}{8} Y^{}_{\rm u} Y^\dagger_{\rm u} \right) \right.\right.
    \nonumber
    \\
    && + \frac{3}{8} g^2_2 T - 12 g^2_3 {\rm Tr} \left( Y^{}_{\rm u} Y^\dagger_{\rm u} + Y^{}_{\rm d} Y^\dagger_{\rm d} \right) - \left( g^2_1  + 6 g^2_2 \right) \lambda + 14 \lambda^2  + 2\lambda T - \frac{1}{4} T^\prime 
    \nonumber
    \\
    && - \left. \frac{3}{2} {\rm Tr} \left( Y^{}_\nu Y^\dagger_\nu Y^{}_l Y^\dagger_l + 3 Y^{}_{\rm u} Y^\dagger_{\rm u} Y^{}_{\rm d} Y^\dagger_{\rm d} \right) \right] C^{}_5 + \left( \frac{35}{32} g^2_1 + \frac{45}{32} g^2_2 - 2\lambda - \frac{3}{8} T  \right) \left[ Y^{}_l Y^\dagger_l C^{}_5 \right.
    \nonumber
    \\
    && + \left. C^{}_5 \left( Y^{}_l Y^\dagger_l \right)^{\rm T} \right] + \left( \frac{13}{32} g^2_1 + \frac{3}{32} g^2_2 + \frac{3}{2} \lambda + \frac{1}{8} T \right) \left[ Y^{}_\nu Y^\dagger_\nu C^{}_5 + C^{}_5 \left( Y^{}_\nu Y^\dagger_\nu \right)^{\rm T} \right]  
    \nonumber
    \\
    &&  - \frac{1}{8} \left[ Y^{}_l Y^\dagger_l Y^{}_l Y^\dagger_l C^{}_5 + C^{}_5 \left( Y^{}_l Y^\dagger_l Y^{}_l Y^\dagger_l  \right)^{\rm T} \right] + \frac{1}{4}  \left[ Y^{}_\nu Y^\dagger_\nu Y^{}_l Y^\dagger_l C^{}_5 + C^{}_5 \left( Y^{}_\nu Y^\dagger_\nu Y^{}_l Y^\dagger_l  \right)^{\rm T} \right]
    \nonumber
    \\
    && + \left.  \frac{1}{8} \left[ Y^{}_\nu Y^\dagger_\nu Y^{}_\nu Y^\dagger_\nu C^{}_5 + C^{}_5 \left( Y^{}_\nu Y^\dagger_\nu Y^{}_\nu Y^\dagger_\nu  \right)^{\rm T} \right] + Y^{}_l  Y^\dagger_l C^{}_5 \left( Y^{}_l Y^\dagger_l \right)^{\rm T}  \right\} - \delta Z^{(L=1)}_H \delta C^{(L=1)}_5 
    \nonumber
    \\
    && - \frac{1}{2} \delta Z^{(L=1)}_H \delta Z^{(L=1)}_\ell C^{}_5 - \frac{1}{2} \delta Z^{(L=1)}_H \left( \delta Z^{(L=1)}_\ell C^{}_5  \right)^{\rm T} - \frac{1}{2} \delta Z^{(L=1)}_\ell \delta C^{(L=1)}_5 
    \nonumber
    \\
    && - \frac{1}{2} \left( \delta Z^{(L=1)}_\ell \delta C^{(L=1)}_5 \right)^{\rm T} -  \frac{1}{4} \delta Z^{(L=1)}_\ell C^{(L=1)}_5 \delta Z^{(L=1) \rm T}_\ell + \frac{1}{8} \delta Z^{(L=1)}_\ell \delta Z^{(L=1)}_\ell C^{}_5  
    \nonumber
    \\
    && + \frac{1}{8} \left( \delta Z^{(L=1)}_\ell \delta Z^{(L=1)}_\ell C^{}_5 \right)^{\rm T}
    \nonumber
    \\
    && + \looptf  \left\{ \left[ - \frac{129}{32} g^4_1 - \frac{169}{96} g^4_2 - \frac{83}{16} g^2_1 g^2_2 + g^2_1 {\rm Tr} \left( \frac{5}{16} Y^{}_\nu Y^\dagger_\nu + \frac{25}{16} Y^{}_l Y^\dagger_l + \frac{25}{48} Y^{}_{\rm d} Y^\dagger_{\rm d} \right.\right.\right.
    \nonumber
    \\
    && + \left. \frac{85}{48} Y^{}_{\rm u} Y^\dagger_{\rm u} \right) + \frac{15}{16} g^2_2 T + 10 g^2_3 {\rm Tr} \left( Y^{}_{\rm u} Y^\dagger_{\rm u} + Y^{}_{\rm d} Y^\dagger_{\rm d} \right) - \lambda g^2_1 - 7 \lambda^2  - 2\lambda T - \frac{1}{8} T^\prime 
    \nonumber
    \\
    && + \left. \frac{1}{4} {\rm Tr} \left( Y^{}_\nu Y^\dagger_\nu Y^{}_l Y^\dagger_l + 3 Y^{}_{\rm u} Y^\dagger_{\rm u} Y^{}_{\rm d} Y^\dagger_{\rm d} \right) \right] C^{}_5 + \left( - \frac{57}{64} g^2_1 + \frac{33}{64} g^2_2 + \frac{5}{16} T  \right) \left[ Y^{}_l Y^\dagger_l C^{}_5 \right.
    \nonumber
    \\
    && + \left. C^{}_5 \left( Y^{}_l Y^\dagger_l \right)^{\rm T} \right] + \left( \frac{23}{64} g^2_1 - \frac{3}{64} g^2_2 - \frac{13}{2} \lambda - \frac{3}{16} T \right) \left[ Y^{}_\nu Y^\dagger_\nu C^{}_5 + C^{}_5 \left( Y^{}_\nu Y^\dagger_\nu \right)^{\rm T} \right]
    \nonumber
    \\
    &&  - 3 g^{}_1 \left( Y^{}_\nu C^\dagger_{BN} Y^{\rm T}_\nu Y^\ast_\nu Y^{\rm T}_\nu - Y^{}_\nu Y^\dagger_\nu Y^{}_\nu C^\dagger_{BN} Y^{\rm T}_\nu \right) - 4\lambda Y^{}_\nu C^\dagger_{HN} Y^{\rm T}_\nu + Y^{}_\nu Y^\dagger_\nu Y^{}_\nu C^\dagger_{HN} Y^{\rm T}_\nu 
    \nonumber
    \\
    && + Y^{}_\nu C^\dagger_{HN} Y^{\rm T}_\nu Y^\ast_\nu Y^{\rm T}_\nu + \frac{1}{2} Y^{}_l  Y^\dagger_l C^{}_5 \left( Y^{}_l Y^\dagger_l \right)^{\rm T} + \frac{1}{2} Y^{}_\nu  Y^\dagger_\nu C^{}_5 \left( Y^{}_\nu Y^\dagger_\nu \right)^{\rm T} 
    \nonumber
    \\
    && + \frac{19}{16} \left[ Y^{}_l Y^\dagger_l Y^{}_l Y^\dagger_l C^{}_5 + C^{}_5 \left( Y^{}_l Y^\dagger_l Y^{}_l Y^\dagger_l  \right)^{\rm T} \right] - \frac{1}{4}  \left[ Y^{}_l Y^\dagger_l Y^{}_\nu Y^\dagger_\nu C^{}_5 + C^{}_5 \left( Y^{}_l Y^\dagger_l Y^{}_\nu Y^\dagger_\nu  \right)^{\rm T} \right]
    \nonumber
    \\
    && + \left. \frac{15}{16} \left[ Y^{}_\nu Y^\dagger_\nu Y^{}_\nu Y^\dagger_\nu C^{}_5 + C^{}_5 \left( Y^{}_\nu Y^\dagger_\nu Y^{}_\nu Y^\dagger_\nu  \right)^{\rm T} \right]  \right\} \;,
    \nonumber
    \\
    \delta C^{(L=2)}_{HN} &=&  \looptfs \left\{ \left[ -\frac{67}{32} g^4_1 + \frac{141}{32} g^4_2 + \frac{9}{16} g^2_1 g^2_2 - \lambda \left( 6g^2_1 + 18 g^2_2 - 54\lambda - 6T \right) \right.\right.
    \nonumber
    \\
    && + \frac{1}{8} g^2_1 {\rm Tr} \left( Y^{}_\nu Y^\dagger_\nu  - 11 Y^{}_l Y^\dagger_l + 7 Y^{}_{\rm d} Y^\dagger_{\rm d} - 5 Y^{}_{\rm u} Y^\dagger_{\rm u} \right) + \frac{3}{8} g^2_2 T - 12 g^2_3 {\rm Tr} \left( Y^{}_{\rm u} Y^\dagger_{\rm u}  + Y^{}_{\rm d} Y^\dagger_{\rm d} \right) 
    \nonumber
    \\
    && - \left. \frac{9}{4} T^\prime - \frac{3}{2} {\rm Tr} \left( Y^{}_\nu Y^\dagger_\nu Y^{}_l Y^\dagger_l + 3 Y^{}_{\rm u} Y^\dagger_{\rm u} Y^{}_{\rm d} Y^\dagger_{\rm d} \right)  \right] C^{}_{HN} + \left[ -\frac{13}{16} \left( g^2_1 + 3g^2_2 \right) + \frac{15}{2} \lambda + \frac{3}{4} T \right]
    \nonumber
    \\
    && \times \left[ C^{}_{HN} Y^\dagger_\nu Y^{}_\nu + \left( Y^\dagger_\nu Y^{}_\nu \right)^{\rm T} C^{}_{HN} \right] -  \frac{9}{2}  \lambda g^{}_1 \left[ C^{}_{BN} Y^\dagger_\nu Y^{}_\nu -  \left( Y^\dagger_\nu Y^{}_\nu \right)^{\rm T} C^{}_{BN} \right] 
    \nonumber
    \\
    &&  - \frac{3}{4} g^{}_1 \left[ C^{}_{BN} Y^\dagger_\nu Y^{}_l Y^\dagger_l Y^{}_\nu - \left( Y^\dagger_\nu Y^{}_l Y^\dagger_l Y^{}_\nu \right)^{\rm T} C^{}_{BN} + C^{}_{BN} Y^\dagger_\nu Y^{}_\nu Y^\dagger_\nu Y^{}_\nu \right.
    \nonumber
    \\
    && -\left. \left( Y^\dagger_\nu Y^{}_\nu Y^\dagger_\nu Y^{}_\nu \right)^{\rm T} C^{}_{BN} \right] - \frac{7}{8} \left[ C^{}_{HN} Y^\dagger_\nu Y^{}_l Y^\dagger_l Y^{}_\nu + \left( Y^\dagger_\nu Y^{}_l Y^\dagger_l Y^{}_\nu \right)^{\rm T} C^{}_{HN} \right]  
    \nonumber
    \\
    && + \left. \frac{13}{8} \left[ C^{}_{HN} Y^\dagger_\nu Y^{}_\nu Y^\dagger_\nu Y^{}_\nu + \left( Y^\dagger_\nu Y^{}_\nu Y^\dagger_\nu Y^{}_\nu \right)^{\rm T} C^{}_{HN} \right] +  \left( Y^\dagger_\nu Y^{}_\nu \right)^{\rm T} C^{}_{HN} Y^\dagger_\nu Y^{}_\nu \right\} 
    \nonumber
    \\
    && - \delta Z^{(L=1)}_H \delta C^{(L=1)}_{HN} - \frac{1}{2} \delta Z^{(L=1)}_H \delta Z^{(L=1)}_{N_L} C^{}_{HN} - \frac{1}{2} \delta Z^{(L=1)}_H \left( \delta Z^{(L=1)}_{N_L} C^{}_{HN}  \right)^{\rm T} 
    \nonumber
    \\
    && - \frac{1}{2} \delta Z^{(L=1)}_{N_L} \delta C^{(L=1)}_{HN} - \frac{1}{2} \left( \delta Z^{(L=1)}_{N_L} \delta C^{(L=1)}_{HN} \right)^{\rm T} -  \frac{1}{4} \delta Z^{(L=1)}_{N_L} C^{}_{HN} \delta Z^{(L=1) \rm T}_{N_L} 
    \nonumber
    \\
    && + \frac{1}{8} \delta Z^{(L=1)}_{N_L} \delta Z^{(L=1)}_{N_L} C^{}_{HN}  + \frac{1}{8} \left( \delta Z^{(L=1)}_{N_L} \delta Z^{(L=1)}_{N_L} C^{}_{HN} \right)^{\rm T}
    \nonumber
    \\
    && +  \looptf \left\{ \left[ \frac{557}{192} g^4_1 - \frac{145}{64} g^4_2 + \frac{15}{32} g^2_1 g^2_2 + \lambda \left( 6g^2_1 + 18g^2_2 - 15\lambda - 6T \right) \right.\right.
    \nonumber
    \\
    && + \frac{5}{48} g^2_1 {\rm Tr} \left( 3Y^{}_\nu Y^\dagger_\nu  + 15 Y^{}_l Y^\dagger_l + 5 Y^{}_{\rm d} Y^\dagger_{\rm d} + 17 Y^{}_{\rm u} Y^\dagger_{\rm u} \right) + \frac{15}{16} g^2_2 T 
    \nonumber
    \\
    && + \left. 10 g^2_3 {\rm Tr} \left( Y^{}_{\rm u} Y^\dagger_{\rm u}  + Y^{}_{\rm d} Y^\dagger_{\rm d} \right)  - \frac{9}{8} T^\prime - \frac{7}{4} {\rm Tr} \left( Y^{}_\nu Y^\dagger_\nu Y^{}_l Y^\dagger_l + 3 Y^{}_{\rm u} Y^\dagger_{\rm u} Y^{}_{\rm d} Y^\dagger_{\rm d} \right)  \right] C^{}_{HN} 
    \nonumber
    \\
    && - 3\lambda Y^{\rm T}_\nu C^\dagger_5 Y^{}_\nu + \left[ \frac{33}{32} \left( g^2_1 + 3g^2_2 \right) - \frac{21}{2} \lambda - \frac{7}{8} T \right] \left[ C^{}_{HN} Y^\dagger_\nu Y^{}_\nu \right. + \left.  \left( Y^\dagger_\nu Y^{}_\nu \right)^{\rm T} C^{}_{HN} \right] 
    \nonumber
    \\
    && +  g^{}_1 \left[ \frac{9}{8} \left( g^2_1 - g^2_2 \right) + \frac{21}{2} \lambda \right] \left[ C^{}_{BN} Y^\dagger_\nu Y^{}_\nu -  \left( Y^\dagger_\nu Y^{}_\nu \right)^{\rm T} C^{}_{BN} \right] 
    \nonumber
    \\
    && + \frac{7}{4}  g^{}_1 \left[ C^{}_{BN} Y^\dagger_\nu Y^{}_l Y^\dagger_l Y^{}_\nu - \left( Y^\dagger_\nu Y^{}_l Y^\dagger_l Y^{}_\nu \right)^{\rm T} C^{}_{BN} \right] + \frac{13}{4}  g^{}_1 \left[ C^{}_{BN} Y^\dagger_\nu Y^{}_\nu Y^\dagger_\nu Y^{}_\nu \right.
    \nonumber
    \\
    && - \left. \left( Y^\dagger_\nu Y^{}_\nu Y^\dagger_\nu Y^{}_\nu \right)^{\rm T} C^{}_{BN} \right] + \frac{15}{16} \left[ C^{}_{HN} Y^\dagger_\nu Y^{}_l Y^\dagger_l Y^{}_\nu + \left( Y^\dagger_\nu Y^{}_l Y^\dagger_l Y^{}_\nu \right)^{\rm T} C^{}_{HN} \right] 
    \nonumber
    \\
    && - \frac{25}{16} \left[ C^{}_{HN} Y^\dagger_\nu Y^{}_\nu Y^\dagger_\nu Y^{}_\nu + \left( Y^\dagger_\nu Y^{}_\nu Y^\dagger_\nu Y^{}_\nu \right)^{\rm T} C^{}_{HN} \right] + \frac{3}{4} \left( Y^{\rm T}_\nu C^\dagger_5 Y^{}_\nu Y^\dagger_\nu Y^{}_\nu \right.
    \nonumber
    \\
    && + \left.\left. Y^{\rm T}_\nu Y^\ast_\nu Y^{\rm T}_\nu C^\dagger_5 Y^{}_\nu \right)  + \left( Y^\dagger_\nu Y^{}_\nu \right)^{\rm T} C^{}_{HN} Y^\dagger_\nu Y^{}_\nu \right\} \;,
    \nonumber
    \\
    \delta C^{(L=2)}_{BN} &=& \looptfs \left\{ \left[ -\frac{3}{16} \left( g^2_1 + 3 g^2_2 \right) + \frac{1}{4} T \right] \left( C^{}_{BN} Y^\dagger_\nu Y^{}_\nu + Y^{\rm T}_\nu Y^\ast_\nu C^{}_{BN} \right) \right.
    \nonumber
    \\
    && +  \left.   \frac{1}{8} Y^{\rm T}_\nu \left( Y^\ast_\nu Y^{\rm T}_\nu - 3 Y^\ast_l Y^{\rm T}_l \right) Y^{}_\nu C^{}_{BN} + \frac{1}{8} C^{}_{BN} Y^\dagger_\nu \left( Y^{}_\nu Y^\dagger_\nu - 3 Y^{}_l Y^\dagger_l \right) Y^{}_\nu \right\}  
    \nonumber
    \\
    && - \frac{1}{2} \delta Z^{(L=1)}_B \delta C^{(L=1)}_{BN} - \frac{1}{4} \delta Z^{(L=1)}_B \delta Z^{(L=1)}_{N_L} C^{}_{BN} - \frac{1}{4} \delta Z^{(L=1)}_B  C^{}_{BN} \delta Z^{(L=1)\rm T}_{N_L}  
    \nonumber
    \\
    && - \frac{1}{2} \delta Z^{(L=1)}_{N_L} \delta C^{(L=1)}_{BN} - \frac{1}{2} \delta C^{(L=1)}_{BN}  \delta Z^{(L=1) \rm T}_{N_L}  -  \frac{1}{4} \delta Z^{(L=1)}_{N_L} C^{}_{BN} \delta Z^{(L=1) \rm T}_{N_L} 
    \nonumber
    \\
    && + \frac{1}{8} \delta Z^{(L=1)}_{N_L} \delta Z^{(L=1)}_{N_L} C^{}_{BN}  + \frac{1}{8} C^{}_{BN} \left( \delta Z^{(L=1)}_{N_L} \delta Z^{(L=1)}_{N_L}  \right)^{\rm T} + \frac{1}{8} \left( \delta Z^{(L=1)}_B \right)^2 C^{}_{BN}
    \nonumber
    \\
    && + \looptf \left\{ \left[ \frac{199}{72} g^4_1 + \frac{9}{8} g^2_1 g^2_2 + \frac{11}{3} g^2_1 g^2_3 - \frac{1}{24} g^2_1 {\rm Tr} \left( 3 Y^{}_\nu Y^\dagger_\nu + 15 Y^{}_l Y^\dagger_l + 5 Y^{}_{\rm d} Y^\dagger_{\rm d} \right.\right. \right.
    \nonumber
    \\
    && + \left.\left.17 Y^{}_{\rm u} Y^\dagger_{\rm u} \right) \vphantom{\frac{1}{1}} \right] C^{}_{BN} + \left[ \frac{1}{32} \left( 35g^2_1 + 51 g^2_2 \right) - \frac{3}{8} T \right] \left( C^{}_{BN} Y^\dagger_\nu Y^{}_\nu + Y^{\rm T}_\nu Y^\ast_\nu C^{}_{BN} \right) 
    \nonumber
    \\
    && - \frac{1}{16} g^{}_1  \left( C^{}_{HN} Y^\dagger_\nu Y^{}_\nu  - Y^{\rm T}_\nu Y^\ast_\nu C^{}_{HN} \right) - \frac{1}{16} Y^{\rm T}_\nu \left( Y^\ast_\nu Y^{\rm T}_\nu + Y^\ast_l Y^{\rm T}_l \right) Y^{}_\nu C^{}_{BN} 
    \nonumber
    \\
    && - \left. \frac{1}{16} C^{}_{BN} Y^\dagger_\nu \left( Y^{}_\nu Y^\dagger_\nu + Y^{}_l Y^\dagger_l \right) Y^{}_\nu - \left( Y^\dagger_\nu Y^{}_\nu \right)^{\rm T} C^{}_{BN} Y^\dagger_\nu Y^{}_\nu   \right\}\;,
\end{eqnarray}
and
\begin{eqnarray}
    \delta G^{(L=2)}_{DN} &=& - \looptfs \frac{3}{4} Y^{\rm T}_\nu C^\dagger_5 Y^{}_\nu
    \nonumber
    \\
    && + \looptf \left\{ \frac{9}{4}  g^{}_1 \left[ C^{}_{BN} Y^\dagger_\nu Y^{}_\nu - \left( Y^\dagger_\nu Y^{}_\nu \right)^{\rm T} C^{}_{BN} \right] - \frac{1}{4} \left[ C^{}_{HN} Y^\dagger_\nu Y^{}_\nu + \left( Y^\dagger_\nu Y^{}_\nu \right)^{\rm T} C^{}_{HN} \right]  \right\} \;,
    \nonumber
    \\
    \delta G^{(L=2)}_{\ell HN1} &=& \looptfs \left[ \left( -\frac{3}{8} g^2_1 -  \frac{9}{4} g^2_2 + \frac{3}{2} \lambda + \frac{3}{4} T \right) C^{}_5 Y^\ast_\nu  - \left( \frac{3}{2} Y^{}_l Y^\dagger_l - \frac{3}{8} Y^{}_\nu Y^\dagger_\nu \right) C^{}_5 Y^\ast_\nu   \right.
    \nonumber
    \\
    && + \left.  \frac{3}{4}  C^{}_5 Y^\ast_\nu Y^{\rm T}_\nu Y^\ast_\nu - \frac{9}{8} C^{}_5 Y^\ast_l Y^{\rm T}_l Y^\ast_\nu  \right] - \frac{1}{2} \delta Z^{(L=1)}_H \delta G^{(L=1)}_{\ell HN1} - \frac{1}{2} \delta Z^{(L=1)}_\ell \delta G^{(L=1)}_{\ell HN1} 
    \nonumber
    \\
    && - \frac{1}{2}  \delta G^{(L=1)}_{\ell HN1} \delta Z^{(L=1)}_{N_L}
    \nonumber
    \\
    && + \looptf \left[ \left( \frac{45}{16} g^2_1 + 6g^2_2 - \frac{3}{4} \lambda - \frac{9}{8} T \right) C^{}_5 Y^\ast_\nu + g^{}_1 \left( \frac{91}{8} g^2_1 + \frac{27}{8} g^2_2 - \frac{3}{2} T \right) Y^{}_\nu C^\dagger_{BN} \right.
    \nonumber
    \\
    && - 3 \lambda Y^{}_\nu C^\dagger_{HN}  + g^{}_1 \left( \frac{9}{4} Y^{}_l Y^\dagger_l Y^{}_\nu C^\dagger_{BN} + 3 Y^{}_\nu Y^\dagger_\nu Y^{}_\nu C^\dagger_{BN} - \frac{3}{4} Y^{}_\nu C^\dagger_{BN} Y^{\rm T}_\nu Y^\ast_\nu \right) 
    \nonumber
    \\
    && - \frac{3}{4}  Y^{}_l Y^\dagger_l Y^{}_\nu C^\dagger_{HN} - \frac{1}{2}  Y^{}_\nu Y^\dagger_\nu Y^{}_\nu C^\dagger_{HN} + \frac{1}{4} Y^{}_\nu C^\dagger_{HN} Y^{\rm T}_\nu Y^\ast_\nu - \frac{3}{16}  C^{}_5 Y^\ast_l Y^{\rm T}_l Y^\ast_\nu 
    \nonumber
    \\
    && -  \left. \frac{9}{16}  Y^{}_\nu Y^\dagger_\nu C^{}_5 Y^\ast_\nu - \frac{3}{4}  C^{}_5 Y^\ast_\nu Y^{\rm T}_\nu Y^\ast_\nu  \right] \;,
    \nonumber
    \\
    \delta G^{(L=2)}_{\ell HN2} &=&  \looptfs \left\{ \left[  \frac{5}{8} \left( g^2_1 + 3g^2_2 \right) - 3\lambda - \frac{1}{2} T \right] Y^{}_\nu C^\dagger_{HN} + g^{}_1 \left( \frac{19}{2} g^2_1 - \frac{9}{4} g^2_2 \right) Y^{}_\nu C^\dagger_{BN} \right. 
    \nonumber
    \\
    && + g^{}_1 \left[ \left( - \frac{9}{4} Y^{}_l Y^\dagger_l + \frac{3}{2} Y^{}_\nu Y^\dagger_\nu \right) Y^{}_\nu C^\dagger_{BN} + \frac{3}{4} Y^{}_\nu C^\dagger_{BN} Y^{\rm T}_\nu Y^\ast_\nu \right]  + \frac{3}{4}  Y^{}_l Y^\dagger_l Y^{}_\nu C^\dagger_{HN} 
    \nonumber
    \\
    && -  Y^{}_\nu Y^\dagger_\nu Y^{}_\nu C^\dagger_{HN} - \left. \frac{5}{4} Y^{}_\nu C^\dagger_{HN} Y^{\rm T}_\nu Y^\ast_\nu \right\}  - \frac{1}{2} \delta Z^{(L=1)}_H \delta G^{(L=1)}_{\ell HN2} - \frac{1}{2} \delta Z^{(L=1)}_\ell \delta G^{(L=1)}_{\ell HN2} 
    \nonumber
    \\
    && - \frac{1}{2}  \delta G^{(L=1)}_{\ell HN2} \delta Z^{(L=1)}_{N_L}
    \nonumber
    \\
    && + \looptf \left\{ \left[ \frac{9}{16} \left( g^2_1 + g^2_2 \right) + \frac{3}{2} \lambda \right] C^{}_5 Y^\ast_\nu + \left[ - \frac{1}{2} \left( g^2_1 + 3g^2_2 \right) + \frac{3}{2}\lambda + \frac{3}{4} T \right] Y^{}_\nu C^\dagger_{HN} \right.
    \nonumber
    \\
    && + g^{}_1 \left( \frac{97}{48} g^2_1  \right.+ \left. \frac{27}{16} g^2_2 -  \frac{3}{2} T \right) Y^{}_\nu C^\dagger_{BN} + g^{}_1 \left[ \left( \frac{49}{8} Y^{}_l Y^\dagger_l - Y^{}_\nu Y^\dagger_\nu \right) Y^{}_\nu C^\dagger_{BN} \right.
    \nonumber
    \\
    && - \left. \frac{5}{8} Y^{}_\nu C^\dagger_{BN} Y^{\rm T}_\nu Y^\ast_\nu \right] - \frac{3}{8} Y^{}_\nu Y^\dagger_\nu C^{}_5 Y^\ast_\nu - \frac{3}{8}  C^{}_5 Y^\ast_\nu Y^{\rm T}_\nu Y^\ast_\nu   -  \frac{5}{8}  Y^{}_l Y^\dagger_l Y^{}_\nu C^\dagger_{HN} 
    \nonumber 
    \\
    && + \left.  \frac{3}{8} Y^{}_\nu C^\dagger_{HN} Y^{\rm T}_\nu Y^\ast_\nu \right\}  \;.
\end{eqnarray}
Again, the above results are those in the Green's basis and need to be converted into the physical ones by redefining fields before being used to derive RGEs and verify consistency relations.

\end{document}